\documentstyle[12pt,axodraw,graphicx]{article}
\begin{document}
\baselineskip 0.6cm

\def\simgt{\mathrel{\lower2.5pt\vbox{\lineskip=0pt\baselineskip=0pt
           \hbox{$>$}\hbox{$\sim$}}}}
\def\simlt{\mathrel{\lower2.5pt\vbox{\lineskip=0pt\baselineskip=0pt
           \hbox{$<$}\hbox{$\sim$}}}}

\begin{titlepage}

\begin{flushright}
UCB-PTH-07/26
\end{flushright}

\vskip 1.0cm

\begin{center}

{\Large \bf 
Evidence for the Multiverse in the Standard Model \\ and Beyond
}

\vskip 0.6cm

{\large
Lawrence J. Hall and Yasunori Nomura
}

\vskip 0.4cm

{\it Department of Physics, University of California,
           Berkeley, CA 94720} \\
{\it Theoretical Physics Group, Lawrence Berkeley National Laboratory,
           Berkeley, CA 94720} \\

\vskip 0.8cm

\abstract{In any theory it is unnatural if the observed values of 
parameters lie very close to special values that determine the existence 
of complex structures necessary for observers.  A naturalness probability, 
$P$, is introduced to numerically evaluate the degree of unnaturalness. 
If $P$ is very small in all known theories, corresponding to a high 
degree of fine-tuning, then there is an observer naturalness problem. 
In addition to the well-known case of the cosmological constant, we 
argue that nuclear stability and electroweak symmetry breaking represent 
significant observer naturalness problems.  The naturalness probability 
associated with nuclear stability depends on the theory of flavor, 
but for all known theories is conservatively estimated as $P_{\rm nuc} 
\simlt (10^{-3}$~--~$10^{-2})$, and for simple theories of electroweak 
symmetry breaking $P_{\rm EWSB} \simlt (10^{-2}$~--~$10^{-1})$.  This 
pattern of unnaturalness in three different arenas, cosmology, nuclear 
physics, and electroweak symmetry breaking, provides evidence for the 
multiverse, since each problem may be easily solved by environmental 
selection.  In the nuclear case the problem is largely solved even if 
the multiverse distribution for the relevant parameters is relatively 
flat.  With somewhat strongly varying distributions, it is possible 
to understand both the close proximity to neutron stability and the 
values of $m_e$ and $m_d - m_u$ in terms of the electromagnetic mass 
difference between the proton and neutron, $\delta_{\rm EM} \simeq 
1 \pm 0.5~{\rm MeV}$.  It is reasonable that multiverse distributions 
are strong functions of Lagrangian parameters, since they depend not 
only on the landscape of vacua, but also on the population mechanism, 
``integrating out'' other parameters, and on a density of observers 
factor.  In any theory with mass scale $M$ that is the origin 
of electroweak symmetry breaking, strongly varying multiverse 
distributions typically lead either to a little hierarchy, $v/M 
\approx (10^{-2}$~--~$10^{-1})$, or to a large hierarchy, $v \ll M$. 
In certain multiverses, where electroweak symmetry breaking occurs 
only if $M$ is below some critical value, we find that a little 
hierarchy develops with the value of $v^2/M^2$ suppressed by an extra 
loop factor, as well as by the strength of the distribution.  Since 
the correct theory of electroweak symmetry breaking is unknown, our 
estimate for $P_{\rm EWSB}$ is theoretical.  The LHC will lead to 
a much more robust determination of $P_{\rm EWSB}$, and, depending 
on which theory is indicated by the data, the observer naturalness 
problem of electroweak symmetry breaking may be removed or strengthened. 
For each of the three arenas, the discovery of a natural theory would 
eliminate the evidence for the multiverse; but in the absence of 
such a theory, the multiverse provides a provisional understanding 
of the data.}

\newpage

\tableofcontents

\end{center}
\end{titlepage}

\section{Introduction}
\label{sec:intro}

The Standard Model, with three gauge forces and three generations of 
quarks and leptons, has been laboriously constructed from decades of 
data involving collisions of particles at ever higher energies.  It 
represents a triumph for symmetries, but leaves many questions about 
nature unanswered.  These questions fall into three groups.  The first 
revolves around observations of the universe that are not described by 
the Standard Model, such as dark matter, dark energy, baryogenesis 
and inflation.  Secondly there is the question of electroweak symmetry 
breaking: while the Standard Model provides a mathematical description, 
it does not contain the physical description.  The quadratic divergence 
in the Higgs mass-squared parameter implies that electroweak symmetry 
breaking is determined by new physics at a scale $M$ at which the 
Standard Model is embedded into a more fundamental theory.  This 
extraordinary behavior is termed ``unnatural'' --- the Higgs mass-squared 
parameter, $m_h^2$, is very sensitive to changes in some parameter $x$ 
of the fundamental theory, especially as $M$ grows:
\begin{equation}
  \Delta \equiv \frac{\partial \ln m_h^2}{\partial \ln x} 
    \sim \frac{M^2}{m_h^2} \gg 1.
\label{eq:unnat}
\end{equation}
The final group of questions boils down to ``Why the Standard Model?'' 
These include why the forces are the ones we observe, why the quarks and 
leptons have the masses they do, and why there aren't other forces and 
particles.

Since symmetries were the key to the Standard Model, it is natural 
to suppose that all three sets of questions will be answered by 
introducing further symmetries, and this has largely defined beyond 
the Standard Model physics for the last three decades.  Yet there 
are three clouds on the horizon for this symmetry approach, one for 
each set of questions
\begin{itemize}
\item
There is no known symmetry to explain why the cosmological constant is 
either zero or of order the observed dark energy.  It appears to take 
a very unnatural value.
\item
For the Standard Model to be natural, $M$ must be low; we should 
already have observed either signs of supersymmetry or signals of 
new interactions in precision electroweak observables.
\item
Some of the Standard Model parameters take values in a special region 
that happen to yield certain nuclear properties.
\end{itemize}

It is possible that these clouds are the first indications of 
a breakdown in the power of symmetry to determine fundamental physics. 
If our universe is just one among very many in an enormous multiverse, 
then observed universes will be those that contain certain complex 
structures necessary for observers~\cite{Barrow-Tipler}.  Such 
arguments from environmental selection can potentially solve the 
cosmological constant problem, and yield a statistical prediction 
for the dark energy in observed universes~\cite{Weinberg:1987dv}. 
In this paper we consider the extent to which nuclear stability 
and electroweak symmetry breaking provide evidence for environmental 
selection.

Many physicists, however, are reluctant to countenance any form of 
anthropic argument.  Why give up on the traditional, extraordinarily 
successful, methods of physics?  How can we hope to understand the 
conditions for intelligent observers when we struggle even to define 
what life is?  Why make the extraordinary leap of postulating an 
extra-horizon multiverse, which has the smell of a secular form of 
God?  In short, many believe that appeals to the environment are an 
escape from true science and that, in the absence of data confirming 
the conventional symmetry approach, it would be better to change 
fields than to succumb to the philosophy of anthropics.

The case of the cosmological constant demolishes these arguments. 
Traditional methods have not given any satisfactory understanding 
for why the cosmological constant is small.  In contrast, the 
environmental argument not only explains why it must be small, 
but makes a statistical prediction for a non-zero value.  Furthermore, 
this prediction requires essentially no understanding of what it 
takes to make an observer.  Of course, the prediction does require 
a multiverse; but with a landscape of vacua provided by string 
theory~\cite{Bousso:2000xa} and generic inflation eternally creating 
universes~\cite{Vilenkin:1983xq}, a multiverse no longer appears 
theoretically unreasonable, and its consequences can be explored under 
the assumption that we are typical observers~\cite{Vilenkin:1994ua}. 
Indeed, since few doubt that environmental conditions explain why 
we find ourselves on the Earth as opposed to elsewhere in the solar 
system, the resistance to environmental reasoning is hard to understand. 
As for any physical theory, the real issue is whether sufficient 
evidence for it can be found to make it convincing.

Even if there is a multiverse, and environmental selection accounts 
for the value of the cosmological constant, it is far from clear what 
the implications are for the rest of physics.  One extreme possibility 
is that all the other questions left open by the Standard Model will 
be solved within some symmetry framework, for example supersymmetric 
$SO(10)$ with flavor symmetries.  In this case, the environmental 
solution of the cosmological constant problem simply allows us to 
ignore this special problem while solving everything else with symmetries 
--- it justifies the model-building approach of the last 30~years. 
The other extreme is that the Standard Model is as far as we can get 
with symmetries, all of its parameters are either strongly selected 
by environmental effects or just reflect random typical values in 
the multiverse, and similarly for issues outside the Standard Model. 
Of course there are many intermediate cases between these two extremes. 
For example, there could be gauge coupling unification within an 
$SO(10)$ theory, but the flavor sector may involve sufficient scanning 
parameters that the electron, up and down quark masses are determined 
by environmental selection.  The aim of this paper is to seek evidence 
for environmental selection both in the measured values of Standard 
Model parameters, and in electroweak symmetry breaking.

The cosmological constant illustrates an additional crucial point: 
environmental selection predicts parameters that are, from the 
conventional viewpoint of symmetries, {\it unnatural}.  This is a key 
point, because so much effort of the last 30~years has been expended 
in trying to understand how new symmetries could naturally lead to 
an unnatural effective low energy theory, namely the Standard Model. 
Perhaps the unnaturalness is a hint that symmetries are not the answer. 
It was realized some time ago that if only the mass parameter of the 
Standard Model scans, then environmental selection could solve the gauge 
hierarchy problem~\cite{Agrawal:1997gf}.  In the context of supersymmetry, 
differing assumptions about scanning parameters and their distributions 
could lead to a large hierarchy~\cite{ArkaniHamed:2004fb} or a small 
hierarchy~\cite{Giudice:2006sn}.  These papers both illustrate that 
environmental selection leads to unnaturalness, and that measurements 
at the LHC could experimentally confirm the presence of unnaturalness 
in electroweak symmetry breaking.  It is an open question whether 
the landscape favors a low or high supersymmetry breaking 
scale~\cite{Banks:2003es}.

The first part of this paper, sections~\ref{sec:def-natural} to 
\ref{sec:flavor}, are devoted to a consideration of the concept of 
naturalness, and its application to issues of nuclear stability.  We 
are motivated by the belief that environmental selection may lead to 
precise predictions, and that it is important to evaluate the numerical 
significance of such predictions.  We often hear it said that, in 
nuclear physics and elsewhere, nature exhibits ``amazing coincidences'' 
that give rise to life.  How can these be precisely evaluated?  In 
section~\ref{sec:def-natural} we formulate the concept of naturalness 
in a very general way, so that in section~\ref{sec:new-natural} we are 
able to identify naturalness problems associated with the existence 
of complex structures in the universe.  In section~\ref{sec:evid-prob} 
we study the stability of neutrons, deuterons and complex nuclei in 
the parameter space of the Standard Model, identifying how close our 
universe is to these stability boundaries.  In particular, we find 
a closeness to the neutron stability boundary that has not previously 
been elucidated.  The naturalness problems associated with the closeness 
to these stability boundaries are evaluated numerically in a variety 
of theories of flavor in section~\ref{sec:flavor}.  We stress that 
the evidence for unnaturalness depends on the ensemble of theories 
being considered, but we find that there is an irreducible amount 
of unnaturalness no matter what the theory of flavor.

In the second half of the paper, we investigate the consequences 
that follow if this unnaturalness arises from environmental selection 
in a multiverse.  It is here that the real utility of our new formulation 
of naturalness becomes apparent.  In section~\ref{sec:sol-ES} we argue 
that problems of unnaturalness are indeed solved by environmental 
selection, and show how the amount of unnaturalness is connected to 
the probability distribution of the multiverse.  We argue that evidence 
for environmental selection will increase if symmetry arguments fail 
to solve an accumulation of naturalness problems.  In the case of 
nuclear stabilities, we show in section~\ref{sec:u-d-e} how this leads 
to multiverse predictions for the masses of the electron, up and down 
quarks.  In section~\ref{sec:EWSB} we embed the Standard Model Higgs 
sector in a generic theory of electroweak symmetry breaking at mass 
scale $M$.  Allowing parameters of this sector to scan, as well as 
parameters of the Standard Model, we consider environmental selection 
from nuclear stability boundaries.  We find that sharply varying 
multiverse distribution functions generically lead to both large 
and little hierarchies between $M$ and the scale $v$ of electroweak 
symmetry breaking.  From the viewpoint of the multiverse, a discovery 
at the LHC that electroweak symmetry breaking requires some amount 
of fine-tuning would not be surprising.  A similar analysis on 
electroweak symmetry breaking is performed in section~\ref{sec:EWSB-alt}, 
assuming that the relevant boundaries are the phase boundary of 
electroweak symmetry breaking rather than the nuclear stability 
boundaries.  We again find that unnaturalness in electroweak symmetry 
breaking is expected for sharply varying multiverse distribution 
functions.  In section~\ref{sec:LH-cc} we study connections between 
environmental selection for the cosmological constant and for 
electroweak symmetry breaking.  For example, we argue that the 
driving force on the multiverse for unnaturalness in electroweak 
symmetry breaking could arise from the distribution for the 
cosmological constant, the connection being through weakly 
interacting massive particle (WIMP) dark matter.  Finally, 
our conclusions are given in section~\ref{sec:concl}.

\section{New Definition of Naturalness Problems}
\label{sec:def-natural}

The concept of naturalness has often been the driving force for finding 
fundamental mechanisms in nature.  Arguments for naturalness are often 
phrased in terms of the sensitivity of low energy Standard Model parameters 
$c_i$ to variations of the parameters in the more fundamental theory 
$a_j$.  A simple measure of naturalness is then given by $\Delta \equiv 
|\partial \ln c_i/ \partial \ln a_j|$, with a large value of $\Delta$ 
signaling a lack of naturalness~\cite{Barbieri:1987fn}.  This definition, 
however, could miss some of the important aspects of the naturalness 
problem.  Here we introduce a new, quantitative definition of naturalness 
that can be applied in much more general situations.  In particular, 
this allows us to identify certain classes of naturalness problems that 
have not been quantitatively defined.  Our definition is also free from 
some of the problems existing in the simplest definition of naturalness 
based on the sensitivity of parameters.

First of all, it is very important to notice that in talking about 
naturalness, we are dealing, either explicitly or implicitly, with an 
ensemble in which parameters of the theory are varied according to some 
definite distribution.  Consider, for example, that the Higgs mass-squared 
parameter, $m_h^2$, is given by the difference of two mass scales of order 
the fundamental scale, $m_h^2 = M_1^2 - M_2^2$.  We say that the theory 
is unnatural if $|m_h^2| \ll M_1^2, M_2^2$.  This statement, however, 
already assumes that it is unlikely for $M_1$ and $M_2$ to be very close 
or for both to be very small; more specifically, the distribution of 
possible values for $M_1$ and $M_2$ is assumed to be almost structureless 
in the $M_1$-$M_2$ (or $M_1^2$-$M_2^2$) plane.  This illustrates that 
the concept of naturalness is closely related to the distribution of 
parameters in an ensemble.

{\it A particular member of an ensemble is unnatural if it has parameters 
very close to special values that are not explained by the symmetries 
of the theory.}  The parameters are special if some physical property 
arises that is {\it not} a generic feature of the members in the 
ensemble, or if they separate regions of parameter space that have 
differing generic features.  Depending on the property considered, we 
encounter various classes of naturalness problems, whose solutions could 
point to various different mechanisms in nature.  With this definition, 
the degree of unnaturalness is given by how close the parameters are to 
the special values, which generically form a special hypersurface in 
multi-dimensional parameter space.  As we will see below, we can quantify 
this degree in terms of the distribution of parameters within the ensemble. 
In this section, after presenting a new definition of naturalness, we 
apply it to well-known situations.  In the next section we use it to 
introduce new types of naturalness problems.

\subsection{Definition}
\label{subsec:def}

Let us start describing our precise definition of naturalness by 
introducing the distribution function $f(x)$ for an ensemble.  The 
function is defined such that the number of members with parameters 
$x_i$ ($i=1,\cdots,N$) taking a value between $x_i$ and $x_i+dx_i$ 
is given by
\begin{equation}
  d{\cal N} = f(x_1,x_2,\cdots,x_N)\, dx_1 dx_2 \cdots dx_N.
\label{eq:dN}
\end{equation}
Here, $x_i$ represent continuous parameters of the theory, e.g. masses 
and coupling constants.%
\footnote{In principle, we can define a distribution function that also 
 has discrete labels representing, e.g., particle content and symmetries 
 of the theories.  Here we consider a distribution function for each 
 theory that has a definite symmetry, matter content, number of spacetime 
 dimensions and so on, and restrict its arguments $x_i$ to be continuous 
 parameters associated with that particular theory.}
The overall normalization of $f$ becomes relevant if we are interested 
in the total number of members, which will be the case when we discuss 
relative likelihoods of different theories.  For the present purpose 
of discussing naturalness of a given theory, however, the overall 
normalization of $f$ is not important.

The distribution function depends on the choice of $x_i$.  What variables 
should we choose as $x_i$?  In general we can choose any variables as 
$x_i$, depending on the context.  To discuss naturalness of the low energy 
theory, however, it is often most convenient to take ``fundamental'' 
parameters, such as masses and coupling constants of the ultraviolet 
theory, as $x_i$.  Now, suppose we have only one such variable $x$. 
If the observed value of $x$, $x_o$, is very close to a special value 
$\bar{x}$, the level of unnaturalness is given by the following 
``naturalness probability''
\begin{equation}
  P = \left| \frac{\int_{\bar{x}}^{x_o}\! f(x)\, dx}
    {\int^{x_{\rm max}}_{x_{\rm min}}\! f(x)\, dx} \right|,
\label{eq:P_xo}
\end{equation}
where $x_{\rm min} \leq x \leq x_{\rm max}$ gives the range of $x$ values 
in the theory under consideration.%
\footnote{With $f(x) = 0$ for $x < x_{\rm min}$ and $x > x_{\rm max}$, the 
 range of the integration in the denominator can be taken from $-\infty$ 
 to $+\infty$.  Practically, if $f(x)$ has a sharp drop-off, it is useful 
 to restrict the range of the variable by a sharp cutoff, as $x_{\rm min}$ 
 and $x_{\rm max}$ in Eq.~(\ref{eq:P_xo}).}
Here, we have included only members on one side of the special point, 
but depending on the situation, a factor of $2$ should be added to the 
numerator to include members on both sides of the special point.  With 
this definition, $P \ll 1$ signals the existence of a naturalness problem.

The form of the distribution function is modified if we redefine the 
variable $x$.  For example, if $f(x) = 1/x$ for a variable $x$, the 
change of the variable $x' = \ln x$ can make the distribution function 
flat for $x'$: $f(x') = 1$.  It is often convenient to go to the basis 
in which the distribution function is constant.  In that basis, the 
naturalness probability of Eq.~(\ref{eq:P_xo}) becomes
\begin{equation}
  P = \left| \frac{x_o - \bar{x}}{x_{\rm max} - x_{\rm min}} \right|,
\label{eq:P_xo-2}
\end{equation}
which is simply given by the distance between $x_o$ and $\bar{x}$ 
divided by the available parameter space.%
\footnote{This interpretation is also possible for a general distribution 
 function $f(x)$ if we consider $f(x)$ to be a sort of metric, or volume 
 factor, in parameter space.}
As should be the case, this quantity becomes smaller as $x_o$ approaches 
$\bar{x}$.

The naturalness probability of Eqs.~(\ref{eq:P_xo},~\ref{eq:P_xo-2}) 
can be extended to the case of multiple parameters $x_i$.  The precise 
definition depends on the dimensionality of $x_i$ and the special 
surface consisting of $\bar{x}_i$.  If the co-dimension of the special 
surface is $1$, we can choose $x$ to be a linear combination of 
$x_i$ perpendicular to the surface, and then use the definition of 
Eq.~(\ref{eq:P_xo-2}) (in the basis where the distribution function 
is constant).  In the case that the co-dimension is higher, we must use 
an appropriate generalization of Eqs.~(\ref{eq:P_xo},~\ref{eq:P_xo-2}) 
defined using multi-dimensional volumes, rather than simple distances. 
A useful definition, in the basis where $f(x_i)$ is constant, is
\begin{equation}
  P = \left| \frac{c_n \{\Sigma_{a=1}^{n} (x_{a,o} - \bar{x}_a)^2\}^{n/2}}
    {\Pi_{a=1}^{n} (x_{a,{\rm max}} - x_{a,{\rm min}})} \right|
  = \frac{v_n}{V_n},
\label{eq:P_multi}
\end{equation}
where $a = 1,\cdots,n$ runs over variables whose observed values, 
$x_{a,o}$, are close to the special values, $\bar{x}_a$, and $c_n = 
\pi^{n/2}/\Gamma(n/2+1)$ is the volume of the unit ball in $n$ dimensions. 
Here, the expression in the numerator, $v_n$, is a measure of the 
accidentally small volume of parameter space required for $x_{a,o}$ 
to be close to $\bar{x}_a$, while that in the denominator, $V_n$, is 
the total volume of parameter space available.  The restriction of 
the variables $x_a$ to those with $|(x_{a,o} - \bar{x}_a)/(x_{a,{\rm max}} 
- x_{a,{\rm min}})| \ll 1$, e.g. $\leq 1/5$, is important to avoid obtaining 
$P \ll 1$ simply as a result of high dimensionality of the parameter space. 
In the case that the special surface in question arises as an intersection 
of co-dimension~1 surfaces, we may restrict the volume of the numerator 
to be one side of the co-dimension~1 surfaces, depending on the situation. 
This definition reduces to that of Eq.~(\ref{eq:P_xo-2}) in the special 
case of $n=1$.

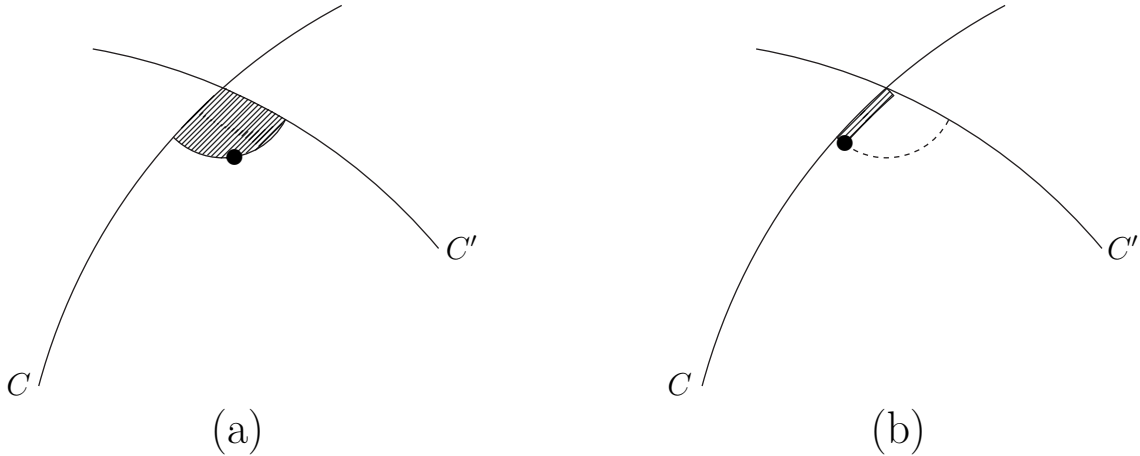
\begin{figure}[t]
\begin{center}
\begin{picture}(450,168)(1,-38)
%
%
  \Text(100,-19)[t]{\Large (a)}
  \CArc(247,-70)(230,118,165) \Text(22,-10)[r]{$C$}
  \CArc(7,-100)(220,40,80) \Text(179,42)[l]{$C'$}
  \CArc(94.3,102)(26.4,224.5,332.5)
  \put(99,76.3){\circle*{6}}
  \put(79.0,87.0){\line(1,1){12.3}}
  \put(77.0,83.0){\line(1,1){18.8}}
  \put(78.0,82.0){\line(1,1){19.2}}
  \put(79.1,81.1){\line(1,1){19.5}}
  \put(80.3,80.3){\line(1,1){19.6}}
  \put(81.5,79.5){\line(1,1){19.8}}
  \put(82.8,78.8){\line(1,1){19.8}}
  \put(84.2,78.2){\line(1,1){19.8}}
  \put(85.6,77.6){\line(1,1){19.7}}
  \put(87.1,77.1){\line(1,1){19.5}}
  \put(88.8,76.8){\line(1,1){19.2}}
  \put(90.4,76.4){\line(1,1){18.9}}
  \put(92.1,76.1){\line(1,1){18.5}}
  \put(94.0,76.0){\line(1,1){17.9}}
  \put(96.0,76.0){\line(1,1){17.2}}
  \put(98.2,76.2){\line(1,1){16.3}}
  \put(100.5,76.5){\line(1,1){15.3}}
  \put(103.5,77.5){\line(1,1){13.6}}
  \put(107.0,79.0){\line(1,1){11.3}}
%
%
  \Text(350,-19)[t]{\Large (b)}
  \CArc(497,-70)(230,118,165) \Text(272,-10)[r]{$C$}
  \CArc(257,-100)(220,40,80) \Text(429,42)[l]{$C'$}
  \DashCArc(344.3,102)(26.4,224.5,332.5){2}
  \Line(344.3,102)(347.2,98.9) \Line(325.8,83.4)(328.7,80.7)
  \CArc(497,-70)(226,131.6,138.2)
  \put(329.9,81.2){\circle*{6}}
  \put(327.5,83.5){\line(1,1){18.3}}
  \put(328.5,82.5){\line(1,1){18.3}}
  \put(329.6,81.6){\line(1,1){18.2}}
\end{picture}
\caption{Illustrations of the definition of the naturalness 
 probability $P$ in multi-dimensional parameter space.}
\label{fig:def}
\end{center}
\end{figure}

The definition of Eq.~(\ref{eq:P_multi}) is illustrated for a 2-dimensional 
parameter space $(x_1,x_2)$ in Fig.~\ref{fig:def}(a). 
The curve $C$ represents special values of these parameters corresponding 
to some physical phenomenon, and the curve $C'$ similarly represents 
special values for some other phenomenon.  If the observed values 
$(x_{1,o}, x_{2,o})$, denoted by the dot, are close to the intersection 
point $(\bar{x}_1, \bar{x}_2)$, the naturalness probability is given by 
Eq.~(\ref{eq:P_multi}), with $a=1,2$.  The numerator represents the area 
of the shaded region, and the denominator represents a much larger area 
corresponding to the range of the parameters.  Here, we have restricted 
the area of the numerator to one side (the lower side) of the curves 
$C$ and $C'$, anticipating an application in later sections.

In Fig.~\ref{fig:def}(b), it is apparent that the observed point is much 
closer to $C$ than $C'$; but the definition of Eq.~(\ref{eq:P_multi}) does 
not capture the additional unnaturalness associated with this.  (The area 
of the numerator in this expression is represented by the dotted line.) 
In this situation, we must first consider the naturalness probability 
$P_C$ associated with $C$, using the definition of Eq.~(\ref{eq:P_xo-2}), 
where the axis $x$ is taken normal to $C$ and passing through the observed 
point at $x = x_o$.  We then take the coordinate $x'$ along the curve 
$C$ and consider the naturalness probability $P_{C'}$ associated with 
$C'$ in this coordinate.  The resulting naturalness probability is then 
$P = P_C P_{C'}$, which is given by the area of the shaded region in 
the figure divided by that corresponding to the range of the parameters. 
This tends to zero as the observed point approaches the curve $C$, 
as expected.  In general, if we find several unnatural features that 
have a hierarchy in their degrees of unnaturalness, we can obtain the 
correct estimate for the naturalness probability $P$ by considering 
it as a product of several naturalness probabilities, each associated 
with an unnaturalness of some fixed degree.  The decomposition of $P$ 
can be made along the lines presented here.

\subsection{Illustrations}
\label{subsec:illust}

\begin{figure}[t]
  \center{\includegraphics[width=.5\textwidth]{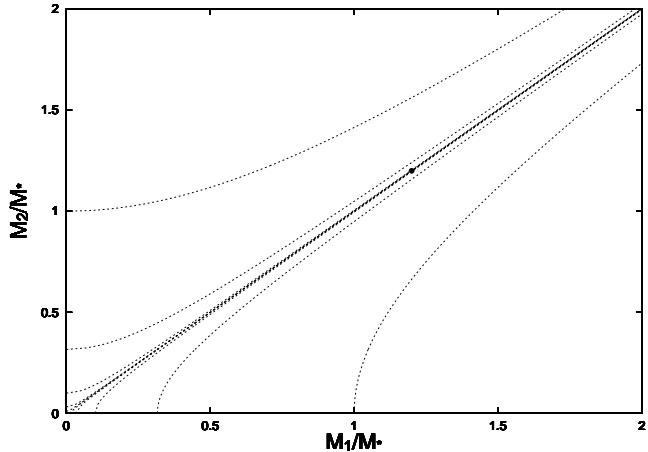}}
\caption{The contour of $m_h^2/M_*^2 = \pm 1, \pm 0.1, \pm 0.01, \cdots$ 
 in the $M_1$-$M_2$ plane.  The special line of $m_h^2 = 0$ is visible 
 at $M_1 = M_2$.  The observed value of $M_2 = M_1 + O(10^{-32})$ is 
 denoted by the little dot for an arbitrary value of $M_1 = 1.2 M_*$.}
\label{fig:v-1}
\end{figure}

Let us now illustrate the use of our definition in the case of the 
conventional gauge hierarchy problem.  Consider that the Standard Model 
is embedded into the fundamental theory at a high scale of $M_1 \approx 
M_2 \approx M_*$ and that the Higgs mass-squared parameter is given in 
terms of $M_1$ and $M_2$ by $m_h^2 = M_1^2 - M_2^2$.  In general, we 
expect that the distribution function is roughly flat in terms of $M_i$ 
($i=1,2$) and that the range of parameters is $0 \leq M_i \leq O(M_*)$.%
\footnote{Here, we have assumed that $M_i$ are mass parameters associated 
 with ``fermions'' (including mass parameters in supersymmetric theories). 
 The argument, however, is not affected if they are associated with 
 scalars, in which case the distribution is expected to be roughly 
 flat in $M_i^2$ with the range $-O(M_*^2) \leq M_i^2 \leq O(M_*^2)$.}
In Fig.~\ref{fig:v-1}, we plot the contour of $m_h^2$ in the $M_1$-$M_2$ 
plane in units of $M_*$. 
We clearly see that there is a special line in this plane, $M_1 = M_2$, 
where $m_h^2 = 0$.  The physical phenomenon making this line special is 
that members very close to this line have a low energy effective field 
theory that contains a scalar excitation whose mass is hierarchically 
smaller than $M_*$.  For $M_* \approx 10^{18}~{\rm GeV}$, the observed 
value of $m_h^2 \simeq -O(100~{\rm GeV})^2$ is denoted by the little dot 
for an arbitrary value of $M_1 = 1.2 M_*$.  We find that it is located 
very close to the special line compared with the expected range of 
$M_i$.  In fact, this is always the case regardless of the value of 
$M_1$ we choose, and is a manifestation of the lack of naturalness 
in the Standard Model.  The problem is that the special line with 
$m_h^2 = 0$ is {\it not} special in terms of the symmetry structure 
of the theory.  The points with $m_h^2 = 0$ are no more symmetric than 
those with $m_h^2 \neq 0$.  By choosing a variable $x$ to be a linear 
combination of $M_1$ and $M_2$ perpendicular to the special line, and 
then using the definition of Eq.~(\ref{eq:P_xo-2}), we obtain $P \approx 
10^{-32}$.  A similar analysis can also be made for the cosmological 
constant problem.  In general, the cosmological constant $\Lambda$ is 
given by the sum of terms of order $M_*^4$ taking the form of $M_i^2 M_j^2$. 
The observed value of $\Lambda$ is then very close to the special 
hypersurface with $\Lambda = 0$, and the points with $\Lambda = 0$ 
are no more symmetric than those with $\Lambda \neq 0$.  The physical 
property making this hypersurface special is that members very close 
to this hypersurface allow an arbitrary large observable universe. 
From Eq.~(\ref{eq:P_xo-2}), the degree of unnaturalness we obtain in 
this case is $P \approx 10^{-120}$.

Here we note that our definition does not suffer from the problem existing 
in the simplest definition of naturalness based on the logarithmic 
derivative, $\Delta \equiv |\partial \ln c_i/ \partial \ln a_j|$. 
Suppose that a dimensionful parameter $\mu$ is given by two dimensionless 
constants $g^2$ and $b$ as $\mu = M_*\, e^{-8\pi^2/g^2 b}$, where $M_*$ 
is the fundamental scale, and that a natural range for $g^2$ and $b$ in 
the fundamental theory is $g^2 b = O(0.1)$.  We then naturally obtain 
$\mu/M_* = e^{-O(1000)}$; there is nothing unnatural with this.  The 
simplest definition, however, gives $\Delta = |\partial \ln \mu/\partial 
\ln g^2| = O(1000)$, signaling (incorrectly) the existence of unnaturalness. 
Our definition does not lead to such a fake signal, since any point with 
$g^2 b = O(0.1)$ is generic --- there is no special value of $g^2 b$ that 
can be singled out as $\bar{x}$ in Eq.~(\ref{eq:P_xo},~\ref{eq:P_xo-2}).

\begin{figure}[t]
  \center{\includegraphics[width=.5\textwidth]{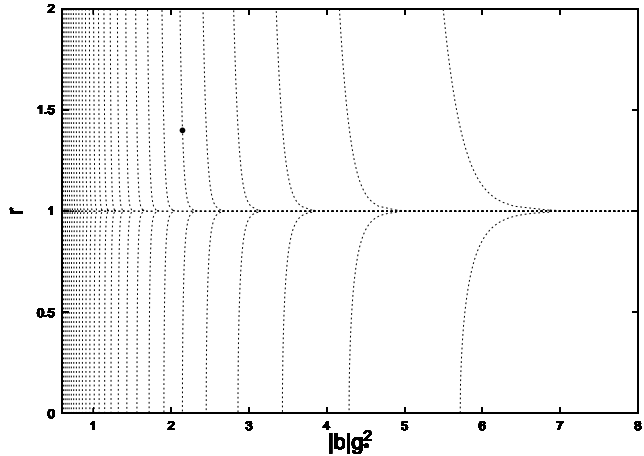}}
\caption{The contour of $m_h^2/M_*^2 = 10^{-4n}$ with $n=3,4,5,\cdots$ in 
 the $|b|g_*^2$-$r$ plane.  The observed value of $v \approx \sqrt{-m_h^2} 
 \approx 100~{\rm GeV}$ can correspond to a completely generic point 
 in the plane, which is indicated by the little dot for an arbitrary 
 value of $r = 1.4$.}
\label{fig:v-2}
\end{figure}

We now see how theories beyond the Standard Model solve the gauge 
hierarchy problem in our language.  As we have seen, the Standard 
Model embedded into more fundamental theory at $M_* \approx 
10^{18}~{\rm GeV}$ leads to the probability of having a small 
weak scale $v \approx 100~{\rm GeV}$ to be $P \approx (v/M_*)^2 
\approx 10^{-32}$.  This is because the Higgs mass-squared parameter, 
$m_h^2$, is given by the sum of various contributions $M_i^2$ of 
order $M_*^2$, with the distribution function expected to be roughly 
flat in $M_i$ (or $M_i^2$).  The distribution function, however, varies 
from one theory to another.  For example, if the weak scale arises 
from dimensional transmutation associated with some gauge coupling 
$g$ becoming strong, as in technicolor~\cite{Weinberg:1975gm} and 
certain supersymmetric theories~\cite{Witten:1981nf}, then $v \approx 
M_* \exp(-8\pi^2/|b|g_*^2)$.  Here, $b$ is the one-loop beta function 
coefficient for the new gauge force and $g_*=g(M_*)$.  The distribution 
function is expected to be roughly flat in the variable $g_*^2$,%
\footnote{The conclusion below does not change if the distribution 
 is flat in $1/g_*^2$, $g_*$, and so on.  The distribution, however, 
 could affect arguments on certain naturalness problems; see 
 section~\ref{sec:evid-prob}.}
giving an approximately flat distribution for $1/\ln(v/M_*)$.  To 
illustrate this situation, we plot in Fig.~\ref{fig:v-2} the contour 
of $m_h^2 = M_1^2 - M_2^2$ with $M_1 = M_* \exp(-8\pi^2/|b|g_*^2)$ 
and $M_2/M_1 \equiv r$ in the $|b|g_*^2$-$r$ plane. 
This can be regarded as a simplified model for the situation in, e.g., 
dynamical supersymmetry breaking.  Typically, there is considerable 
uncertainty in the range of $|b|g_*^2$.  Here we take $0 \leq |b|g_*^2 
\leq O(10)$.  As is seen from the figure, the observed value of $v$ can 
correspond to a completely generic (not special) point in the plane, 
which is indicated by the little dot for an arbitrary value of $r = 1.4$. 
This, therefore, provides a solution to the gauge hierarchy problem; 
see Fig.~\ref{fig:v-1} for comparison.  The important point here is that 
different theories can lead to very different distribution functions 
$f(x)$ for physical parameters $x_i$.  A distribution flat in $v^2$ is 
the origin of the gauge hierarchy problem, while one flat in $\ln v$ 
(or $1/\ln v$) provides the solution.  The situation with the cosmological 
constant is similar, although the problem is much more severe because 
all known theories give distributions that are essentially flat in 
$\Lambda$, while none are known that are flat in $\ln \Lambda$ 
(or $1/\ln \Lambda$).

\subsection{Generality of the framework}
\label{subsec:gen}

So far, we have illustrated our definition of naturalness by looking 
at naturalness problems that have a conventional description: the 
observed values of parameters are (very) close to special values where 
the sensitivity of a low energy parameter to high energy ones diverges, 
for example $\partial \ln m_h^2/ \partial \ln M_i, \partial \ln\Lambda/ 
\partial \ln M_i \rightarrow \infty$.  This is one class of naturalness 
problems, which we may call ``fine-tuning naturalness problems.''  The 
real power of our formalism, however, lies in the fact that we can discuss 
many different types of naturalness problems {\it in a unified manner}, 
simply by extending what we mean by ``special values'' for parameters. 
Any physical property that is {\it not} a generic feature of the members 
in the ensemble is a candidate for identifying special values.  There 
can be many classes of naturalness problems, depending on the property 
considered.  The closeness of the observed values to the special values, 
signaled by $P \ll 1$, can then be used as evidence for a new mechanism 
to understand such accidents.

One application of this general idea is to use naturalness arguments 
as evidence for the presence of some symmetry beyond the Standard Model. 
For this purpose it is often convenient to take parameters of the low energy 
theory to be $x_i$.  Imagine that the observed values $x_{i,o}$ of $N$ 
dimensionless Standard Model parameters $x_i$ ($i=1,\cdots,N$) take values 
close to the special surface that defines a symmetry relation
\begin{equation}
  S(x_i) = 0.
\label{eq:S}
\end{equation}
If the symmetry were in fact absent, nature would be described by some 
member of an ensemble giving the $x_i$ parameters distributed according 
to $f(x_i)$, which is not generically concentrated on the surface of 
Eq.~(\ref{eq:S}).  After redefining the parameters to make the distribution 
function flat, we can introduce an axis $x$ normal to the symmetry surface 
that passes through the observed point at $x_i = x_{i,o}$.  The degree 
of unnaturalness is then given by Eq.~(\ref{eq:P_xo-2}).  The closer the 
observed point is to the special value $\bar{x}$ on the symmetry surface, 
the less probable it becomes that nature is described by this ensemble, 
i.e. by the theory without the symmetry.  A small value for $P$ can thus 
be taken as evidence that the symmetry is indeed present in nature, at 
least in an approximate form.%
\footnote{Here we assume that conversions from low energy parameters 
 to high energy ones, e.g. through renormalization group evolution, 
 do not significantly affect the size of $P$.  This is generically 
 a good assumption.}
It could be that $x_o - \bar{x}$ is dominated by the experimental 
uncertainty.  In this case improved experiments may provide further 
evidence for the symmetry.  A well known example is provided by a grand 
unified symmetry, such as $SU(5)$, that gives a symmetry relation between 
the Standard Model gauge coupling constants $S(g_1^2, g_2^2, g_3^2) 
= 0$~\cite{Georgi:1974sy}.  If in fact there is no unified symmetry, 
the observed values of $\alpha$, $\alpha_s$ and $\sin^2\!\theta_W$ 
give a naturalness probability of $P \approx 0.1$ in an ensemble 
without supersymmetry, and $P \approx 0.01$ in an ensemble with 
supersymmetry~\cite{Dimopoulos:1981zb}.  Here, $\alpha$, $\alpha_s$ 
and $\theta_W$ are the fine structure constant, the effective QCD 
coupling, and the Weinberg angle, respectively.  This lack of naturalness 
can then be regarded as evidence for some form of a unified symmetry 
in theories with supersymmetry.

In the next section we introduce classes of naturalness problems that 
arise from the existence of complex structures.  We will see that the 
formalism developed here elucidates the identification of these naturalness 
problems.  As in other naturalness problems, successful solutions to 
these problems could lead us to find new mechanisms or dynamics in nature, 
which we consider in later sections.

\section{Complexity and Observer Naturalness Problems}
\label{sec:new-natural}

The definition of naturalness introduced in the previous section allows us 
to identify new classes of naturalness problems.  A member in the ensemble 
is unnatural if it has parameters unusually close to ``special'' values; 
but clearly there are many reasons that parameters could be ``special.'' 
We frequently stress special values that lead to a large hierarchy of mass 
scales, but in this section we consider special values that lead to the 
existence of relatively long-lived complex structures, such as nuclei, 
stars and galaxies.

\subsection{Complexity naturalness problem}
\label{subsec:complexity}

As the parameters $x_i$ vary, moving from one member of an ensemble to 
another, suppose that a physical threshold is crossed that is crucial 
for the existence of some complex structure.  This defines a special 
surface in the parameter space
\begin{equation}
  C(x_i) = 1,
\label{eq:C}
\end{equation}
that divides the volume of parameter space into two regions, one that 
supports the complex structure, $C(x_i) < 1$, and one that does not, 
$C(x_i) > 1$.  In general this surface is not one with enhanced symmetry. 
Therefore, a member in the ensemble having parameters unusually close 
to this surface has a ``complexity naturalness problem.''  The degree 
of unnaturalness is given numerically by Eq.~(\ref{eq:P_xo}), where 
the single variable $x$ is normal to the ``complexity surface'' of 
Eq.~(\ref{eq:C}), where it takes the value $\bar{x}$, and the particular 
unnatural member in question has a nearby value for this parameter, $x_o$.

One caveat is that we have assumed that the physical threshold relevant 
for the existence of the complex structure is sharp in the parameter space. 
This can be verified in any particular case, and is certainly true in the 
examples discussed in the next section.  For example, the stability of any 
particular nucleus gives a sharp boundary corresponding to values of the 
coupling strengths and quark masses that lead to a surface of zero binding 
energy.  In general a lack of sharpness is due to the time evolution of 
complex structures in universes corresponding to the different members 
of an ensemble.  As parameters vary from one member to another, complex 
structures could gradually become less stable.  For example, the stability 
of large scale structure is not completely sharp --- as the cosmological 
constant is gradually increased only the regions with larger statistical 
fluctuations in the density perturbations are able to collapse.  Still, 
the relevant parameter space for the cosmological constant spans over 
100~orders of magnitude, and over this space the transition for the 
existence of large scale structure is very sharp.

Another caveat is the hidden assumption that these complexity surfaces 
are not distributed so densely throughout the entire parameter space that 
a typical member in the ensemble is expected to be close to one or more 
surfaces.   There may indeed be many complexity surfaces; for example, 
in the Standard Model there are several hundred relatively stable nuclei, 
each with its own complexity surface in an ensemble that contains the 
Standard Model.  However, if the relevant parameters --- the Yukawa 
couplings, $y$, the weak scale, $v$, and the QCD scale, $\Lambda_{\rm QCD}$ 
--- vary by many orders of magnitudes in the ensemble, then these complexity 
surfaces will all be tightly clustered in a ``complexity zone,'' $y v \sim 
\alpha \Lambda_{\rm QCD} \sim O(0.01) \Lambda_{\rm QCD}$.  A complexity 
naturalness problem now arises because this zone is itself small compared 
with the entire volume of parameter space.  Most members in the ensemble 
lie in voids far from the complexity zone, and the closeness of the observed 
parameters to one or more of the complexity surfaces implies that the 
member describing our universe lies in a special region.  There may be 
several complexity zones for nuclear physics; for example ones with four, 
five or six quark flavors lighter than the QCD scale, but in certain 
parameter directions each zone will be small.   In the case of a single 
variable, the degree of unnaturalness can be taken to be $P = \Delta 
x_{\rm z}/(x_{\rm max} - x_{\rm min})$ where $\Delta x_{\rm z}$ is the 
width of the complexity zone.  If the density of complexity surfaces is 
very high it might be that a member in the complexity zone is very close 
to some surface; but this could be a reflection of the density of the 
surfaces rather than any additional unnaturalness beyond that of being 
in the complexity zone.

In summary, a theory possesses a complexity naturalness problem if, in 
the parameter space of an ensemble, the member describing our universe 
lies very close to a surface corresponding to a physical threshold that 
allows the existence of some relatively long-lived complex structure. 
For the problem to exist, there should not be many surfaces of a similar 
character distributed densely and almost uniformly over the parameter 
space; otherwise, the closeness to one of these surfaces would simply 
be a generic phenomenon for the members in the ensemble.

\subsection{Observer naturalness problem}
\label{subsec:observer}

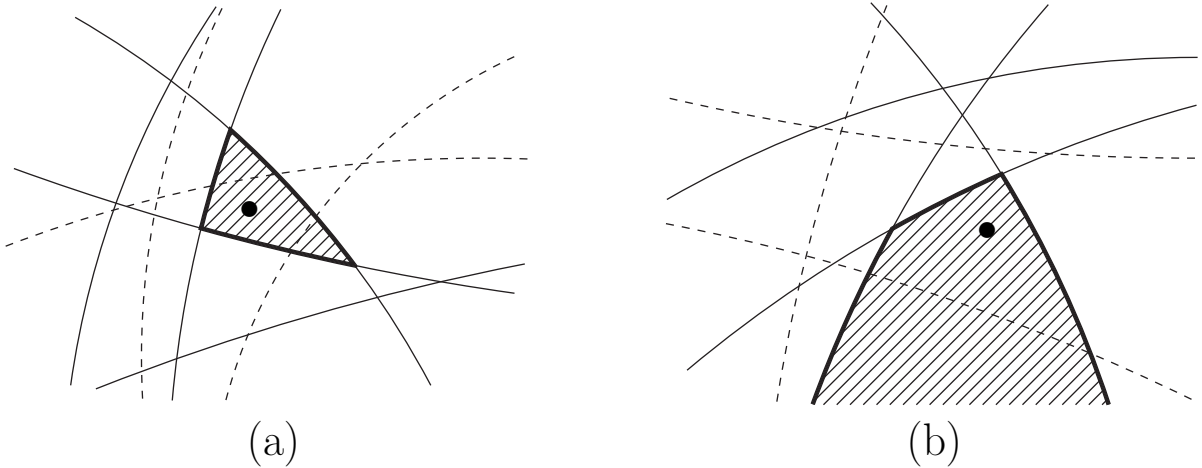
\begin{figure}[t]
\begin{center}
\begin{picture}(450,168)(1,-38)
%
%
  \Text(100,-19)[t]{\Large (a)}
  \CArc(360,-53)(340,146,172)
  \CArc(480,-48)(420,154,175)
  \CArc(-150,-170)(350,28,60)
  \CArc(320,950)(930,250,262)
  \CArc(370,-863)(920,101,111.5)
  \DashCArc(350,5)(300,154,183){3}
  \DashCArc(178,-420)(500,88,111){3}
  \DashCArc(275,-63)(200,115,165){3}
  \Line(82.6,87.6)(84.7,89.7)
  \Line(80.1,80.1)(87.3,87.3)
  \Line(77.8,72.8)(90.0,85.0)
  \Line(75.7,65.7)(92.5,82.5)
  \Line(73.8,58.8)(95.0,80.0)
  \Line(73.3,53.3)(97.4,77.4)
  \Line(77.2,52.2)(100.0,75.0)
  \Line(81.2,51.2)(102.5,72.5)
  \Line(85.2,50.2)(105.0,70.0)
  \Line(89.2,49.2)(107.3,67.3)
  \Line(93.2,48.2)(109.8,64.8)
  \Line(97.2,47.2)(112.0,62.0)
  \Line(101.2,46.2)(114.5,59.5)
  \Line(105.2,45.2)(116.8,56.8)
  \Line(109.2,44.2)(119.0,54.0)
  \Line(113.2,43.2)(121.1,51.1)
  \Line(117.3,42.3)(123.5,48.5)
  \Line(121.4,41.4)(125.7,45.7)
  \Line(125.5,40.5)(127.9,42.9)
  \Line(129.7,39.7)(130.1,40.1)
  \put(91,61){\circle*{6}}
  \CArc(480,-48)(419.7,160.7,166.0) \CArc(480,-48)(420.3,160.59,166.0)
  \CArc(480,-48)(419.4,160.7,166.0) \CArc(480,-48)(420.6,160.57,166.0)
  \CArc(-150,-170)(349.7,36.9,48.2) \CArc(-150,-170)(350.3,36.78,48.2)
  \CArc(-150,-170)(349.4,36.9,48.2) \CArc(-150,-170)(350.6,36.68,48.2)
  \CArc(320,950)(929.7,254.51,258.2) \CArc(320,950)(930.3,254.51,258.30)
  \CArc(320,950)(929.4,254.51,258.2) \CArc(320,950)(930.6,254.51,258.29)
%
%
  \Text(350,-19)[t]{\Large (b)}
  \CArc(754,-177)(480,139,160)
  \CArc(25,-141)(410,18.2,43)
  \CArc(577,-383)(500,105,130)
  \CArc(448,-282)(400,90,120)
  \DashCArc(966,-119)(685,158,171){3}
  \DashCArc(447,960)(880,257,270){3}
  \DashCArc(95,-730)(800,64,79){3}
  \Line(332.5,52.5)(333.9,53.9)
  \Line(326.9,41.9)(344.8,59.8)
  \Line(322.0,32.0)(355.0,65.0)
  \Line(317.4,22.4)(364.5,69.5)
  \Line(313.5,13.5)(373.9,73.9)
  \Line(309.8,4.8)(376.0,71.0)
  \Line(306.5,-3.5)(377.8,67.8)
  \Line(303.4,-11.6)(379.6,64.6)
  \Line(307.0,-13.0)(381.5,61.5)
  \Line(312.0,-13.0)(383.2,58.2)
  \Line(317.0,-13.0)(385.0,55.0)
  \Line(322.0,-13.0)(386.8,51.8)
  \Line(327.0,-13.0)(388.6,48.6)
  \Line(332.0,-13.0)(390.2,45.2)
  \Line(337.0,-13.0)(392.0,42.0)
  \Line(342.0,-13.0)(393.5,38.5)
  \Line(347.0,-13.0)(395.1,35.1)
  \Line(352.0,-13.0)(396.9,31.9)
  \Line(357.0,-13.0)(398.5,28.5)
  \Line(362.0,-13.0)(400.0,25.0)
  \Line(367.0,-13.0)(401.3,21.3)
  \Line(372.0,-13.0)(402.8,17.8)
  \Line(377.0,-13.0)(404.3,14.3)
  \Line(382.0,-13.0)(405.9,10.9)
  \Line(387.0,-13.0)(407.3,7.3)
  \Line(392.0,-13.0)(408.7,3.7)
  \Line(397.0,-13.0)(410.0,0.0)
  \Line(402.0,-13.0)(411.2,-3.8)
  \Line(407.0,-13.0)(412.7,-7.3)
  \Line(412.0,-13.0)(413.9,-11.1)
  \put(370,53){\circle*{6}}
  \CArc(754,-177)(479.7,151.2,160) \CArc(754,-177)(480.3,151.26,160)
  \CArc(754,-177)(479.4,151.2,160) \CArc(754,-177)(480.6,151.28,160)
  \CArc(25,-141)(409.7,18.2,31.6) \CArc(25,-141)(410.3,18.2,31.72)
  \CArc(25,-141)(409.4,18.2,31.6) \CArc(25,-141)(410.6,18.2,31.72)
  \CArc(577,-383)(499.7,113.9,119.2) \CArc(577,-383)(500.3,113.88,119.21)
  \CArc(577,-383)(499.4,113.9,119.3) \CArc(577,-383)(500.6,113.88,119.23)
\end{picture}
\caption{Characteristic situations for complexity and observer boundaries 
 in 2-dimensional parameter space.  Complexity boundaries that are and are 
 not relevant for the existence of observers are depicted by solid and 
 dashed lines, respectively.  The shaded region indicates an observer 
 region ${\cal O}$, which (a) may or (b) may not be a small region around 
 the observed point, which is denoted by the dot.  The observer boundary, 
 $O(x_i) = 1$, is given by the border of ${\cal O}$ and is represented 
 by the thick solid line.}
\label{fig:boundaries}
\end{center}
\end{figure}

Now we introduce another, closely related, naturalness problem.  Complex 
structures are required for the existence of ``observers.''  This implies 
that some complexity boundaries may also act as boundaries that divide 
the parameter space of the ensemble into those members that may support 
certain observers and those that can not.  These boundaries are harder 
to define than general complexity boundaries, since we are unable to 
give a precise definition of an observer.  Moreover, the parameter region 
may not simply be divided into the two regions ``with'' and ``without'' 
observers --- the expectation value for the number of observers may in 
general be a complicated function over the parameter space, with the 
value significantly varying across a complexity boundary.  Nevertheless, 
since the changes of the expectation value across some of these boundaries 
are expected to be drastic, caused by drastic changes of complex structures, 
for certain purposes we may approximate this function to be step-like. 
This leads to the concept of the ``observer boundary''
\begin{equation}
  O(x_i) = 1,
\label{eq:O}
\end{equation}
which divides the parameter space into one ``with'' observers, $O(x_i) < 1$, 
and one ``without,'' $O(x_i) > 1$.  In general, there are many complexity 
boundaries, $C_a(x) = 1$, that are relevant for the existence of observers. 
The region of parameter space that allows observers, ${\cal O}$, is then 
given by the common set of $C_a(x) < 1$ for all $a$, and the observer 
boundary, $O(x_i) = 1$, is the border of this region.  As more boundaries 
are added as relevant ones for observers, the region ${\cal O}$ shrinks, 
but the number of observer boundaries does not generically increase.  This 
is the crucial difference between complexity and observer boundaries. 
For illustration of these boundaries in an example of 2-dimensional space, 
see Fig.~\ref{fig:boundaries}.

While the identification of observer boundaries suffers from some 
ambiguities, there are certain advantages to focusing on these boundaries 
rather than on general complexity boundaries.  In general there can be 
several disconnected regions, ${\cal O}_I$, that can support observers. 
However, since the existence of observers undoubtedly requires certain 
complex structures, we can be convinced relatively easily that these 
regions are not distributed densely throughout the entire parameter space. 
For example, if we take the existence of stable complex nuclei to be one 
of the required conditions for observers, then we find that almost the entire 
region with $v/\Lambda_{\rm QCD} \simgt 10^4$, with the other parameters 
of the Standard Model fixed, is outside ${\cal O}_I$~\cite{Agrawal:1997gf}, 
regardless of the existence of any other physical thresholds.  Increasing 
the size of the up or down Yukawa coupling more than a factor of a few, 
with the other parameters fixed, also leads to a sterile universe in which 
both proton and neutron cannot be stable inside nuclei.  The situation 
is similar for other conditions.  For example, the requirement for 
the existence of complex structures in the universe, such as galaxies, 
makes the entire region with $\Lambda/Q^3 T_{\rm eq}^4 \simgt 1$ outside 
${\cal O}_I$~\cite{Weinberg:1987dv}, where $Q$ is the primordial scale 
of density perturbations and $T_{\rm eq}$ the temperature of matter 
radiation equality.  These imply that the observer regions ${\cal O}_I$ 
fill only a small fraction of the parameter space.  They are small 
islands in the entire parameter space of the ensemble, and so the 
observer boundaries are not distributed densely and uniformly over 
the parameter space.

The sparseness of ${\cal O}_I$ allows us to focus on the region ${\cal O}$ 
($\subset \{ {\cal O}_I \}$) that contains our observed point.  The unusual 
closeness of the observed parameters to the relevant observer boundary 
(the boundary of ${\cal O}$) then signals unnaturalness, raising an 
``observer naturalness problem.''  The degree of unnaturalness can be 
quantified using the definition given in the previous section.  The 
estimate for the degree will be conservative if we use only crucial 
requirements for the existence of observers, such as the existence of 
nuclei, stars, and galaxies, since this will give a large allowed region 
for observers, ${\cal O}$.  Imposing more and more subtle conditions, 
such as the existence of carbon, oxygen and so on, will only decrease 
${\cal O}$ and therefore increase the degree of unnaturalness.  A relevant 
question is whether we find an observer naturalness problem in the 
Standard Model and beyond, keeping only relatively robust conditions. 
This question will be addressed in the next section.

\section{Evidence for Observer Naturalness Problems}
\label{sec:evid-prob}

In this section we study if there exist observer naturalness problems 
in the Standard Model and beyond.  In general it is easier to study 
the existence of an observer naturalness problem than that of a complexity 
naturalness problem, since it is easier there to see that the relevant 
boundaries are {\it not} uniformly and densely distributed over the 
parameter space.  In order to identify and quantify the observer 
naturalness problem reliably, we take the following approach.  We first 
list complexity boundaries that are candidates for composing the observer 
boundary, in the relevant parameter space of the Standard Models of 
particle physics and cosmology.  These boundaries have effects of various 
degrees on the environment, and we take only the ones giving fairly drastic 
effects.  This leads to a large observer volume ${\cal O}$ and hence to 
a {\it conservative} estimate for the degree of an observer naturalness 
problem.  The actual degree of unnaturalness could only be more severe if 
other complexity boundaries are important for the development of observers, 
since this would reduce ${\cal O}$.

\subsection{The relevant parameters}
\label{subsec:rel_param}

Let us consider some ensemble that leads at low energies to the Standard 
Model of particle physics (taken to include neutrino masses) and to the 
standard cosmological model.  Different members of this ensemble give 
different values for the parameters of the Standard Model, $x_{\rm SM}$, 
and cosmology, $x_{\rm cos}$.  Some members may give values for this set 
$x = \{ x_{\rm SM}, x_{\rm cos} \}$ that are so distant from the observed 
values, $x_o$, that the corresponding physics and astrophysics is completely 
different from that observed --- for example, if five quark flavors were 
lighter than the QCD scale, or if the baryon asymmetry were of order unity. 
Here we restrict our discussion to a subset of the ensemble in which 
variations in the parameters about $x_o$ are limited.  In fact, we are 
interested in determining whether small deviations in $x$ from $x_o$ 
could dramatically change certain relatively stable complex structures 
(atoms, stars ...) that we observe.

The physics and astrophysics of these complex structures depends on only 
a subset of $x$; for example, variations in the bottom quark mass, or its 
mixing to the charm quark, by a factor of $2$ have no effect on the complex 
structures of interest.   Hence, we now restrict $x$ to include only
\begin{equation}
  x_{\rm SM}: \;\;\; \alpha,\, \alpha_s,\, 
    y_u,\, y_d,\, y_t,\, y_e,\, \lambda_h,\, m_h^2,
\label{eq:x_SM}
\end{equation}
and
\begin{equation}
  x_{\rm cos}: \;\;\; M_{\rm Pl},\, T_{\rm eq},\, Q,\, \eta_B,\, \Lambda.
\label{eq:x_cos}
\end{equation}
Here, $y_{u,d,t,e}$ are the Yukawa couplings for the up, down and top 
quarks and the electron, $\lambda_h$ the Higgs quartic coupling, $m_h^2$ 
the Higgs mass-squared parameter, $M_{\rm Pl}$ the reduced Planck mass, 
and $\eta_B$ the baryon asymmetry.  In this 13-dimensional parameter 
space there are special surfaces that represent complexity boundaries
\begin{equation}
  C_A(x) = 1,
\label{eq:C_A}
\end{equation}
such that the complex structure $A$ differs drastically from one side 
of the boundary to the other.

Several comments are in order for our choice of the parameter set $x$.
\begin{itemize}
\item
$\lambda_h$ and $m_h^2$ can be traded for the electroweak vacuum 
expectation value, $v$, and the Higgs boson mass.  We include $\lambda_h$ 
and $y_t$ because they typically play an important role in electroweak 
symmetry breaking.  For example, in the Standard Model there is 
a boundary corresponding to the existence of a electroweak symmetry 
breaking vacuum with $v \ll M_{\rm Pl}$~\cite{Feldstein:2006ce}
\begin{equation}
  C_v (\lambda_h,y_t) = 1.
\label{eq:C_v}
\end{equation}
\item
The $SU(2)$ gauge coupling of the Standard Model is omitted from 
Eq.~(\ref{eq:x_SM}), since charged current weak interactions at low 
energies are described by $v$.  Neutral current, and therefore the 
weak mixing angle, play little role on relevant complex structures.
\item
We have omitted the strange quark Yukawa coupling $y_s$ for simplicity. 
It does play a role in nuclear physics.
\item
The only parameter from the lepton sector is $y_e$.  Other lepton sector 
parameters could affect $\eta_B$ via leptogenesis~\cite{Fukugita:1986hr}. 
Since the source of $\eta_B$ is unknown, we prefer to list it as an 
independent cosmological parameter.
\item
The 13-dimensional parameter set could be further reduced, since only 
certain combinations appear in $C_A$ of Eq.~(\ref{eq:C_A}).  For example, 
only dimensionless combinations appear, so that $M_{\rm Pl}$ could be 
removed by using it as the unit of mass.  For the complexity boundaries 
arising from atomic and nuclear structures, $v$ appears only in the 
combinations $m_f \equiv y_f v$ ($f=u,d,e$) as fermion masses, so that 
for these boundaries $x_{\rm SM}$ can be reduced to
\begin{equation}
  x'_{\rm SM}: \;\;\; \alpha,\, \frac{m_u}{\Lambda_{\rm QCD}},\, 
    \frac{m_d}{\Lambda_{\rm QCD}},\, \frac{m_e}{\Lambda_{\rm QCD}},
\label{eq:x_SM'}
\end{equation}
where $\Lambda_{\rm QCD}$ is the QCD scale, which takes a value of 
$\approx 100~{\rm MeV}$ in our universe.  For naturalness arguments, 
however, we often prefer to use the more basic set of Eq.~(\ref{eq:x_SM}).
\item
In a theory with two Higgs doublets, such as the minimal supersymmetric 
standard model, the basic set of Eq.~(\ref{eq:x_SM}) should be expanded. 
For most purposes, we simply have to replace the electroweak vacuum 
expectation value $v$ by two vacuum expectation values; for type-II two 
Higgs doublet theories, we have $v_u$ for the up-type Higgs doublet and 
$v_d$ for the down-type Higgs doublet.
\end{itemize}

\subsection{The relevant complex structures}
\label{subsec:rel_com_str}

A crucial question for the complexity boundaries is what are the 
relevant complex structures.  Our interest in a particular boundary 
depends on how important the corresponding complex structure is for 
explaining the structure of the physical world.  Consider two extremes. 
The size of the cosmological constant, relative to the matter density 
of the universe when density perturbations become non-linear, leads 
to a boundary that determines whether large scale structure forms:
\begin{equation}
  \frac{\Lambda}{Q^3 T_{\rm eq}^4} \approx 1.
\label{eq:E_CC}
\end{equation}
A member on one side of the boundary will lead to formation of galaxies, 
while one on the other side leads to an inflating universe of isolated 
elementary particles.  This clearly has important effects on the basic 
structure of the universe.  As a second extreme example, consider the 
complex mesons with a $b$ quark constituent.  As the $b$ quark mass, 
$m_b$, is increased above the $W$ boson mass, $m_W$, the $b$ quark decays 
so rapidly that $B$ mesons cease to exist.  This boundary of $m_b/m_W 
\approx 1$, however, has a negligible effect on our environment, which 
is why we did not include the $b$-quark Yukawa coupling, $y_b$, in 
$x_{\rm SM}$.

Somewhat arbitrarily, we divide the relevant complexity boundaries into 
three classes according to how dramatic the environmental change is 
across the boundary:
\begin{itemize}
\item
{\it Catastrophic} boundaries change our universe into one that is 
essentially unrecognizable.  In addition to Eq.~(\ref{eq:E_CC}), and 
Eq.~(\ref{eq:C_v}), we would include the case that electroweak symmetry 
is broken dominantly by the Higgs potential
\begin{equation}
  \frac{v}{\Lambda_{\rm QCD}} \approx 1.
\label{eq:E_ewsb}
\end{equation}
This is required so that the baryon asymmetry of the universe is not 
washed out by the sphaleron effects~\cite{Arkani-Hamed:2005yv}.  We also 
consider that the absence of any complex nuclei is catastrophic. In the 
simplified case that the only parameter that is varied from its Standard 
Model value is $v$, this boundary is~\cite{Agrawal:1997gf}
\begin{equation}
  \frac{v}{10^4\, \Lambda_{\rm QCD}} \approx 1.
\label{eq:E_cn}
\end{equation}
For $v$ larger than this boundary, the only stable nucleus is either 
$p = uud$ or $\Delta^{++} = uuu$.
\item
{\it Violent} boundaries separate members where a crucial complex 
structure of our universe is absent.  For example, across the boundary
\begin{equation}
  \frac{m_n}{m_p + m_e} = 1,
\label{eq:E_H}
\end{equation}
where $m_p$ and $m_n$ are the masses for the proton and neutron, the 
neutron becomes stable and hydrogen unstable.  Such a neutron-stable 
world would not have dense astrophysical objects fueled by nuclear energy 
release, such as main-sequence stars in our universe~\cite{Hogan:2006xa}. 
Another example of violent boundaries is that of vanishing deuteron 
binding energy
\begin{equation}
  B_D(m_u, m_d, \Lambda_{\rm QCD}, \alpha) = 0.
\label{eq:E_D}
\end{equation}
Across this boundary, no nuclei form during big bang nucleosynthesis, 
so that the universe protonizes (or neutronizes).  Stars could only 
burn via exotic triple proton reactions, with extremely high central 
densities.
\item
{\it Substantial} boundaries separate members where a crucial complex 
structure of our universe is drastically changed.  For example, across 
certain boundaries stars may exist but are very different from those we 
see.  For example, if the $pp$ reaction $pp \rightarrow De^+\nu$ is not 
available to ignite stars, then protostars would collapse to a higher 
temperature before igniting via the $pep$ reaction, $ppe^- \rightarrow 
D\nu$.  A more substantial change to stars occurs if the deuteron is 
beta unstable, $D \rightarrow ppe^-\bar{\nu}$; stars could still burn, 
but only by using the helium produced during big bang nucleosynthesis. 
The existence of a stable diproton, the $^2$He nucleus, would also 
change stellar nuclear reactions, shortening lifetimes of hydrogen 
burning stars.  Changes of some nuclear energy levels, controlled by 
quark masses and $\alpha$, could also lead to substantial changes of 
the abundances of various nuclear species, such as carbon and oxygen.
\end{itemize}

Each boundary gives a surface of special values for the parameters, 
$\bar{x}$, which does not correspond to a surface of an enhanced symmetry. 
Among the ones listed, we find that most boundaries are clustered 
around a zone $y_{u,d,e} v \sim \alpha \Lambda_{\rm QCD} \sim O(0.01) 
\Lambda_{\rm QCD}$, where our universe also resides.  This strongly 
suggests the existence of a complexity naturalness problem.  Here, however, 
we focus more on the observer naturalness problem, which arises if the 
member describing our universe lies unusually close to the observer boundary. 
There are certain ambiguities in identifying which of the complexity 
boundaries compose the observer boundary.  An important point, however, 
is that by using only the boundaries that certainly have disastrous effects 
on the environment, we can be on the conservative side in evaluating the 
existence of an observer naturalness problem.  For this reason, we take 
only boundaries that are catastrophic or violent, rather than just 
substantial, to compose our observer boundary.  We also select only the 
boundaries that allow us to derive reasonably accurate values for $\bar{x}$, 
allowing a reliable estimate of the naturalness probability $P$.

\subsection{Stability boundaries for neutrons, deuterons and complex nuclei}
\label{subsec:stab_bound}

Following work by others, we focus on the boundaries across which we lose 
complex nuclei, Eq.~(\ref{eq:E_cn}), neutron instability, Eq.~(\ref{eq:E_H}), 
and the deuteron, Eq.~(\ref{eq:E_D}).  We take $y_u$, $y_d$, $y_e$, 
$v/\Lambda_{\rm QCD}$ and $\alpha$ to be our parameters $x_i$.  The 
choice is motivated by the expectation that the relation of these 
parameters to those in the ultraviolet theory is relatively direct.

We now represent the boundaries of Eq.~(\ref{eq:E_cn},~\ref{eq:E_H},%
~\ref{eq:E_D}) in terms of the deviations of $x_i$ from the observed 
values $x_{i,o}$.  Let us first discuss the neutron stability boundary 
of Eq.~(\ref{eq:E_H}).  The instability of a neutron, or equivalently 
the stability of hydrogen, requires
\begin{equation}
  m_n - m_p - m_e > 0,
\label{eq:boundary-H-1}
\end{equation}
where we have neglected the neutrino mass.  The neutron-proton mass 
difference, $m_n-m_p$, arises from both the strong isospin violating 
effect, $\delta_{d-u}$, and the electromagnetic contribution to 
the proton mass, $\delta_{\rm EM}$: $m_n-m_p = \delta_{d-u} - 
\delta_{\rm EM}$.  In our universe, $\delta_{d-u} \simeq 2.26 \pm 
0.51~{\rm MeV}$~\cite{Beane:2006fk} and $\delta_{\rm EM} = \delta_{d-u} 
- (m_n-m_p) \simeq (2.26 \pm 0.51) - 1.29~{\rm MeV}$.  To a first 
approximation, these quantities scale as $\delta_{d-u} \propto m_d-m_u$ 
and $\delta_{\rm EM} \propto \alpha$, so that
\begin{equation}
  m_n - m_p \simeq \frac{m_d-m_u}{m_{d,o}-m_{u,o}} (2.26~{\rm MeV}) 
    - \frac{\alpha}{\alpha_o} (0.97~{\rm MeV}).
\label{eq:mn-mp}
\end{equation}
Here we have taken the central value for $\delta_{d-u}$, and the 
variables with and without the subscript $o$ represent, respectively, 
the values in our observed universe and those in an arbitrary 
member of the ensemble.  In terms of the parameters $x_i$, 
Eq.~(\ref{eq:boundary-H-1}) can be written as
\begin{equation}
  \frac{1}{0.77} \left( \frac{y_d-y_u}{y_{d,o}-y_{u,o}} 
    - 0.23 \frac{y_e}{y_{e,o}} \right) 
    \frac{v/\Lambda_{\rm QCD}}{(v/\Lambda_{\rm QCD})_o} 
  > 0.56 \frac{\alpha}{\alpha_o}.
\label{eq:boundary-H-2}
\end{equation}
Here, we have neglected a small dependence of $\Lambda_{\rm QCD}$ on $v$ 
as well as logarithmic evolution of the Yukawa couplings.  For two Higgs 
doublet theories,%
\footnote{Here and below we assume type-II two Higgs doublet models 
 when we discuss two Higgs doublet theories.}
the Yukawa couplings in Eq.~(\ref{eq:boundary-H-2}) should be replaced as
\begin{eqnarray}
  && y_u \rightarrow y_u \sin\beta, \quad
  y_d \rightarrow y_d \cos\beta, \quad
  y_e \rightarrow y_e \cos\beta,
\label{eq:2HDM-1} \\
  && y_{u,o} \rightarrow y_{u,o} \sin\beta_o, \quad
  y_{d,o} \rightarrow y_{d,o} \cos\beta_o, \quad
  y_{e,o} \rightarrow y_{e,o} \cos\beta_o,
\label{eq:2HDM-2}
\end{eqnarray}
where $v \equiv \sqrt{v_u^2 + v_d^2}$ and $\tan\beta \equiv v_u/v_d$.

We next discuss the boundary of Eq.~(\ref{eq:E_D}).  This boundary 
corresponds to the stability of a deuteron under strong interactions, 
which requires
\begin{equation}
  B_D = m_p + m_n - m_D < 0,
\label{eq:boundary-D-1}
\end{equation}
where $m_D$ is the deuteron mass.  The binding energy $B_D$ depends on 
the quark masses as
\begin{equation}
  B_D - B_{D,o} = -a \left( \frac{m_u+m_d}{m_{u,o}+m_{d,o}} - 1 \right),
\label{eq:B_D}
\end{equation}
where $B_{D,o} \simeq 2.2~{\rm MeV}$ is the observed deuteron binding 
energy.  The parameter $a$ is uncertain, but it is estimated in 
Ref.~\cite{Agrawal:1997gf} to be $a \simeq (1.3$~--~$5.5)~{\rm MeV}$ 
using models of nucleon binding.  This allows us to write 
Eq.~(\ref{eq:boundary-D-1}) in terms of $x_i$ as
\begin{equation}
  \frac{y_u+y_d}{y_{u,o}+y_{d,o}} 
    \frac{v/\Lambda_{\rm QCD}}{(v/\Lambda_{\rm QCD})_o} 
  < 1 + \frac{2.2~{\rm MeV}}{a}.
\label{eq:boundary-D-2}
\end{equation}
For two Higgs doublet theories, the Yukawa couplings must be replaced as 
in Eqs.~(\ref{eq:2HDM-1},~\ref{eq:2HDM-2}).

We finally consider the condition for the existence of complex nuclei. 
The stability of complex nuclei requires the energy release for the $\beta$ 
decay, $n \rightarrow p e^- \bar{\nu}$, to be smaller than the binding 
energy of nuclei per nucleon, $E_{\rm bin}$:
\begin{equation}
  m_n - m_p - m_e \simlt E_{\rm bin},
\label{eq:boundary-cn-1}
\end{equation}
where we have neglected the neutrino mass.  Precisely speaking, $E_{\rm 
bin}$ varies with a nucleus, receiving contributions both from nuclear 
forces and the Coulomb repulsion between protons.  To a first approximation, 
however, $E_{\rm bin}$ can be regarded as the same for all nuclei and 
being controlled purely by $\Lambda_{\rm QCD}$, taking the value of 
$E_{\rm bin} \simeq 8~{\rm MeV}$ in our universe.%
\footnote{We neglect the dependence of $E_{\rm bin}$ on $m_u$ and $m_d$, 
 as this will not affect our conclusions.}
Substituting this into Eq.~(\ref{eq:boundary-cn-1}), and using 
Eq.~(\ref{eq:mn-mp}), we obtain
\begin{equation}
  \frac{1}{0.77} \left( \frac{y_d-y_u}{y_{d,o}-y_{u,o}} 
    - 0.23 \frac{y_e}{y_{e,o}} \right) 
    \frac{v/\Lambda_{\rm QCD}}{(v/\Lambda_{\rm QCD})_o} 
  \simlt 4.6 + 0.56 \frac{\alpha}{\alpha_o}.
\label{eq:boundary-cn-2}
\end{equation}
The expression should be modified according to Eqs.~(\ref{eq:2HDM-1},%
~\ref{eq:2HDM-2}) for two Higgs doublet theories.

\subsection{Observer naturalness problem in the Standard Model and beyond}
\label{subsec:obs_nat_prob}

\begin{figure}
  \center{\includegraphics[width=.36\textwidth]{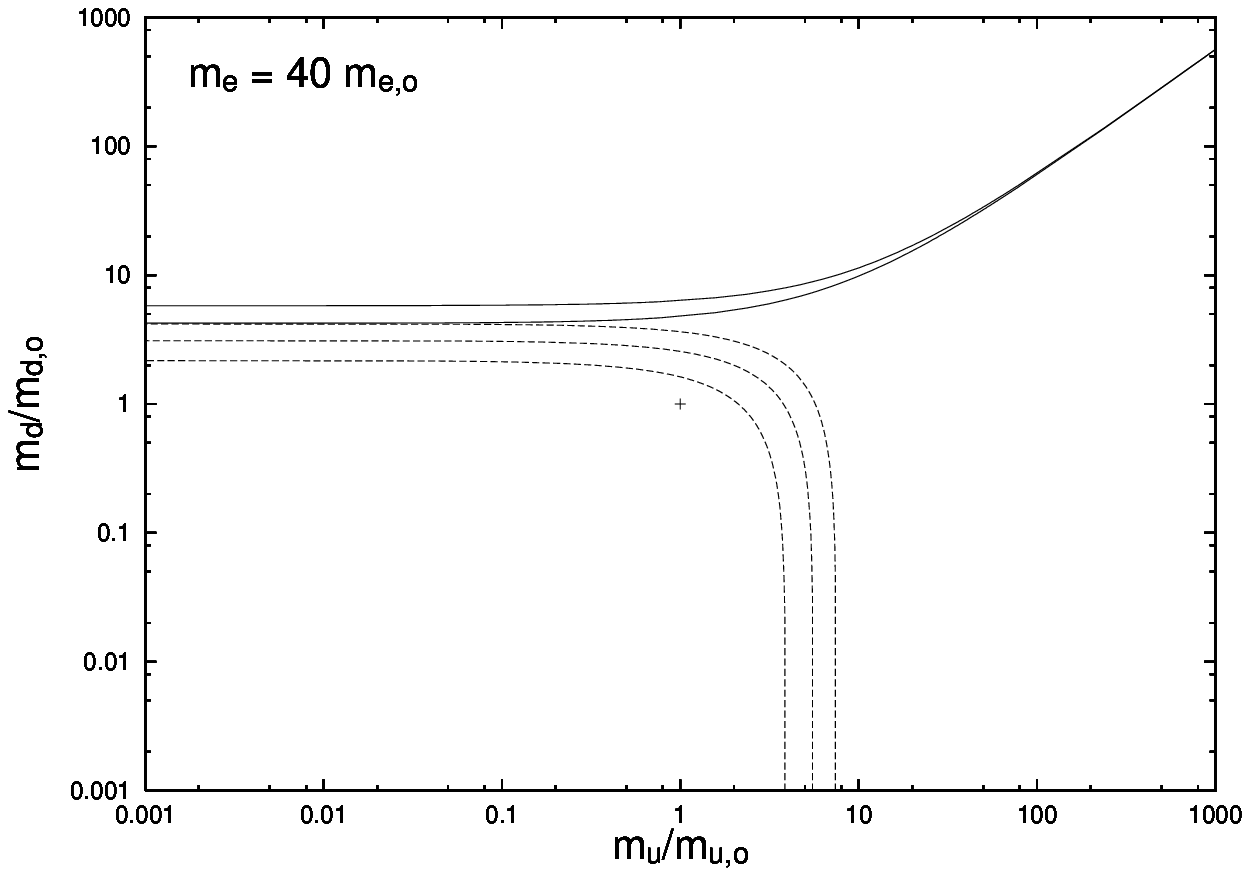}
  \hspace{1.5cm} \includegraphics[width=.36\textwidth]{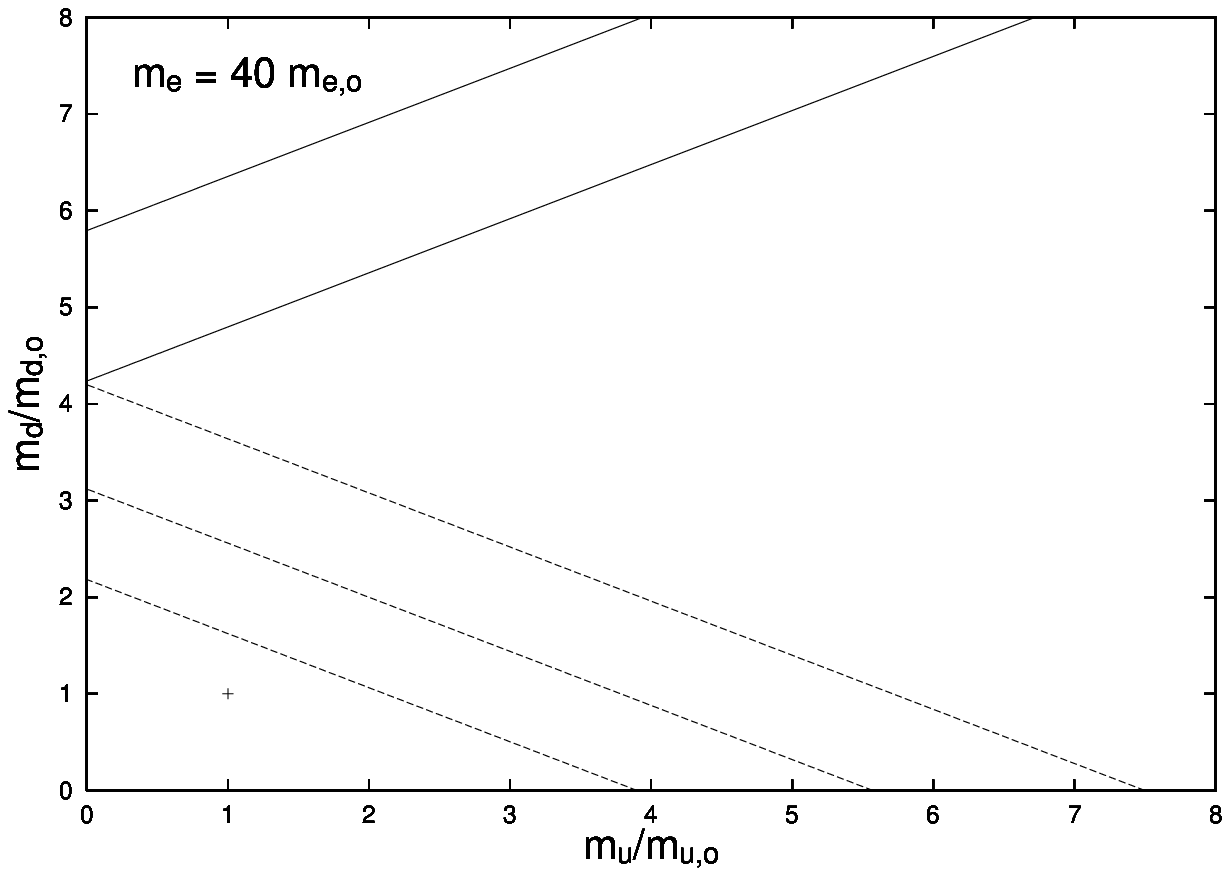}}
  \center{\includegraphics[width=.36\textwidth]{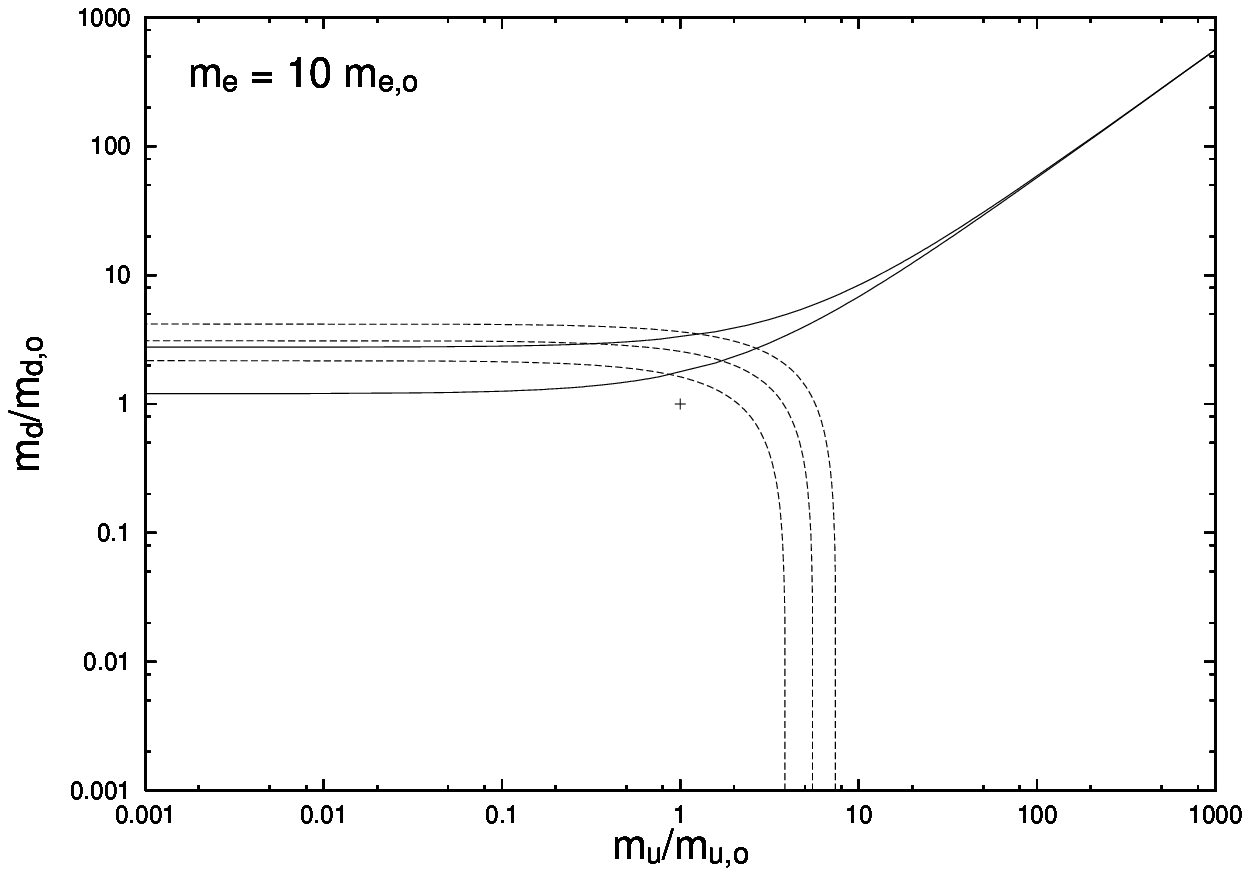}
  \hspace{1.5cm} \includegraphics[width=.36\textwidth]{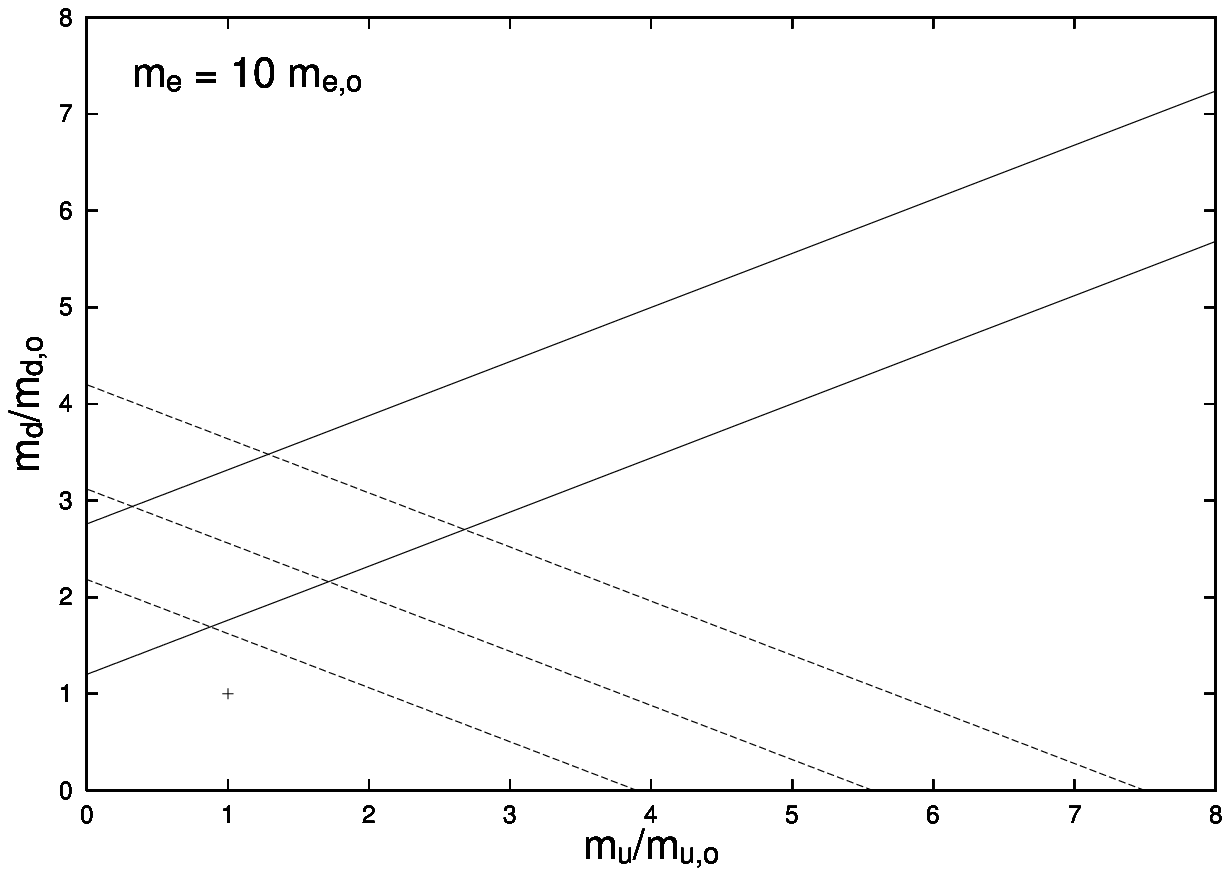}}
  \center{\includegraphics[width=.36\textwidth]{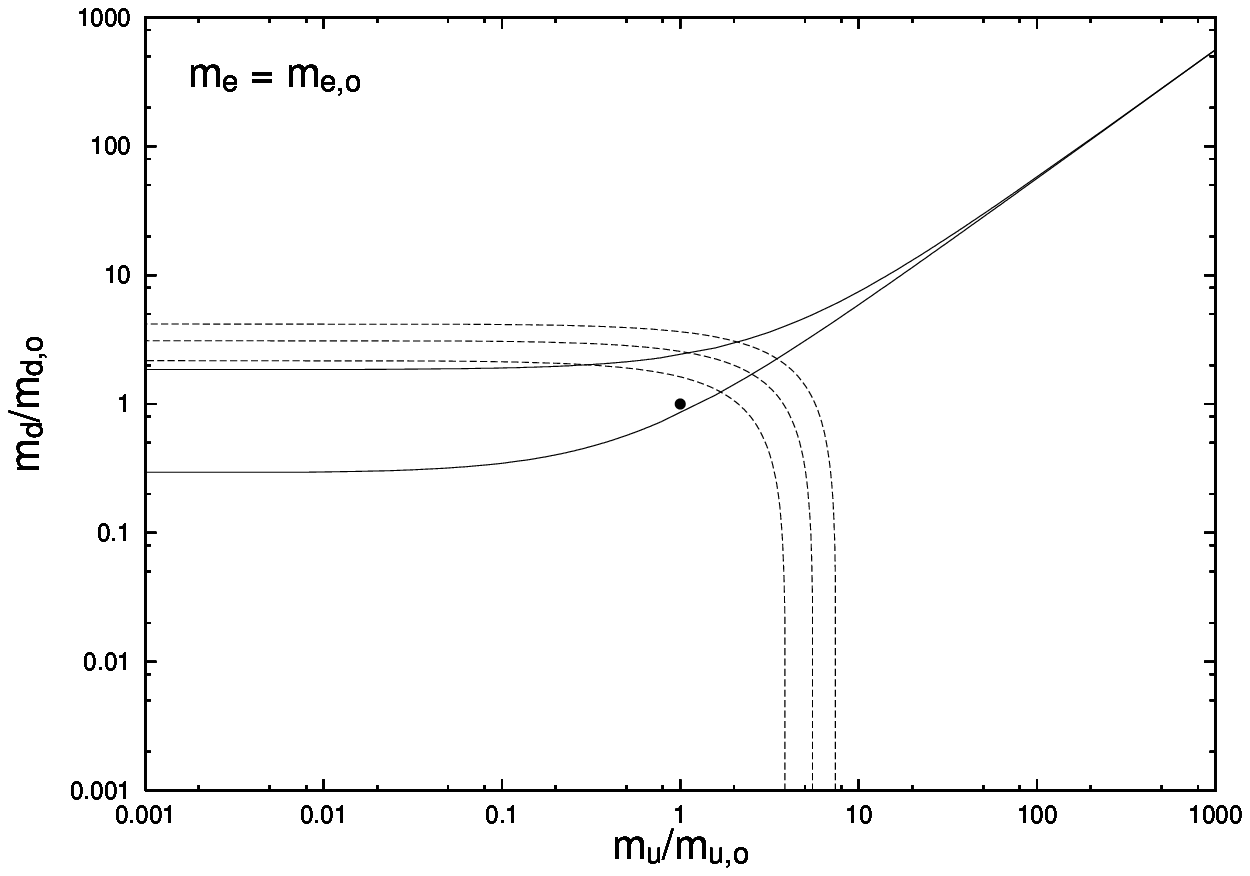}
  \hspace{1.5cm} \includegraphics[width=.36\textwidth]{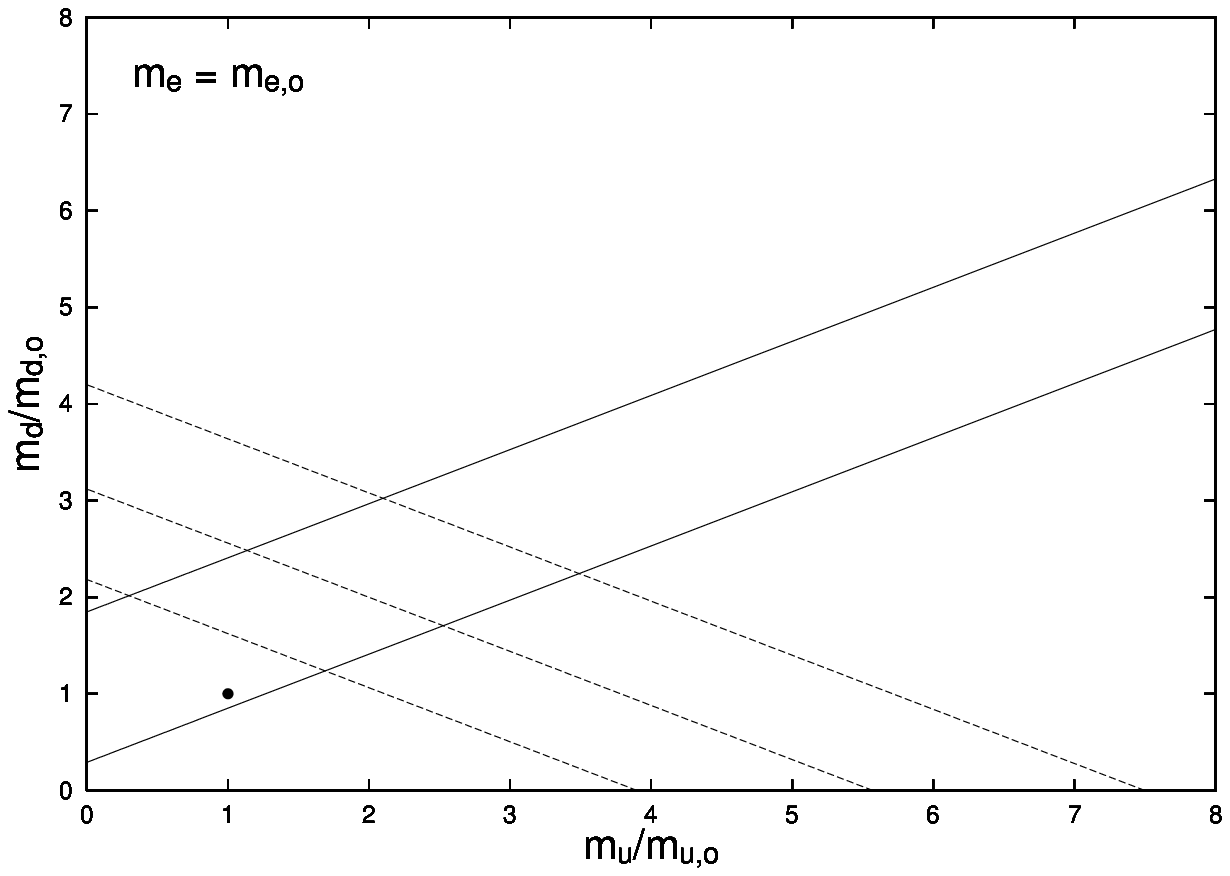}}
  \center{\includegraphics[width=.36\textwidth]{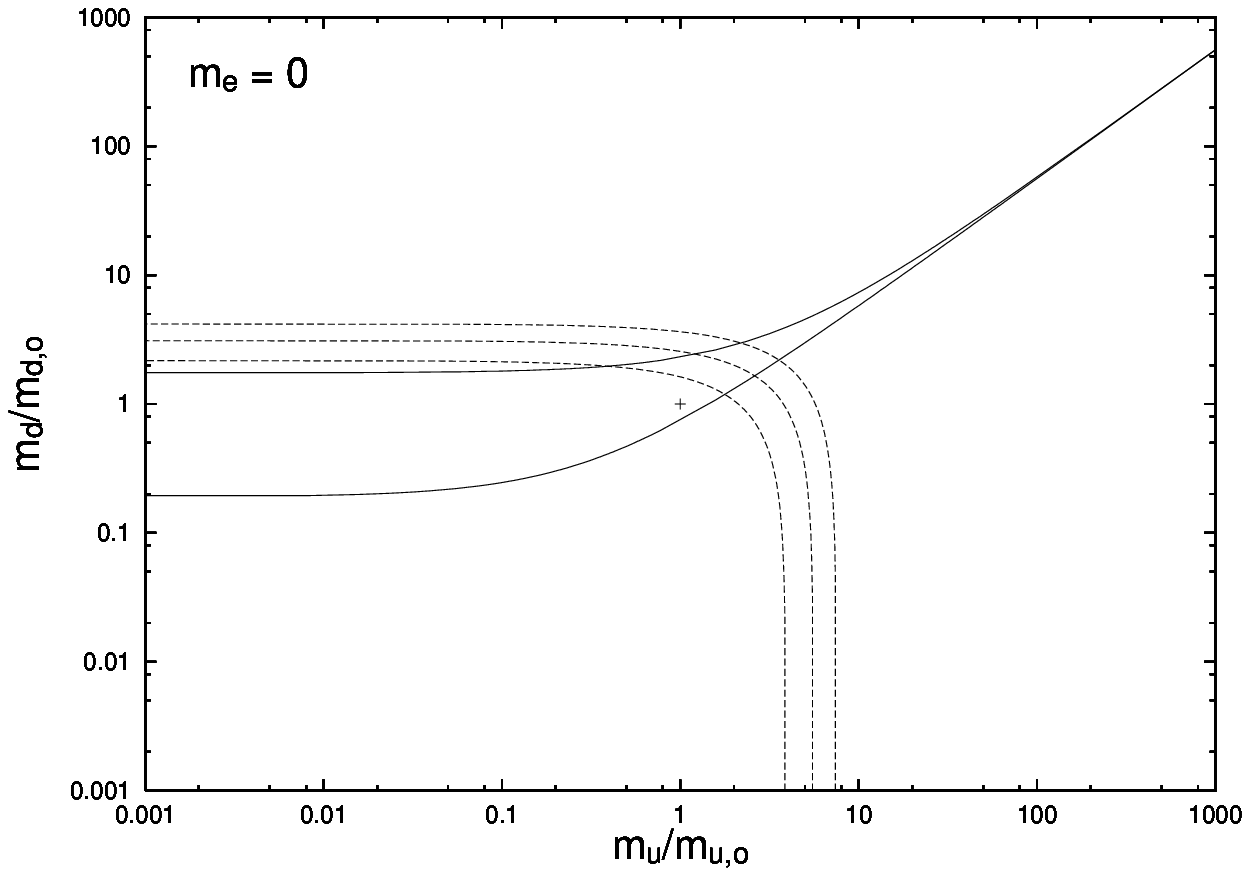}
  \hspace{1.5cm} \includegraphics[width=.35\textwidth]{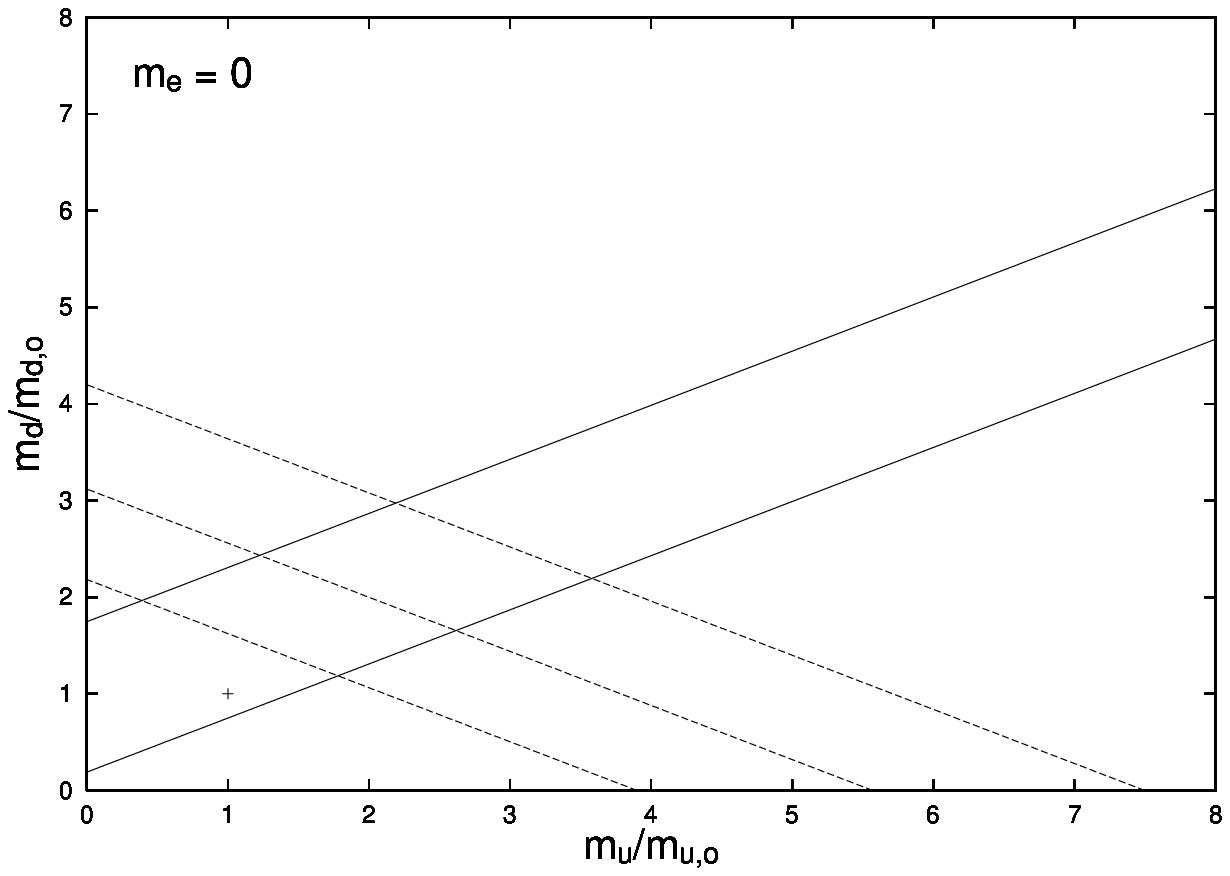}}
\caption{The location of the observer boundary in $m_u$-$m_d$-$m_e$ space. 
 The neutron and complex nuclei boundaries of Eqs.~(\ref{eq:boundary-H-2}) 
 and (\ref{eq:boundary-cn-2}) are depicted by solid lines (below and 
 above, respectively).  Dashed lines represent the deuteron boundary of 
 Eq.~(\ref{eq:boundary-D-2}) (for $a = 5.5$, $2.2$ and $1.3~{\rm MeV}$ 
 from below).  The observed point is represented by little dots.}
\label{fig:region}
\end{figure}

The region inside our observer boundary, ${\cal O}$, is defined by 
Eqs.~(\ref{eq:boundary-H-2},~\ref{eq:boundary-D-2},~\ref{eq:boundary-cn-2}), 
with modification by Eqs.~(\ref{eq:2HDM-1},~\ref{eq:2HDM-2}) for 
two Higgs doublet theories.  To visualize the overall shape of this 
region, in Fig.~\ref{fig:region} we depict the three boundaries of 
Eqs.~(\ref{eq:boundary-H-2},~\ref{eq:boundary-D-2},~\ref{eq:boundary-cn-2}) 
in $m_u$-$m_d$-$m_e$ space. 
The plots in the left column adopt a logarithmic scale in $m_u$ and $m_d$, 
while those in the right a linear scale in $m_u$ and $m_d$.  The fine 
structure constant is fixed to be $\alpha = \alpha_o$, and we have used 
the leading-order chiral perturbation value of $m_{u,o}/m_{d,o} = 0.56$ 
to draw these plots.  (The qualitative features of the plots are not 
affected if we vary $\delta_{d-u}$ and $m_{u,o}/m_{d,o}$ in the range 
of $\delta_{d-u} \simeq 2.26 \pm 0.51~{\rm MeV}$~\cite{Beane:2006fk} 
and $m_{u,o}/m_{d,o} = 0.3$~--~$0.6$~\cite{Yao:2006px}.)  From the figure, 
we find that the observed parameters are close to the three boundaries, 
especially when viewed on a logarithmic scale.  In particular, they are 
very close to the neutron stability bound of Eq.~(\ref{eq:boundary-H-1}). 
This implies that we have an observer naturalness problem.

The degree of unnaturalness $P$ depends on a theory, since the relation 
between $\{m_u, m_d, m_e, \alpha\}$ and the fundamental parameters, as 
well as the form of the distribution function, depend on the ensemble 
we consider.  However, we can still make a conservative estimate on 
$P$, based on the observation that the naturalness probability $\tilde{P}$ 
obtained by fixing all the fundamental parameters except for one 
generally satisfies $\tilde{P} \simgt P$.  In the present context, 
we can fix $y_u$, $y_d$, $y_e$, $\alpha$ and $\Lambda_{\rm QCD}$, as 
well as $\tan\beta$ for two Higgs doublet theories.  The conditions of 
Eqs.~(\ref{eq:boundary-H-2},~\ref{eq:boundary-D-2},~\ref{eq:boundary-cn-2}) 
then lead to
\begin{equation}
  0.5 \simlt \frac{v}{v_o} \simlt 2,
\label{eq:cond-v}
\end{equation}
where we have used $a = 2.2~{\rm MeV}$ for illustrative purpose.  The 
probability of $v$ falling in this range, $\tilde{P}_v$, then gives 
a conservative estimate of the naturalness probability $P$.  Alternatively, 
we can fix $v$ instead of $\Lambda_{\rm QCD}$.  In this case, we have
\begin{equation}
  0.5 \simlt \frac{\Lambda_{\rm QCD}}{\Lambda_{{\rm QCD},o}} \simlt 2.
\label{eq:cond-Lambda}
\end{equation}
The probability of $\Lambda_{\rm QCD}$ falling in this range, 
$\tilde{P}_{\Lambda_{\rm QCD}}$, can also give a conservative 
estimate of $P$, since $P \simlt {\rm min}\{ \tilde{P}_v, 
\tilde{P}_{\Lambda_{\rm QCD}} \}$.  Which of $\tilde{P}_v$ and 
$\tilde{P}_{\Lambda_{\rm QCD}}$ we use is determined by which is 
easier to estimate and, in case both are easy to estimate, which 
gives a stronger bound on $P$.

Let us first consider an ensemble in which the distribution of $v$ 
is flat in $v^2$, as in the Standard Model embedded into more fundamental 
theory at a high scale.  In this case, we obtain an extremely small 
value of $\tilde{P}_v \approx v_o^2/M_*^2$, where $M_* \gg v_o$ is 
the cutoff scale.  For $M_* \approx 10^{18}~{\rm GeV}$, we obtain 
$\tilde{P}_v \approx 10^{-32}$.  The origin of this small value, however, 
is precisely the existence of the gauge hierarchy problem.  We thus see 
the existence of a severe observer naturalness problem in these theories, 
but it is hard to disentangle from the fine-tuning naturalness problem.

We therefore focus on theories in which the conventional gauge 
hierarchy problem is solved.  In these theories, the distribution 
of $v$ is expected to be flat in $\ln v$ or $1/\ln v$ within an ensemble. 
To estimate $\tilde{P}_v$, however, we need to know the range of $v$, 
which depends on how $v$ is related to the fundamental parameters of 
the theory.  To avoid model dependence coming from this, we here consider 
$\tilde{P}_{\Lambda_{\rm QCD}}$, instead of $\tilde{P}_v$.  The value of 
the QCD scale, $\Lambda_{\rm QCD}$, is determined from the strong gauge 
coupling constant at the cutoff scale, $g_3(M_*)$, through renormalization 
group evolution.  The probability $\tilde{P}_{\Lambda_{\rm QCD}}$ can 
then be estimated if (i) the theory between $\Lambda_{\rm QCD}$ and $M_*$, 
together with the value of $M_*$, is specified, and (ii) the distribution 
function for $g_3(M_*)$, including the range of $g_3(M_*)$, is given.%
\footnote{Here we neglect higher order effects, which are expected to 
 be small.}
Specifying the theory between $\Lambda_{\rm QCD}$ and $M_*$ is important 
because the existence of colored states whose masses are associated with 
$v$ may increase the correlation between $v$ and $\Lambda_{\rm QCD}$, 
enhancing $\tilde{P}_{\Lambda_{\rm QCD}}$.  The distribution and the 
range of $g_3(M_*)$ are also important.  For example, if the distribution 
of $g_3(M_*)$ were flat in $1/g_3^2(M_*)$ within the range $1/g_3^2(M_*) 
\simgt b_3/16\pi^2$, the QCD scale is distributed almost flat in $\ln 
\Lambda_{\rm QCD}$ for all $\Lambda_{\rm QCD} \simlt M_*$.  Here, $b_3$ 
is the one-loop beta function coefficient for $g_3$ evaluated at $M_*$. 
In this case, we would obtain $\tilde{P}_{\Lambda_{\rm QCD}}$ estimated 
using Eq.~(\ref{eq:cond-Lambda}) to be infinitely small.  To avoid this 
unphysical conclusion, we can introduce an arbitrary cutoff $c$ on the 
distribution, $1/g_3^2(M_*) \simlt c$.  Alternatively, we can consider 
that the distribution of $g_3(M_*)$ is flat in $g_3^2(M_*)$.  In this case, 
restricting the range to be $g_3^2(M_*) \simlt c'$ gives a finite answer 
to $\tilde{P}_{\Lambda_{\rm QCD}}$.  The value of $\tilde{P}_{\Lambda_{\rm 
QCD}}$ depends on $c'$, but we can make a reasonable conjecture on the 
value of $c'$, e.g. $c' \approx 1$ or $\approx 16\pi^2/b_3$.

\begin{figure}
  \center{\includegraphics[width=.41\textwidth]{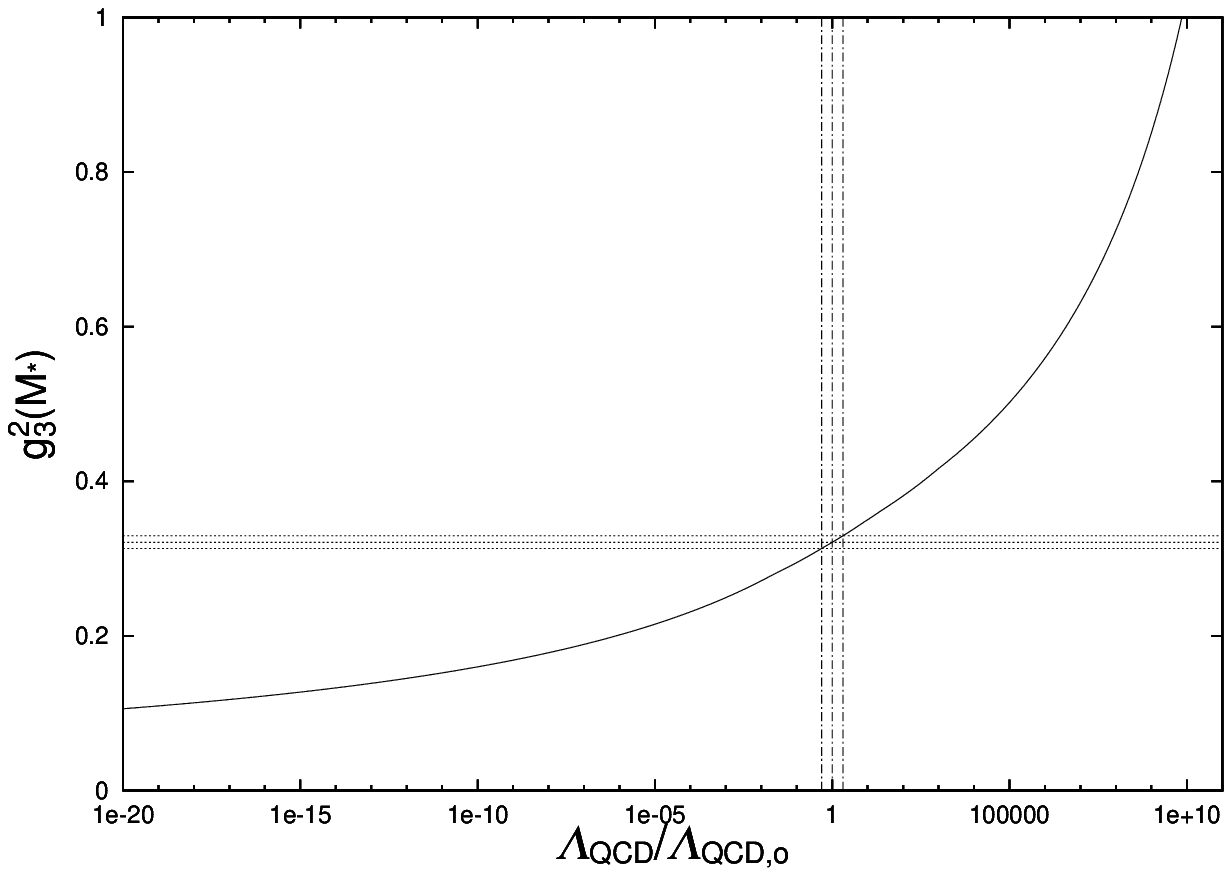}
  \hspace{1.8cm} \includegraphics[width=.41\textwidth]{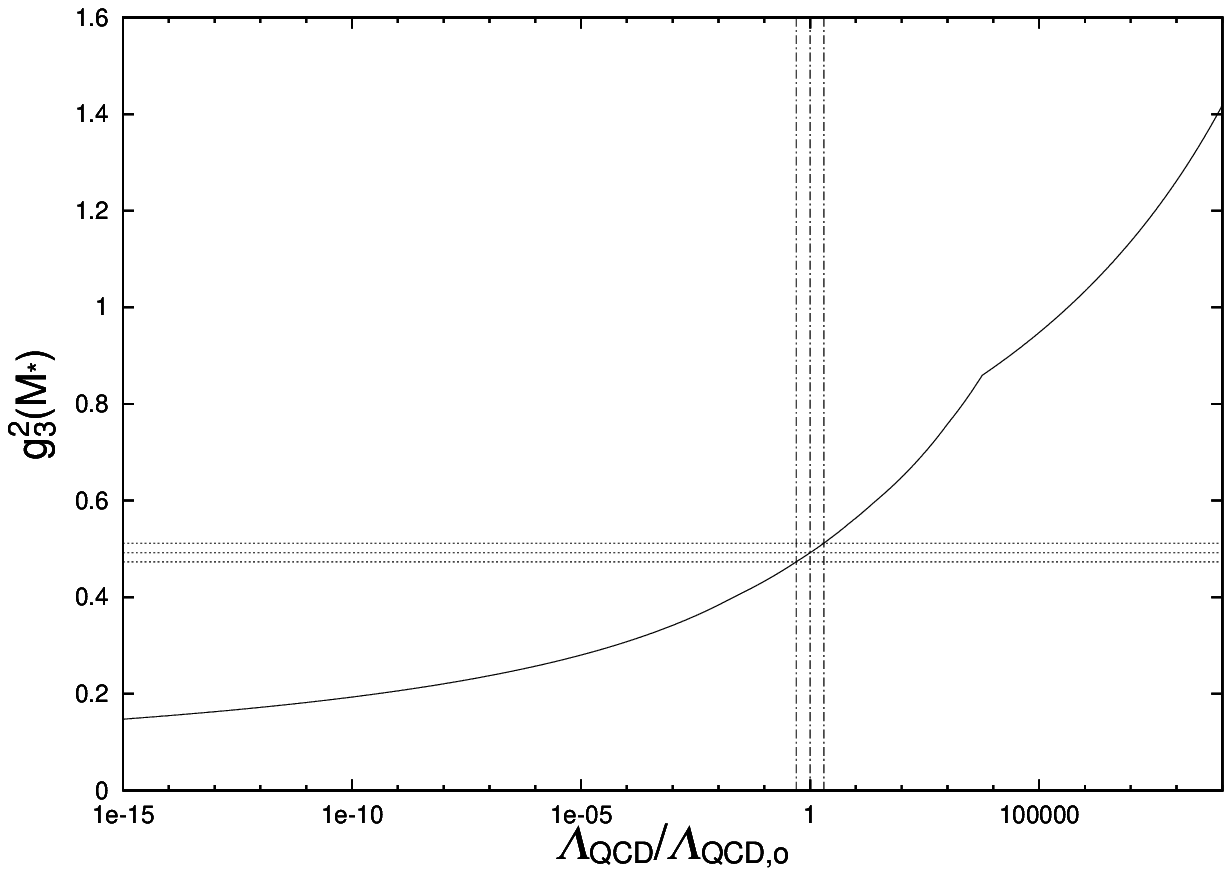}}
\caption{The value of $g_3^2(M_*)$ as a function of 
 $\Lambda_{\rm QCD}/\Lambda_{{\rm QCD},o}$ for non-supersymmetric 
 theories (left) and supersymmetric theories (right).  The range of 
 Eq.~(\ref{eq:cond-Lambda}) is depicted by the vertical lines, while 
 the corresponding range of $g_3^2(M_*)$ by the horizontal lines.}
\label{fig:P_Lambda}
\end{figure}

In Fig.~\ref{fig:P_Lambda}, we plot the value of $g_3^2(M_*)$ as 
a function of $\Lambda_{\rm QCD}/\Lambda_{{\rm QCD},o}$ obtained 
using one-loop renormalization group equations for non-supersymmetric 
theories (left) and supersymmetric theories (right). 
In non-supersymmetric theories, $M_*$ is taken to be $10^{14}~{\rm GeV}$, 
which is the scale where the $SU(3)$ and $U(1)$ gauge couplings almost 
meet.  (This is the scale where the three gauge couplings would almost 
meet if five Higgs doublet fields or a vector-like fermion with the Higgs 
quantum numbers were introduced at the weak scale.)  In supersymmetric 
theories, we take $M_*$ to be the unification scale, $2 \times 
10^{16}~{\rm GeV}$, and the scale of superparticle masses to be 
$m_{\rm SUSY} \simeq 1~{\rm TeV}$.  Dependence of the results 
on these parameters, however, is weak.  From the figure, we find 
the probability of $\Lambda_{\rm QCD}$ falling in the range of 
Eq.~(\ref{eq:cond-Lambda}) to be
\begin{equation}
  \tilde{P}_{\Lambda_{\rm QCD}} \simeq 0.016\, \frac{1}{c'} 
    \simeq 61^{-1}\, \frac{1}{c'},
\label{eq:P-nonSUSY}
\end{equation}
for non-supersymmetric theories and
\begin{equation}
  \tilde{P}_{\Lambda_{\rm QCD}} \simeq 0.038\, \frac{1}{c'} 
    \simeq 26^{-1}\, \frac{1}{c'},
\label{eq:P-SUSY}
\end{equation}
for supersymmetric theories.  Here, we have taken the distribution 
function to be flat in $g_3^2(M_*)$.  The value of $c'$ is uncertain, 
but we expect $c' \simgt 1$.  The numbers in Eqs.~(\ref{eq:P-nonSUSY},%
~\ref{eq:P-SUSY}) depend on $a$ in Eq.~(\ref{eq:B_D}), through the 
dependence of the lower value of Eq.~(\ref{eq:cond-Lambda}) on $a$. 
For $a = 1.3~{\rm MeV}$ ($5.5~{\rm MeV}$), $0.016$ and $0.038$ in 
these equations become $0.020$ and $0.046$ ($0.012$ and $0.029$), 
respectively.  Note that the analysis here provides the {\it most 
conservative} estimate for the level of an observer naturalness problem 
existing in theories beyond the Standard Model (in which the conventional 
gauge hierarchy problem is solved, so that $\tilde{P}_v$ can be larger 
than $\tilde{P}_{\Lambda_{\rm QCD}}$):
\begin{equation}
  P \simlt \tilde{P}_{\Lambda_{\rm QCD}}.
\label{eq:P-PQCD}
\end{equation}
In theories where the gauge hierarchy problem is not solved, the degree 
of the observer naturalness problem is much more severe, $P \simlt 
\tilde{P}_v \ll 1$.  We thus find that the bounds on $P$ derived here 
provide evidence for an observer naturalness problem in the Standard 
Model and beyond.

To obtain a conservative estimate on $P$, we have used here only 
$P \simlt {\rm min}\{ \tilde{P}_v, \tilde{P}_{\Lambda_{\rm QCD}} 
\}$.  Physically, the case with $P \approx {\rm min}\{ \tilde{P}_v, 
\tilde{P}_{\Lambda_{\rm QCD}} \}$ corresponds to the situation 
where we have the ``best'' model of flavor, i.e. the observed quark 
and lepton masses in units of the weak scale are reproduced essentially 
without any free parameter.  This is, however, not the case for many 
theories of flavor, which will have $P < {\rm min}\{ \tilde{P}_v, 
\tilde{P}_{\Lambda_{\rm QCD}} \}$, as discussed in the next section.

\section{Naturalness Probabilities and Theories of Flavor}
\label{sec:flavor}

The observed values of the first generation masses, $m_{u,d,e}$, 
are special: Fig.~\ref{fig:region} shows that they are close to the 
boundaries of neutron, deuteron and complex nuclei stability.  To 
judge whether this is likely to be accidental we need to evaluate 
the naturalness probability, but this depends on the theory of flavor. 
We begin by considering ensembles of theories in which $\alpha$ and 
$v/\Lambda_{\rm QCD}$ are fixed, but the flavor observables $x_F$ 
have some probability distribution $f(x_F)$.

In the Standard Model the relevant flavor observables are the Yukawa 
couplings: $x_F = y_{u,d,e}$.  While they are all very small, of order 
$10^{-6}$~--~$10^{-5}$ at unified scales, they are all close to the 
maximal values allowed by the neutron, deuteron and complex nuclei 
boundaries.  Of particular important, the observed values are especially 
close to the neutron (hydrogen) stability boundary.  While $y_e$ is 
about a factor of $3$ from this boundary, in the $y_u$-$y_d$ plane 
the distance from the observed point to this boundary corresponds 
to a variation in the coupling only of
\begin{equation}
  \sqrt{\frac{(1-z)^2(1-A)^2}{1+z^2}} = 13^{+17}_{-3}\%,
\label{eq:y-var}
\end{equation}
where $z \equiv m_u/m_d$ and $A \equiv (\delta_{d-u} - m_n + m_p 
+ m_e)/\delta_{d-u}$ evaluated in our universe.  Here, the central 
value is obtained with $z = 0.56$ and $\delta_{d-u} = 2.26~{\rm MeV}$. 
The electromagnetic interaction raises the proton mass above the 
neutron mass by about an MeV.  Given that quarks can have masses up 
to $100~{\rm GeV}$ or more, it is remarkable that the $m_d - m_u$ 
mass difference just overcompensates the electromagnetic shift to 
make the hydrogen mass only $\simeq 0.78~{\rm MeV}$ smaller than the 
neutron mass.  The deuteron and complex nuclei stability boundaries 
are also close by, corresponding to changes in the Yukawa couplings 
by factors of $\approx (1.5$~--~$3)$ and $\approx 2$, respectively.

It is important to numerically evaluate these accidents; for 
illustration, we consider two simple distribution functions. 
If $f(y_{u,d,e})$ are flat and non-zero in the range of $y_{u,d,e}$ 
from $0$ to $1$, then the corresponding naturalness probability for 
this ensemble, using Eq.~(\ref{eq:P_multi}), is $P_F \approx 10^{-16}$; 
while if $f(\log_{10}y_{u,d,e})$ are flat and non-zero in the range 
of $\log_{10}y_{u,d,e}$ from $-6$ to $0$, then $P_F \approx 10^{-4}$. 
As expected from Fig.~\ref{fig:region}, the volume of parameter space 
closer to the special points on the observer boundary than the measured 
point is very small compared with the expected total volume of parameter 
space in the ensemble.  From the viewpoint of neutron instability and 
deuteron and complex nuclei stability, the Standard Model description 
of flavor is highly unnatural.

Theories of flavor that go beyond the simple Yukawa coupling 
parameterization of the Standard Model typically involve further 
symmetries, such as flavor or unified gauge symmetries.  While the 
set of flavor observables $x_F$ in these theories can be smaller than 
that in the Standard Model, all known models do involve free parameters. 
If a theory could be found in which $y_{u,d,e}$ are precisely predicted, 
i.e. if variations in $y_{u,d,e}$ are less than the distances to the 
special points on the observer boundary when $x_F$ vary, then such 
a theory would have $P_F \approx 1$.  However, if $y_{u,d,e}$ vary 
significantly (and independently) as $x_F$ vary in the ensemble, as 
in most theories of flavor, then the theory is likely to have $P_F$ 
(much) smaller than of order unity.  The best hope for a significant 
improvement in naturalness is then to obtain a successful prediction 
for the ratios of the first generation Yukawa couplings --- symmetries 
that successfully predict the $y_u$, $y_d$, and $y_e$ ratios could 
show that our universe lying so close to the observer boundary is 
just accidental.  This is, however, not so easy, especially because 
of the extreme closeness of the observed point to the neutron stability 
boundary.  Moreover, since the observer boundary involves the combinations 
$y_{u,d,e} v/\Lambda_{\rm QCD}$, the normalization of $y_{u,d,e}$ is 
coupled to the value of $v/\Lambda_{\rm QCD}$.  Below, we do not take 
$v/\Lambda_{\rm QCD}$ as fixed, but consider it to vary in the ensemble 
as discussed in the previous section.

In certain theories of flavor, for example with Abelian flavor symmetries, 
the smallness of $y_f$ ($f = u,d,e$) follows from a single small parameter 
of the theory
\begin{equation}
  y_f = c_f\, \epsilon^{p_f},
\label{eq:AFS}
\end{equation}
where $\epsilon \ll 1$ while $c_f$ are of order unity.  The powers 
$p_f$ are not free parameters of the theory, but result from a certain 
judicious choice of Abelian charges for all the quark and lepton 
fields~\cite{Froggatt:1978nt}.  The ratios of the Yukawa couplings 
can then be just numbers of order unity for a fixed value of $\epsilon$ 
(and $\tan\beta$ for two Higgs doublet theories).  Let us consider, 
for example, that the coefficients $c_f$ vary in the range $0 \leq c_f 
\leq 2 c_{f,o}$ with a flat distribution on a linear scale, or in the 
range $\log_{10}c_{f,o}-1/2 \leq \log_{10}c_f \leq \log_{10}c_{f,o}+1/2$ 
with a flat distribution on a logarithmic scale.  In this case, the 
extreme closeness of the neutron stability boundary gives $P_F \approx 
1/18$ and $1/36$, respectively, for a fixed value of $\epsilon$ (and 
$\tan\beta$).  In addition to this, there is the issue of the normalization 
of the masses $m_{u,d,e}$, arising from a variation of $v/\Lambda_{\rm QCD}$. 
Assuming that the distribution function is flat in $g_3^2(M_*)$, this 
leads to a further reduction of the probability at least by a factor of 
$\tilde{P} \simlt 1/35$ and $1/15$ for non-supersymmetric and supersymmetric 
theories, respectively.  (These numbers are obtained by requiring that 
$v/\Lambda_{\rm QCD}$ should be within a factor of $\approx (2$~--~$3)$ 
from the observed value so that its variation can be absorbed into 
changes of $c_f$, without a large extra cost in $P_F$.  The range of 
$g_3^2(M_*)$ here has been chosen to be $0 \leq g_3^2(M_*) \leq 1$.) 
Possible variations of $\epsilon$ (and $\tan\beta$) may or may not lead 
to further reductions of the probability.  We thus conclude that theories 
with Abelian flavor symmetries considered here have
\begin{equation}
  P \simlt P_F \tilde{P} \simlt (10^{-3}~\mbox{--}~10^{-2}),
\label{eq:final_P}
\end{equation}
from physics of flavor alone.

To increase $P$ from flavor physics, it is necessary to predict ratios 
of $y_{u,d,e}$ to better than the factor of $\approx 2$ we used in 
the above estimates.  A more elaborate charge assignment for the quark 
and lepton fields under an Abelian flavor symmetry is not sufficient, 
since it will leave the relative coefficients of order unity undetermined. 
A non-Abelian flavor symmetry, or unified gauge symmetry, could improve 
$P$ by successfully predicting $y_u:y_d:y_e$ as Clebsch factors.  For 
example, $y_u:y_d:y_e = 1:2:1$ at the unified scale agrees with inferred 
values of $m_{u,d,e}$, providing the eigenvalues are not affected 
much by mixing to heavier generations.  A simple theory of flavor 
incorporating such a relation could lessen the significance of the 
observer naturalness problem of Eq.~(\ref{eq:final_P}).  However, 
uncertainties from QCD are too large for us to know if the relation 
really puts us sufficiently close to the neutron stability boundary, 
so that the problem is simply an accidental consequence of symmetry. 
Moreover, in conventional theories of flavor we expect simple Clebsch 
factors to apply to the heaviest generation not the lightest.  This 
is because the lightest generation typically gets contributions from 
nontrivial matrix diagonalizations, so that the relation would have 
to involve more than one generation.  For example, the inferred values 
of $m_{u,d,e}$ are consistent with unified boundary conditions $y_u = 
y_e$ and $y_d = 0$, provided the down quark mass arises from diagonalizing 
a symmetric $2 \times 2$ matrix that fully accounts for the Cabibbo 
angle.  This requires, however, experimental inputs of the strange quark 
mass and the Cabibbo angle, so that it does not provide a real symmetry 
solution to the problem.  We find it rather difficult to have a symmetry 
understanding for $y_u:y_d:y_e$, and even if we had, unnaturalness 
still arises from the coincidence between $v$ and $\Lambda_{\rm QCD}$, 
given by Eqs.~(\ref{eq:P-nonSUSY},~\ref{eq:P-SUSY}).  We conclude that, 
in the absence of a convincing and significant progress in flavor physics, 
we have unnaturalness associated with flavor at the level given by 
Eq.~(\ref{eq:final_P}).

\section{Environmental Selection}
\label{sec:sol-ES}

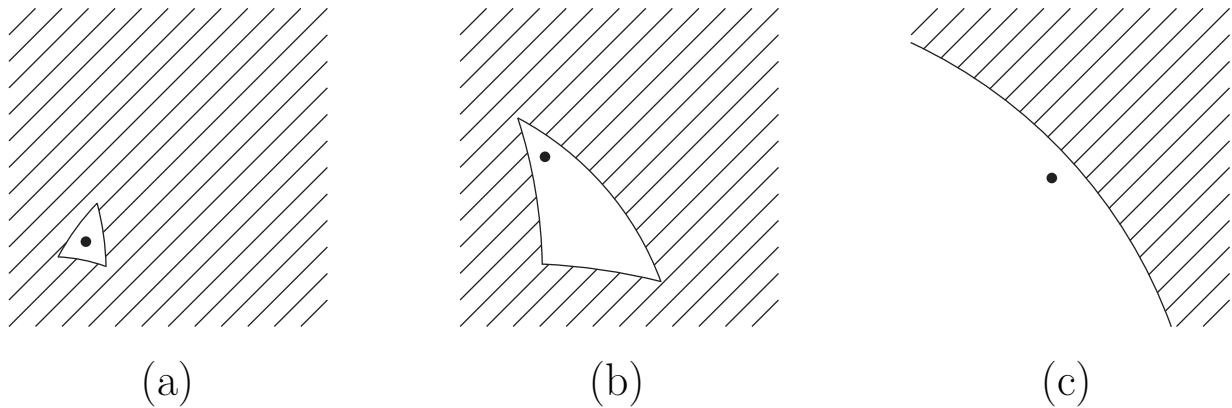
\begin{figure}[t]
\begin{center}
\begin{picture}(440,147)(31,-32)
%
%
  \Text(80,-13)[t]{\Large (a)}
  \CArc(118,-16)(90,136,152)
  \CArc(-43.4,20.7)(100,1,15)
  \CArc(35.9,-33.7)(60,70,87.6)
  \Line(20,110)(30,120)
  \Line(20,100)(40,120)
  \Line(20,90)(50,120)
  \Line(20,80)(60,120)
  \Line(20,70)(70,120)
  \Line(20,60)(80,120)
  \Line(20,50)(90,120)
  \Line(20,40)(100,120)
  \Line(20,30)(110,120)
  \Line(20,20)(120,120)
  \Line(20,10)(41.6,31.6) \Line(53.8,43.8)(130,120)
  \Line(20,0)(45.5,25.5)  \Line(55.5,35.5)(140,120)
  \Line(30,0)(53.6,23.6)  \Line(56.4,26.4)(140,110)
  \Line(40,0)(140,100)
  \Line(50,0)(140,90)
  \Line(60,0)(140,80)
  \Line(70,0)(140,70)
  \Line(80,0)(140,60)
  \Line(90,0)(140,50)
  \Line(100,0)(140,40)
  \Line(110,0)(140,30)
  \Line(120,0)(140,20)
  \Line(130,0)(140,10)
  \Vertex(49,32){2}
%
%
  \Text(250,-13)[t]{\Large (b)}
  \CArc(31,20)(190,1,18)
  \CArc(157.8,-22.9)(115,20.2,62)
  \CArc(210,-206.2)(230,76,87.3)
  \Line(190,110)(200,120)
  \Line(190,100)(210,120)
  \Line(190,90)(220,120)
  \Line(190,80)(230,120)
  \Line(190,70)(240,120)
  \Line(190,60)(250,120)
  \Line(190,50)(213.3,73.3) \Line(216.1,76.1)(260,120)
  \Line(190,40)(215.5,65.5) \Line(222.2,72.2)(270,120)
  \Line(190,30)(217.3,57.3) \Line(228.1,68.1)(280,120)
  \Line(190,20)(218.8,48.8) \Line(233.5,63.5)(290,120)
  \Line(190,10)(219.9,39.9) \Line(238.7,58.7)(300,120)
  \Line(190,0)(220.7,30.7)  \Line(243.6,53.6)(310,120)
  \Line(200,0)(223.4,23.4)  \Line(248.1,48.1)(310,110)
  \Line(210,0)(232.7,22.7)  \Line(252.4,42.4)(310,100)
  \Line(220,0)(241.6,21.6)  \Line(256.3,36.3)(310,90)
  \Line(230,0)(250.2,20.2)  \Line(259.9,29.9)(310,80)
  \Line(240,0)(258.6,18.6)  \Line(263.1,23.1)(310,70)
  \Line(250,0)(310,60)
  \Line(260,0)(310,50)
  \Line(270,0)(310,40)
  \Line(280,0)(310,30)
  \Line(290,0)(310,20)
  \Line(300,0)(310,10)
  \Vertex(222,64){2}
%
%
  \Text(420,-13)[t]{\Large (c)}
  \CArc(279.5,-65.1)(190,20,65)
  \Line(360,110)(370,120)
  \Line(364.6,104.6)(380,120)
  \Line(371.2,101.2)(390,120)
  \Line(377.5,97.5)(400,120)
  \Line(383.6,93.6)(410,120)
  \Line(389.6,89.6)(420,120)
  \Line(395.3,85.3)(430,120)
  \Line(400.9,80.9)(440,120)
  \Line(406.3,76.3)(450,120)
  \Line(411.4,71.4)(460,120)
  \Line(416.4,66.4)(470,120)
  \Line(421.2,61.2)(480,120)
  \Line(425.9,55.9)(480,110)
  \Line(430.3,50.3)(480,100)
  \Line(434.5,44.5)(480,90)
  \Line(438.5,38.5)(480,80)
  \Line(442.4,32.4)(480,70)
  \Line(446.1,26.1)(480,60)
  \Line(449.5,19.5)(480,50)
  \Line(452.7,12.7)(480,40)
  \Line(455.7,5.7)(480,30)
  \Line(460,0)(480,20)
  \Line(470,0)(480,10)
  \Vertex(413,56){2}
\end{picture}
\caption{The situations that lead to an observer naturalness problem. 
 The observed point is denoted by the little dot, while the region outside 
 ${\cal O}$ is shaded.}
\label{fig:obs_prob}
\end{center}
\end{figure}

We have seen that theories of particle physics and cosmology are likely 
to have observer naturalness problems, signaled by $P \ll 1$.  The 
problems arise in an ensemble if the observed point is close to the 
observer boundary.  The situations in which this happens can be classified 
into the following three cases: the observed point is (a) a typical 
point within a small observer region ${\cal O}$, (b) close to the 
boundary of a small observer region ${\cal O}$, and (c) close to the 
boundary of a large observer region ${\cal O}$, which may or may not 
be a closed region in the parameter space.  These cases are depicted 
schematically in Fig.~\ref{fig:obs_prob}.

The heart of the observer naturalness problem is the coincidence: why 
is the observed point so close to the observer boundary, which is a very 
special region in the parameter space?  To solve this problem, we must 
make the observer boundary {\it really} special --- the existence and 
location of the observer boundary should somehow affect the process of 
selecting a member in the ensemble as the one describing our universe. 
This leads us to consider environmental selection, effects sensitive 
to the existence of an observer.  In this section we study how 
environmental selection on a multiverse can solve observer naturalness 
problems in general, and under what circumstances (and in what sense) 
we can identify evidence for it.  We also discuss possible implications 
of this solution in identifying the correct theory describing our 
universe.

The nature of the observer naturalness problem is dramatically altered 
if the members in an ensemble are physically realized in a multiverse. 
Until now, the ensemble has been a useful fictitious mathematical device 
to study problems of naturalness; now we assume that each member of 
the ensemble is physically realized as a universe in spacetime.  The 
fundamental theory of nature is assumed to have a huge number of vacua, 
the landscape of vacua, so that physics at low energy may be described 
by many possible effective theories.  In fact, we are interested in 
a subset of these theories, $T$, that lead to physics described by 
the parameters $x_i$ discussed in section~\ref{sec:evid-prob}, with 
the parameters $x_i$ taking different values in different universes. 
We assume that it is possible to define a distribution function of 
the multiverse, $\tilde{f}_T(x_i)$, such that, averaging over the 
entire multiverse, we obtain $\langle x_i \rangle = \Sigma_T \int\! 
\tilde{f}_T(x_i)\, x_i\, dx_i$.  We stress that this is the average 
value on the entire multiverse, independent of whether universes 
contain observers.  If the population mechanism, including any relevant 
volume factors, were independent of $x_i$, then $\tilde{f}_T(x_i)$ 
would represent the distribution of vacua in the theory underlying 
the landscape.  However, in general it includes both relevant 
population and volume factors.%
\footnote{For subtleties in defining the distribution function in the 
 multiverse, see e.g.~\cite{Linde:2007nm} and references therein.}
We first focus on a single theory, so that we can drop the index $T$ 
until later subsections.

Environmental selection on the multiverse involves two key points 
that give a central role to measurements made by observers
\begin{itemize}
\item
Universes that have parameters outside the region ${\cal O}$ do not 
contain observers, hence values of $x_i$ outside ${\cal O}$ cannot be 
measured.  When discussing naturalness of the observed universe, we 
should not be asking questions about the entire multiverse but only 
about the observer region ${\cal O}$.
\item
The number of observers in universes in ${\cal O}$ with $x_i$ in the 
region $x_i$ to $x_i + dx_i$ is given by $\tilde{f}(x_i)\, n(x_i)\, 
dx_i \equiv f(x_i)\, dx_i$.  Each universe is to be weighted by an 
observer distribution $n(x_i)$, the factor associated with the number 
of observers that develop in a universe with parameters $x_i$.  We 
make no attempt to define an ``observer.''
\end{itemize}
These two points are really two aspects of a single selection process, 
with $n(x_i)$ defined over the entire space of $x_i$.  We, however, 
find it useful to consider that $n(x_i)$ rapidly drops to zero at 
certain observer boundaries, so that the region outside these boundaries 
does not affect the naturalness of a multiverse.

In the approximation that the observer boundary is sharp, the naturalness 
probability of Eq.~(\ref{eq:P_xo}) is then replaced by
\begin{equation}
  P_{\cal O} = \left| \frac{\int_{\bar{x}}^{x_o}\! f(x)\, dx} 
    {\int_{\cal O}\! f(x)\, dx} \right|,
\label{eq:P_xo_ES}
\end{equation}
where the integral in the denominator implies that $x$ is integrated 
only in the region ${\cal O}$.  The expression for higher dimensional 
$x$ space is also obtained similarly.  We stress that here and throughout 
the rest of the paper
\begin{equation}
  f(x_i) = \tilde{f}(x_i)\, n(x_i).
\label{eq:f}
\end{equation}
The entire region of parameter space of the multiverse is now irrelevant; 
the only question is whether we are typical observers in universes in 
${\cal O}$.  To calculate $\tilde{f}$ it is necessary to know both the 
landscape of vacua and the population mechanism.  On the other hand, $n$ 
is independent of the landscape and, with sufficient understanding of 
the physical environment for observers, could be calculated in principle 
from the low energy theory.  We take the practical viewpoint that both 
$\tilde{f}$ and $n$ are unknown, and hence give ourselves the freedom 
to assume any reasonable smooth distribution for $f$.  It is then clear 
that distributions can be found that make $P_{\cal O} \approx 1$, solving 
the observer naturalness problems whether of the form of (a), (b) or 
(c) of Fig.~\ref{fig:obs_prob}.  A more detailed description of each 
of these three types of solution is given in section~\ref{subsec:3-manif}. 
In section~\ref{subsec:cut_factor} we consider the relative probability 
of different theories, $T$, solving an observer naturalness problem.

It is quite clear that there are many origins for a significant $x_i$ 
dependence of $f(x_i) = \tilde{f}(x_i)\, n(x_i)$ in the observer 
region ${\cal O}$.  For $\tilde{f}$ these are the distribution of 
vacua in the underlying theory as well as relevant volume and population 
factors, while for $n$ a variation in $x_i$ could change the density 
of galaxies, stars, nuclei and so on that are relevant for observers. 
In section~\ref{subsec:force} we investigate a further effect: it could 
be that $f$ or the shape of ${\cal O}$ depends on more variables, $x_b$, 
than the ones we focus on, $x_a$, and that when these extra variables 
are integrated out the resulting effective distribution acquires further 
$x_a$ dependence.  Finally, a crucial issue is how to evaluate evidence 
for environmental selection.  In section~\ref{subsec:evid} we argue that, 
for any observer naturalness problem, a low value for the naturalness 
probability, $P$, provides evidence for environmental selection.  More 
precisely, for the known set of (simple) symmetry theories, $T$, the 
evidence is determined by the largest value of $P_T$
\begin{equation}
  P_{\rm max} = {\rm max}_T \{ P_T \}.
\label{eq:P_max}
\end{equation}
If $P_{\rm max} \approx 1$ there is no evidence, but as $P_{\rm max}$ 
decreases so the evidence becomes more substantial.

\subsection{Three manifestations of environmental selection}
\label{subsec:3-manif}

Using a set of Lagrangian parameters $x_i$ of some low energy theory, 
an observer naturalness problem might appear in the three ways 
illustrated in Fig.~\ref{fig:obs_prob}.  In each case environmental 
selection may be at work, but the description of the solution to 
the problem is different.

\begin{figure}[t]
\begin{center}
\begin{picture}(390,161)(11,-36)
%
%
  \Text(85,-23)[t]{\large $P \ll 1$}
  \Line(10,0)(160,0) \Line(160,0)(153,4) \Line(160,0)(153,-4)
  \Line(20,-10)(20,130) \Line(20,130)(16,123) \Line(20,130)(24,123)
  \CArc(118,-16)(90,136,152)
  \CArc(-43.4,20.7)(100,1,15)
  \CArc(35.9,-33.7)(60,70,87.6)
  \Vertex(49,32){2}
%
%
  \Line(190,63)(215,63) \Line(210,66)(220,60)
  \Line(190,57)(215,57) \Line(210,54)(220,60)
  \Line(192,63)(190,61) \Line(194,63)(190,59) \Line(196,63)(190,57)
  \Line(198,63)(192,57) \Line(200,63)(194,57) \Line(202,63)(196,57)
  \Line(204,63)(198,57) \Line(206,63)(200,57) \Line(208,63)(202,57)
  \Line(210,63)(204,57) \Line(212,63)(206,57) \Line(214,63)(208,57)
  \Line(215.7,62.7)(210,57) \Line(217.0,62.0)(212,57)
  \Line(218.2,61.2)(214,57) \Line(219.4,60.4)(217.6,58.6)
%
%
  \Text(325,-23)[t]{\large $P_{\cal O} \sim 1$}
  \Line(250,0)(400,0) \Line(400,0)(393,4) \Line(400,0)(393,-4)
  \Line(260,-10)(260,130) \Line(260,130)(256,123) \Line(260,130)(264,123)
  \CArc(358,-16)(90,136,152)
  \CArc(196.4,20.7)(100,1,15)
  \CArc(275.9,-33.7)(60,70,87.6)
  \Line(260,120)(265,125)
  \Line(260,115)(270,125)
  \Line(260,110)(275,125)
  \Line(260,105)(280,125)
  \Line(260,100)(285,125)
  \Line(260,95)(290,125)
  \Line(260,90)(295,125)
  \Line(260,85)(300,125)
  \Line(260,80)(305,125)
  \Line(260,75)(310,125)
  \Line(260,70)(315,125)
  \Line(260,65)(320,125)
  \Line(260,60)(325,125)
  \Line(260,55)(330,125)
  \Line(260,50)(335,125)
  \Line(260,45)(340,125)
  \Line(260,40)(345,125)
  \Line(260,35)(350,125)
  \Line(260,30)(355,125)
  \Line(260,25)(360,125)
  \Line(260,20)(365,125)
  \Line(260,15)(370,125)
  \Line(260,10)(281.6,31.6) \Line(293.8,43.8)(375,125)
  \Line(260,5)(281.1,26.1)  \Line(294.5,39.5)(380,125)
  \Line(260,0)(285.5,25.5)  \Line(295.5,35.5)(385,125)
  \Line(265,0)(289.7,24.7)  \Line(295.8,30.8)(390,125)
  \Line(270,0)(293.6,23.6)  \Line(296.4,26.4)(395,125)
  \Line(275,0)(395,120)
  \Line(280,0)(395,115)
  \Line(285,0)(395,110)
  \Line(290,0)(395,105)
  \Line(295,0)(395,100)
  \Line(300,0)(395,95)
  \Line(305,0)(395,90)
  \Line(310,0)(395,85)
  \Line(315,0)(395,80)
  \Line(320,0)(395,75)
  \Line(325,0)(395,70)
  \Line(330,0)(395,65)
  \Line(335,0)(395,60)
  \Line(340,0)(395,55)
  \Line(345,0)(395,50)
  \Line(350,0)(395,45)
  \Line(355,0)(395,40)
  \Line(360,0)(395,35)
  \Line(365,0)(395,30)
  \Line(370,0)(395,25)
  \Line(375,0)(395,20)
  \Line(380,0)(395,15)
  \Line(385,0)(395,10)
  \Line(390,0)(395,5)
  \Vertex(289,32){2}
\end{picture}
\caption{Replacement of the probability $P \rightarrow P_{\cal O}$, 
 illustrated for the case of (a) of Fig.~\ref{fig:obs_prob}.}
\label{fig:sol_prob}
\end{center}
\end{figure}
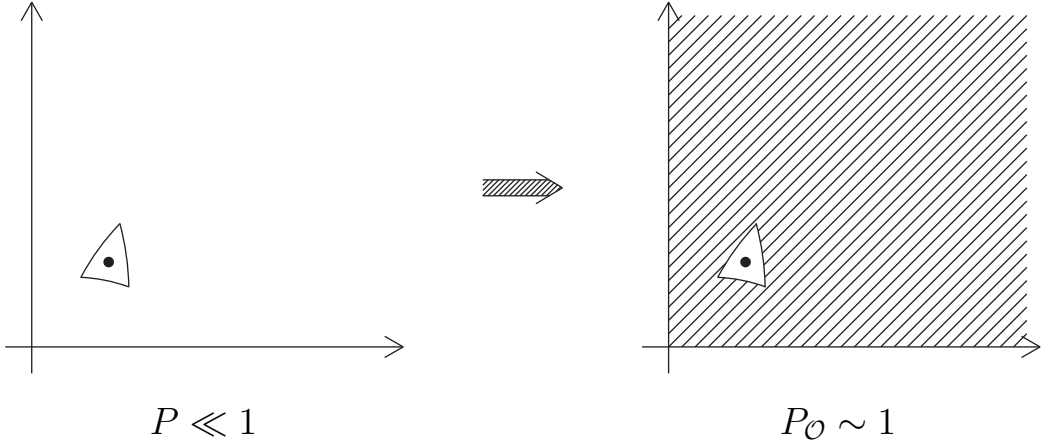

Suppose we find an observer naturalness problem of the form of (a) 
in Fig.~\ref{fig:obs_prob}.  This suggests that the distribution 
function, $f$, is (approximately) constant in the Lagrangian basis 
$x_i$, since the replacement of $P \rightarrow P_{\cal O}$ then 
completely eliminates the observer naturalness problem.  Assuming 
a constant $f$, the naturalness probability $P_{\cal O}$ of 
Eq.~(\ref{eq:P_xo_ES}) becomes
\begin{equation}
  P_{\cal O} = \left| \frac{x_o - \bar{x}}{\Delta x} \right|,
\label{eq:P_xo-2_ES}
\end{equation}
where $\Delta x$ is the range of $x$ in the observer region ${\cal O}$. 
Similarly, for multiple parameters the naturalness probability of 
Eq.~(\ref{eq:P_multi}) is replaced by
\begin{equation}
  P_{{\cal O}} = \left| \frac{c_n \{\Sigma_{a=1}^{n} 
    (x_{a,o} - \bar{x}_a)^2\}^{n/2}}{V_n({\cal O})} \right| 
  = \frac{v_n}{V_n({\cal O})},
\label{eq:P_multi_ES}
\end{equation}
where $V_n({\cal O})$ is the volume of parameter space in the observer 
region ${\cal O}$.  The elimination of observer naturalness problems 
by the replacement $P \rightarrow P_{\cal O}$ is illustrated in 
Fig.~\ref{fig:sol_prob}.
The enormous increase in the probability here arises simply because 
all of the universes that lie outside the region ${\cal O}$ are cut 
out of the denominator.

The situation in Fig.~\ref{fig:sol_prob} is the case of the overall 
picture for the cosmological constant $\Lambda$.  For conventional 
naturalness, consider an ensemble with a distribution function $f$ 
that is constant in $\Lambda$, which is the case in most theories. 
The probability that a member of this ensemble has a small value of 
$\Lambda_o$ near the special value zero is $P(\Lambda) = \Lambda_o/M_*^4$, 
where $M_*$ is the fundamental scale.  On the other hand, environmental 
selection, arising from a multiverse with $\tilde{f}$ and $n$ assumed 
flat in $\Lambda$, replaces $\Lambda_{\rm max} \approx M_*^4$ by 
$\Delta\Lambda \approx Q^3 T_{\rm eq}^4$, giving
\begin{equation}
  P_{\cal O}(\Lambda) \approx \frac{\Lambda_o}{Q^3 T_{\rm eq}^4}.
\label{eq:ES_for_CC}
\end{equation}
The vast majority of universes have a huge cosmological constant; but 
that is irrelevant because they contain only a dilute gas of elementary 
particles undergoing inflation.  We should cut all these universes out 
of the measure, and consider only those where large scale structure 
forms, allowing the possibility of complex observers.  This eliminates 
(or greatly ameliorates) the problem associated with the cosmological 
constant: $P(\Lambda) \ll 1 \rightarrow P_{\cal O}(\Lambda) \sim 1$. 
A more refined analysis includes the effects of nontrivial $n(\Lambda)$ 
near the observer boundary~\cite{Martel:1997vi}.

Let us now consider observer naturalness problems in the forms of (b) 
or (c) of Fig.~\ref{fig:obs_prob}.  In these cases, even after selection 
we are apparently left with some residual naturalness problem $P_{\cal O} 
\ll 1$ in some particular theory $T$.  What does this imply?  Does this 
mean that the observer naturalness problem in these forms cannot be 
solved by environmental selection?

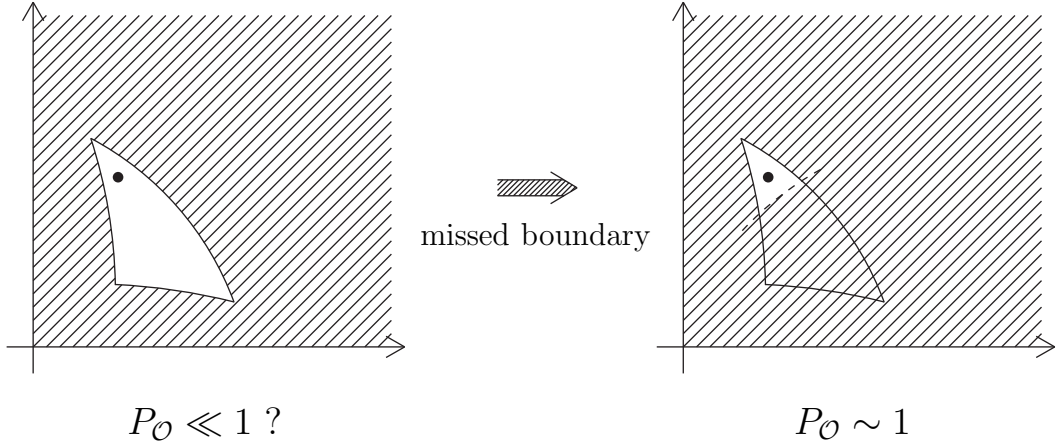
\begin{figure}[t]
\begin{center}
\begin{picture}(395,161)(6,-36)
%
%
  \Text(80,-23)[t]{\large $P_{\cal O} \ll 1$ ?}
  \Line(5,0)(155,0) \Line(155,0)(148,4) \Line(155,0)(148,-4)
  \Line(15,-10)(15,130) \Line(15,130)(11,123) \Line(15,130)(19,123)
  \CArc(-144,20)(190,1,18)
  \CArc(-17.2,-22.9)(115,20.2,62)
  \CArc(35,-206.2)(230,76,87.3)
  \Line(15,120)(20,125)
  \Line(15,115)(25,125)
  \Line(15,110)(30,125)
  \Line(15,105)(35,125)
  \Line(15,100)(40,125)
  \Line(15,95)(45,125)
  \Line(15,90)(50,125)
  \Line(15,85)(55,125)
  \Line(15,80)(60,125)
  \Line(15,75)(65,125)
  \Line(15,70)(70,125)
  \Line(15,65)(75,125)
  \Line(15,60)(80,125)
  \Line(15,55)(37.2,77.2) \Line(37.9,77.9)(85,125)
  \Line(15,50)(38.3,73.3) \Line(41.1,76.1)(90,125)
  \Line(15,45)(39.5,69.5) \Line(44.2,74.2)(95,125)
  \Line(15,40)(40.5,65.5) \Line(47.2,72.2)(100,125)
  \Line(15,35)(41.5,61.5) \Line(50.2,70.2)(105,125)
  \Line(15,30)(42.3,57.3) \Line(53.1,68.1)(110,125)
  \Line(15,25)(43.1,53.1) \Line(55.8,65.8)(115,125)
  \Line(15,20)(43.8,48.8) \Line(58.5,63.5)(120,125)
  \Line(15,15)(44.4,44.4) \Line(61.2,61.2)(125,125)
  \Line(15,10)(44.9,39.9) \Line(63.7,58.7)(130,125)
  \Line(15,5)(45.4,35.4)  \Line(66.2,56.2)(135,125)
  \Line(15,0)(45.7,30.7)  \Line(68.6,53.6)(140,125)
  \Line(20,0)(45.9,25.9)  \Line(70.9,50.9)(145,125)
  \Line(25,0)(48.4,23.4)  \Line(73.1,48.1)(150,125)
  \Line(30,0)(53.0,23.0)  \Line(75.3,45.3)(150,120)
  \Line(35,0)(57.7,22.7)  \Line(77.4,42.4)(150,115)
  \Line(40,0)(62.2,22.2)  \Line(79.3,39.3)(150,110)
  \Line(45,0)(66.6,21.6)  \Line(81.3,36.3)(150,105)
  \Line(50,0)(71.0,21.0)  \Line(83.1,33.1)(150,100)
  \Line(55,0)(75.2,20.2)  \Line(84.9,29.9)(150,95)
  \Line(60,0)(79.5,19.5)  \Line(86.6,26.6)(150,90)
  \Line(65,0)(83.6,18.6)  \Line(88.1,23.1)(150,85)
  \Line(70,0)(87.7,17.7)  \Line(89.6,19.6)(150,80)
  \Line(75,0)(150,75)
  \Line(80,0)(150,70)
  \Line(85,0)(150,65)
  \Line(90,0)(150,60)
  \Line(95,0)(150,55)
  \Line(100,0)(150,50)
  \Line(105,0)(150,45)
  \Line(110,0)(150,40)
  \Line(115,0)(150,35)
  \Line(120,0)(150,30)
  \Line(125,0)(150,25)
  \Line(130,0)(150,20)
  \Line(135,0)(150,15)
  \Line(140,0)(150,10)
  \Line(145,0)(150,5)
  \Vertex(47,64){2}
%
%
  \Line(190,63)(215,63) \Line(210,66)(220,60)
  \Line(190,57)(215,57) \Line(210,54)(220,60)
  \Line(192,63)(190,61) \Line(194,63)(190,59) \Line(196,63)(190,57)
  \Line(198,63)(192,57) \Line(200,63)(194,57) \Line(202,63)(196,57)
  \Line(204,63)(198,57) \Line(206,63)(200,57) \Line(208,63)(202,57)
  \Line(210,63)(204,57) \Line(212,63)(206,57) \Line(214,63)(208,57)
  \Line(215.7,62.7)(210,57) \Line(217.0,62.0)(212,57)
  \Line(218.2,61.2)(214,57) \Line(219.4,60.4)(217.6,58.6)
  \Text(205,47)[t]{missed boundary}
%
%
  \Text(325,-23)[t]{\large $P_{\cal O} \sim 1$}
  \Line(250,0)(400,0) \Line(400,0)(393,4) \Line(400,0)(393,-4)
  \Line(260,-10)(260,130) \Line(260,130)(256,123) \Line(260,130)(264,123)
  \CArc(101,20)(190,1,18)
  \CArc(227.8,-22.9)(115,20.2,62)
  \CArc(280,-206.2)(230,76,87.3)
  \DashCArc(370,-38)(120,119,137){3}
  \Line(260,120)(265,125)
  \Line(260,115)(270,125)
  \Line(260,110)(275,125)
  \Line(260,105)(280,125)
  \Line(260,100)(285,125)
  \Line(260,95)(290,125)
  \Line(260,90)(295,125)
  \Line(260,85)(300,125)
  \Line(260,80)(305,125)
  \Line(260,75)(310,125)
  \Line(260,70)(315,125)
  \Line(260,65)(320,125)
  \Line(260,60)(325,125)
  \Line(260,55)(282.2,77.2) \Line(282.9,77.9)(330,125)
  \Line(260,50)(283.3,73.3) \Line(286.1,76.1)(335,125)
  \Line(260,45)(284.5,69.5) \Line(289.2,74.2)(340,125)
  \Line(260,40)(285.5,65.5) \Line(292.2,72.2)(345,125)
  \Line(260,35)(286.5,61.5) \Line(295.2,70.2)(350,125)
  \Line(260,30)(287.3,57.3) \Line(298.1,68.1)(355,125)
  \Line(260,25)(288.1,53.1) \Line(300.8,65.8)(360,125)
  \Line(260,20)(297.5,57.5) \Line(303.5,63.5)(365,125)
  \Line(260,15)(370,125)
  \Line(260,10)(375,125)
  \Line(260,5)(380,125)
  \Line(260,0)(385,125)
  \Line(265,0)(390,125)
  \Line(270,0)(395,125)
  \Line(275,0)(395,120)
  \Line(280,0)(395,115)
  \Line(285,0)(395,110)
  \Line(290,0)(395,105)
  \Line(295,0)(395,100)
  \Line(300,0)(395,95)
  \Line(305,0)(395,90)
  \Line(310,0)(395,85)
  \Line(315,0)(395,80)
  \Line(320,0)(395,75)
  \Line(325,0)(395,70)
  \Line(330,0)(395,65)
  \Line(335,0)(395,60)
  \Line(340,0)(395,55)
  \Line(345,0)(395,50)
  \Line(350,0)(395,45)
  \Line(355,0)(395,40)
  \Line(360,0)(395,35)
  \Line(365,0)(395,30)
  \Line(370,0)(395,25)
  \Line(375,0)(395,20)
  \Line(380,0)(395,15)
  \Line(385,0)(395,10)
  \Line(390,0)(395,5)
  \Vertex(292,64){2}
\end{picture}
\caption{Illustration of the situation in which a missed boundary (shown 
 dashed) leads to a fictitious problem of $P_{\cal O} \ll 1$ after 
 environmental selection.}
\label{fig:missed_bound}
\end{center}
\end{figure}

The answer is clearly no.  To understand what can be going on, it is 
important to realize that our knowledge can often be very incomplete. 
Imagine that we have assumed some distribution function $f$ for 
an ensemble and found an observer naturalness problem in the form 
of (b) or (c) of Fig.~\ref{fig:obs_prob} in the basis where $f$ 
is constant.  This seems to imply that the observer naturalness 
problem cannot be solved by a simple cutout procedure illustrated 
in Fig.~\ref{fig:sol_prob}.  However, do we really know that we 
have identified all the relevant complexity boundaries, and thus 
identified the correct observer boundary?  It could be that we have 
missed some relevant boundary and, after taking it into account, 
$P_{\cal O}$ may become order unity.  This is illustrated in 
Fig.~\ref{fig:missed_bound} for the case of (b) of Fig.~\ref{fig:obs_prob}.

Another possibility is that our initial assumption on the distribution 
function may not be correct.  In practice, when we consider a theory, 
we start by assuming a ``natural'' distribution function $f(x_i)$, 
which is often taken to be constant in ``Lagrangian parameters'' $x_i$. 
However, there are {\it many} reasons why the distribution function in 
this parameter basis may not be constant even approximately.  First, 
the Lagrangian parameters $x_i$ of the low energy theory may be functions 
of the Lagrangian parameters $y_j$ in the fundamental theory, $x_i = 
x_i(y_j)$, and the distribution function may be constant in $y_j$ rather 
than $x_i$.  Second, in the multiverse picture, a mismatch between the 
naive and true distribution functions may also result from the population 
mechanism depending on $x_i$, or may be induced by environmental 
selection via a nontrivial $n(x_i)$.  Finally, environmental selection 
may depend not only on $x_i$ but also on other variables, inducing 
an extra dependence of $f$ on $x_i$, as we will discuss in the next 
subsection.  Hence, the distribution function $f(x_i)$ may have 
a strong dependence on the Lagrangian parameters $x_i$ of the low 
energy theory, and in this case the situations of (b) or (c) in 
Fig.~\ref{fig:obs_prob} actually correspond to typical observers 
with $P_{\cal O} \approx 1$, expected from environmental selection.

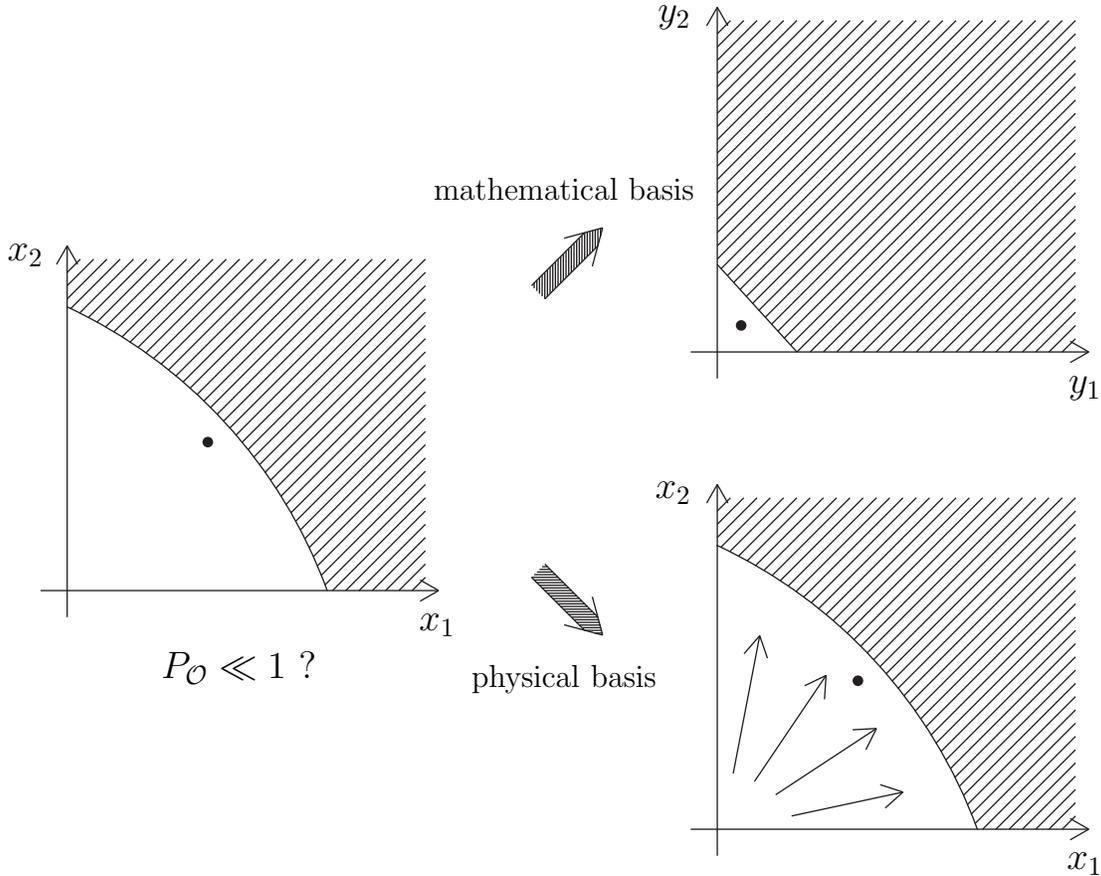
\begin{figure}[t]
\begin{center}
\begin{picture}(395,323)(6,-108)
%
%
  \Text(80,-23)[t]{\large $P_{\cal O} \ll 1$ ?}
  \Line(5,0)(155,0) \Line(155,0)(148,4) \Line(155,0)(148,-4)
  \Line(15,-10)(15,130) \Line(15,130)(11,123) \Line(15,130)(19,123)
  \Text(155,-9)[t]{\large $x_1$} \Text(6,127)[r]{\large $x_2$}
  \CArc(-65.5,-65.1)(190,20,65)
  \Line(15,120)(20,125)
  \Line(15,115)(25,125)
  \Line(15,110)(30,125)
  \Line(16.3,106.3)(35,125)
  \Line(19.6,104.6)(40,125)
  \Line(22.9,102.9)(45,125)
  \Line(26.2,101.2)(50,125)
  \Line(29.4,99.4)(55,125)
  \Line(32.5,97.5)(60,125)
  \Line(35.6,95.6)(65,125)
  \Line(38.6,93.6)(70,125)
  \Line(41.7,91.7)(75,125)
  \Line(44.6,89.6)(80,125)
  \Line(47.5,87.5)(85,125)
  \Line(50.3,85.3)(90,125)
  \Line(53.2,83.2)(95,125)
  \Line(55.9,80.9)(100,125)
  \Line(58.6,78.6)(105,125)
  \Line(61.3,76.3)(110,125)
  \Line(63.9,73.9)(115,125)
  \Line(66.4,71.4)(120,125)
  \Line(69.0,69.0)(125,125)
  \Line(71.4,66.4)(130,125)
  \Line(73.9,63.9)(135,125)
  \Line(76.2,61.2)(140,125)
  \Line(78.6,58.6)(145,125)
  \Line(80.9,55.9)(150,125)
  \Line(83.1,53.1)(150,120)
  \Line(85.3,50.3)(150,115)
  \Line(87.5,47.5)(150,110)
  \Line(89.5,44.5)(150,105)
  \Line(91.6,41.6)(150,100)
  \Line(93.5,38.5)(150,95)
  \Line(95.5,35.5)(150,90)
  \Line(97.4,32.4)(150,85)
  \Line(99.3,29.3)(150,80)
  \Line(101.1,26.1)(150,75)
  \Line(102.8,22.8)(150,70)
  \Line(104.5,19.5)(150,65)
  \Line(106.2,16.2)(150,60)
  \Line(107.7,12.7)(150,55)
  \Line(109.3,9.3)(150,50)
  \Line(110.7,5.7)(150,45)
  \Line(112.1,2.1)(150,40)
  \Line(115,0)(150,35)
  \Line(120,0)(150,30)
  \Line(125,0)(150,25)
  \Line(130,0)(150,20)
  \Line(135,0)(150,15)
  \Line(140,0)(150,10)
  \Line(145,0)(150,5)
  \Vertex(68,56){2}
%
%
  \Line(190,115)(210,135) \Line(195,110)(215,130)
  \Line(216.8,136.8)(202.3,132.9) \Line(216.8,136.8)(212.9,122.3)
  \Line(191,116)(191,114) \Line(192,117)(192,113)
  \Line(193,118)(193,112) \Line(194,119)(194,111)
  \Line(195,120)(195,110) \Line(196,121)(196,111)
  \Line(197,122)(197,112) \Line(198,123)(198,113)
  \Line(199,124)(199,114) \Line(200,125)(200,115)
  \Line(201,126)(201,116) \Line(202,127)(202,117)
  \Line(203,128)(203,118) \Line(204,129)(204,119)
  \Line(205,130)(205,120) \Line(206,131)(206,121)
  \Line(207,132)(207,122) \Line(208,133)(208,123)
  \Line(209,134)(209,124) \Line(210,135)(210,125)
  \Line(211,135.4)(211,126) \Line(212,135.7)(212,127)
  \Line(213,135.9)(213,128) \Line(214,136.2)(214,129)
  \Line(215,136.5)(215,129.9) \Line(216,136.7)(216,133.7)
  \Text(203,148)[b]{mathematical basis}
  \Line(190,5)(210,-15) \Line(195,10)(215,-10)
  \Line(216.8,-16.8)(202.3,-12.9) \Line(216.8,-16.8)(212.9,-2.3)
  \Line(194,9)(196,9) \Line(193,8)(197,8)
  \Line(192,7)(198,7) \Line(191,6)(199,6)
  \Line(190,5)(200,5) \Line(191,4)(201,4)
  \Line(192,3)(202,3) \Line(193,2)(203,2)
  \Line(194,1)(204,1) \Line(195,0)(205,0)
  \Line(196,-1)(206,-1) \Line(197,-2)(207,-2)
  \Line(198,-3)(208,-3) \Line(199,-4)(209,-4)
  \Line(200,-5)(210,-5) \Line(201,-6)(211,-6)
  \Line(202,-7)(212,-7) \Line(203,-8)(213,-8)
  \Line(204,-9)(214,-9) \Line(205,-10)(215,-10)
  \Line(206,-11)(215.4,-11) \Line(207,-12)(215.7,-12)
  \Line(208,-13)(216.0,-13) \Line(209,-14)(216.2,-14)
  \Line(210,-15)(216.5,-15) \Line(213.7,-16)(216.8,-16)
  \Text(203,-28)[t]{physical basis}
%
%
  \Line(250,90)(400,90) \Line(400,90)(393,94) \Line(400,90)(393,86)
  \Line(260,80)(260,220) \Line(260,220)(256,213) \Line(260,220)(264,213)
  \Text(400,81)[t]{\large $y_1$} \Text(251,217)[r]{\large $y_2$}
  \Line(290,90)(260,123)
  \Line(260,210)(265,215)
  \Line(260,205)(270,215)
  \Line(260,200)(275,215)
  \Line(260,195)(280,215)
  \Line(260,190)(285,215)
  \Line(260,185)(290,215)
  \Line(260,180)(295,215)
  \Line(260,175)(300,215)
  \Line(260,170)(305,215)
  \Line(260,165)(310,215)
  \Line(260,160)(315,215)
  \Line(260,155)(320,215)
  \Line(260,150)(325,215)
  \Line(260,145)(330,215)
  \Line(260,140)(335,215)
  \Line(260,135)(340,215)
  \Line(260,130)(345,215)
  \Line(260,125)(350,215)
  \Line(261.3,121.3)(355,215)
  \Line(263.8,118.8)(360,215)
  \Line(266.1,116.1)(365,215)
  \Line(268.5,113.5)(370,215)
  \Line(270.9,110.9)(375,215)
  \Line(273.3,108.3)(380,215)
  \Line(275.6,105.6)(385,215)
  \Line(278.0,103.0)(390,215)
  \Line(280.4,100.4)(395,215)
  \Line(282.8,97.8)(395,210)
  \Line(285.2,95.2)(395,205)
  \Line(287.6,92.6)(395,200)
  \Line(290,90)(395,195)
  \Line(295,90)(395,190)
  \Line(300,90)(395,185)
  \Line(305,90)(395,180)
  \Line(310,90)(395,175)
  \Line(315,90)(395,170)
  \Line(320,90)(395,165)
  \Line(325,90)(395,160)
  \Line(330,90)(395,155)
  \Line(335,90)(395,150)
  \Line(340,90)(395,145)
  \Line(345,90)(395,140)
  \Line(350,90)(395,135)
  \Line(355,90)(395,130)
  \Line(360,90)(395,125)
  \Line(365,90)(395,120)
  \Line(370,90)(395,115)
  \Line(375,90)(395,110)
  \Line(380,90)(395,105)
  \Line(385,90)(395,100)
  \Line(390,90)(395,95)
  \Vertex(269,100){2}
%
%
  \Line(250,-90)(400,-90) \Line(400,-90)(393,-86) \Line(400,-90)(393,-94)
  \Line(260,-100)(260,40) \Line(260,40)(256,33) \Line(260,40)(264,33)
  \Text(400,-99)[t]{\large $x_1$} \Text(251,37)[r]{\large $x_2$}
  \CArc(179.5,-155.1)(190,20,65)
  \Line(260,30)(265,35)
  \Line(260,25)(270,35)
  \Line(260,20)(275,35)
  \Line(261.3,16.3)(280,35)
  \Line(264.6,14.6)(285,35)
  \Line(267.9,12.9)(290,35)
  \Line(271.2,11.2)(295,35)
  \Line(274.4,9.4)(300,35)
  \Line(277.5,7.5)(305,35)
  \Line(280.6,5.6)(310,35)
  \Line(283.6,3.6)(315,35)
  \Line(286.7,1.7)(320,35)
  \Line(289.6,-0.4)(325,35)
  \Line(292.5,-2.5)(330,35)
  \Line(295.3,-4.7)(335,35)
  \Line(298.2,-6.8)(340,35)
  \Line(300.9,-9.1)(345,35)
  \Line(303.6,-11.4)(350,35)
  \Line(306.3,-13.7)(355,35)
  \Line(308.9,-16.1)(360,35)
  \Line(311.4,-18.6)(365,35)
  \Line(314.0,-21.0)(370,35)
  \Line(316.4,-23.6)(375,35)
  \Line(318.9,-26.1)(380,35)
  \Line(321.2,-28.8)(385,35)
  \Line(323.6,-31.4)(390,35)
  \Line(325.9,-34.1)(395,35)
  \Line(328.1,-36.9)(395,30)
  \Line(330.3,-39.7)(395,25)
  \Line(332.5,-42.5)(395,20)
  \Line(334.5,-45.5)(395,15)
  \Line(336.6,-48.4)(395,10)
  \Line(338.5,-51.5)(395,5)
  \Line(340.5,-54.5)(395,0)
  \Line(342.4,-57.6)(395,-5)
  \Line(344.3,-60.7)(395,-10)
  \Line(346.1,-63.9)(395,-15)
  \Line(347.8,-67.2)(395,-20)
  \Line(349.5,-70.5)(395,-25)
  \Line(351.2,-73.8)(395,-30)
  \Line(352.7,-77.3)(395,-35)
  \Line(354.3,-80.7)(395,-40)
  \Line(355.7,-84.3)(395,-45)
  \Line(357.1,-87.9)(395,-50)
  \Line(360,-90)(395,-55)
  \Line(365,-90)(395,-60)
  \Line(370,-90)(395,-65)
  \Line(375,-90)(395,-70)
  \Line(380,-90)(395,-75)
  \Line(385,-90)(395,-80)
  \Line(390,-90)(395,-85)
  \Vertex(313,-34){2}
  \Line(288,-85)(330,-76)\Line(330,-76)(321.4,-73.2)\Line(330,-76)(323.3,-82.0)
  \Line(282,-77)(320,-52)\Line(320,-52)(311.0,-52.5)\Line(320,-52)(316.0,-60.0)
  \Line(274,-72)(301,-32)\Line(301,-32)(292.9,-35.9)\Line(301,-32)(300.4,-41.0)
  \Line(266,-69)(276,-17)\Line(276,-17)(270.1,-23.8)\Line(276,-17)(278.9,-25.5)
\end{picture}
\caption{Illustration of the situation in which naive and true distribution 
 functions differ significantly.  In the basis where the true distribution 
 function is constant (the mathematical basis), the observed point is 
 a typical point in ${\cal O}$, and the parameters chosen, $y_i$, are in 
 general nontrivial functions of ``Lagrangian parameters'' $x_i$.  In the 
 basis spanned by $x_i$ (the physical basis), the distribution function 
 is nontrivial, giving a ``probability force'' field, defined in 
 Eq.~(\ref{eq:F}).}
\label{fig:diff_distr}
\end{center}
\end{figure}

The situation in which naive and true distribution functions differ 
significantly is illustrated in Fig.~\ref{fig:diff_distr} for the case 
of (c) of Fig.~\ref{fig:obs_prob}. 
(For the discussion here we assume that all relevant parts of the 
observer boundary have been correctly identified.)  There are two 
(equivalent) ways to describe this situation.  One is to go to the 
basis in which the true distribution function is constant.  In this 
basis, which we may call the ``mathematical basis,'' the observed point 
is a typical point in ${\cal O}$, and the parameters chosen, $y_i$, 
can in general be nontrivial functions of the ``Lagrangian parameters'' 
$x_i$.  In practice, the transformation from $x_i$ to $y_i$ should be 
inferred from the observed data and a calculated observer boundary. 
Another way to describe the situation is to keep using Lagrangian 
parameters, or parameters that have the most direct or intuitive 
physical meaning at low energies, $x_i$.  In this basis, which we may 
call the ``physical basis,'' we have a nontrivial distribution function 
$f(x_i)$.  The distribution function is peaked towards the region 
outside ${\cal O}$, so that a typical observer in ${\cal O}$ measures 
$x_i$ close to the observer boundary.

In the physical basis, $x_i$, it is very convenient to introduce a 
``probability force,'' which is simply the gradient of the probability 
distribution:
\begin{equation}
  {\bf F} = \nabla f,
\label{eq:F}
\end{equation}
where $\nabla \equiv (\partial/\partial x_1,\partial/\partial x_2,\cdots)$. 
Within the region ${\cal O}$, this vector field gives an indication of 
what values of physical parameters are most likely to be measured in 
the multiverse.  If the vector field indicates a flow towards an observer 
boundary, observers should be living close to the corresponding boundary. 
An example of the probability force field is depicted in the physical 
basis plot of Fig.~\ref{fig:diff_distr}.

\subsection{Effective distributions from ``integrating out'' parameters}
\label{subsec:force}

What are the possible origins of the probability force?  One is a 
nontrivial prior distribution function $f_{\rm prior}$, which arises 
from a mismatch between the physical and mathematical bases, as discussed 
previously.  In practice, however, this is not the only source of the 
probability force.  The key is that when we discuss physical predictions 
in the multiverse picture, we typically choose a subset, $x_a$, of the 
entire parameter set, $x_i$, and study probabilistic predictions in the 
space of these chosen parameters $x_a$.  An important point here is that 
when we focus on $x_a$, we cannot simply ignore the other variables $x_b$ 
($i = \{ a,b \}$); they must be {\it integrated out}.  This provides an 
effect on the distribution function {\it defined in the subspace $x_a$}, 
and thus modifies the probability force in $x_a$ space.

To illustrate this idea, let us consider the simplest example of a constant 
$f_{\rm prior}$, i.e. the case in which the physical and mathematical bases 
coincide.  In this case the distribution of observers is given by
\begin{equation}
  d{\cal N} = c\, dx_1 \cdots dx_N,
\label{eq:dN-1}
\end{equation}
where $c = f_{\rm prior}$ is a constant, and the probability force in 
$x_i$ space is zero, ${\bf F} = 0$.  This, however, does not mean that the 
probability force is zero if we consider only a subset of the variables, 
e.g. $x_1, \cdots, x_{N-1}$.  When we focus on the variables $x_1, \cdots, 
x_{N-1}$ (as we focused only on $y_u$, $y_d$, $y_e$, $v/\Lambda_{\rm QCD}$ 
and $\alpha$ out of the 13~parameters of Eqs.~(\ref{eq:x_SM},~\ref{eq:x_cos}) 
in sections~\ref{sec:evid-prob} and \ref{sec:flavor}), we must integrate 
out the other variable $x_N$.  Now, in the multiverse picture with 
environmental selection, the only relevant universes are those within 
the observer region ${\cal O}$, implying that integrals should be 
performed only over ${\cal O}$.  The domain of $x_N$ integration 
is then determined by the observer boundary, which is generically 
written as $x_N^{\rm min}(x_1,\cdots,x_{N-1}) \leq x_N \leq 
x_N^{\rm max}(x_1,\cdots,x_{N-1})$,%
\footnote{Depending on the shape of ${\cal O}$, we may have to integrate 
 $x_N$ over multiple domains for some values of $x_1, \cdots, x_{N-1}$.}
giving
\begin{equation}
  d{\cal N} = f_{\rm eff}(x_1,\cdots,x_{N-1})\, dx_1 \cdots dx_{N-1},
\label{eq:dN-2}
\end{equation}
where $f_{\rm eff}(x_1,\cdots,x_{N-1}) \equiv c\,\{ x_N^{\rm 
max}(x_1,\cdots,x_{N-1}) - x_N^{\rm min}(x_1,\cdots,x_{N-1}) \}$. 
The probability force {\it defined in $x_1$-$\cdots$-$x_{N-1}$ space} 
is then obtained by Eq.~(\ref{eq:F}) with $f$ replaced by $f_{\rm eff}$: 
${\bf F}_{\rm eff} = \nabla f_{\rm eff}$, where $\nabla = (\partial/
\partial x_1,\cdots,\partial/\partial x_{N-1})$.  This generically 
gives a nontrivial force, ${\bf F}_{\rm eff} \neq {\bf 0}$.

In general, when we consider physics of the landscape in the subspace 
of $x_a$ ($a = 1,\cdots,n$), we must integrate out the other variables 
$x_b$ ($b=n+1,\cdots,N$) to obtain the effective distribution function, 
and thus the probability force, defined in $x_a$ space.  The domain of 
integration is given by the observer region ${\cal O}$, leading to
\begin{equation}
  f_{\rm eff}(x_1,\cdots,x_n) 
    = \int \cdots \int_{\cal O} f_{\rm prior}(x_1,\cdots,x_N)\, 
      dx_{n+1} \cdots dx_N,
\label{eq:f_eff}
\end{equation}
where $f_{\rm prior} = \tilde{f}_{\rm prior} n$ is the prior distribution 
function defined in the entire parameter space $x_i$, whose $x_b$ 
dependence could arise from both $\tilde{f}_{\rm prior}$ and $n$. 
The important point here is that the effective distribution function 
$f_{\rm eff}$ in $x_a$ space is {\it not} obtained simply by ``neglecting'' 
the other variables $x_b$ in $f_{\rm prior}$, i.e. by setting $x_b$ to 
the observed values in $f_{\rm prior}$:
\begin{equation}
  f_{\rm eff}(x_1,\cdots,x_n) 
    \neq f_{\rm prior}(x_1,\cdots,x_n,x_{n+1,o},\cdots,x_{N,o}).
\label{eq:f_eff-wrong}
\end{equation}
Since the observer region ${\cal O}$ can in general have a complicated 
shape in $x_i$ space, the effective distribution function can have 
a quite different form than the expression in the right-hand-side of 
Eq.~(\ref{eq:f_eff-wrong}).  The effective probability force defined 
in $x_a$ space is then given by
\begin{equation}
  {\bf F}_{\rm eff} = \nabla f_{\rm eff},
\label{eq:F_eff}
\end{equation}
where $\nabla = (\partial/\partial x_1,\cdots,\partial/\partial x_n)$.

The argument here suggests that it is a rather common phenomenon to 
have a nontrivial probability force when we focus only on a subset 
of the entire parameter set $x_i$, which is almost always the case. 
Then, if the resulting probability force is strong, we are likely to 
encounter an observer naturalness problem in the form of (b) or (c) 
in Fig.~\ref{fig:obs_prob}.

\subsection{Cut factor and comparisons between different theories}
\label{subsec:cut_factor}

In general there are many theories $T$ that lead to the low energy 
Lagrangian with parameters $x$ and that are described by a multiverse 
distribution $f_T(x)$.  As an example, $x$ may be the Yukawa couplings 
of the up and down quarks and electron, $y_{u,d,e}$, and $T$ may 
label the various theories of flavor.  The number of observers 
in the multiverse who are governed by theory $T$ is
\begin{equation}
  n_T = \int_{\cal O} \tilde{f}_T(x)\, n(x)\, dx.
\label{eq:n_T}
\end{equation}
A typical observer will be governed by the theory $T$ that has the 
maximum value of $n_T$ (assuming that several $T$ do not have comparable 
$n_T$).  The absolute normalization of $n(x)$ is unimportant, since 
we are only interested in relative numbers of observers governed 
by different theories
\begin{equation}
  \frac{n_T}{n_{T'}} 
  = \frac{\int_{\cal O} \tilde{f}_T(x)\, n(x)\, dx} 
      {\int_{\cal O} \tilde{f}_{T'}(x)\, n(x)\, dx} 
  = \frac{{\cal N}_T}{{\cal N}_{T'}} 
      \frac{\tilde{C}_T^{-1}}{\tilde{C}_{T'}^{-1}},
\label{eq:n_TT'}
\end{equation}
where ${\cal N}_T = \int \tilde{f}_T(x)\, dx$ is the total ``number'' 
of universes (including volume factor weights) described by theory 
$T$ and
\begin{equation}
  \tilde{C}_T = \frac{\int\! \tilde{f}_T(x)\, dx} 
    {\int_{\cal O}\! \tilde{f}_T(x) n(x)\, dx}.
\label{eq:cut}
\end{equation}

Suppose now that $n(x)$ is relatively flat over ${\cal O}$ so that it 
is a good approximation to take it constant.  Multiplying this constant 
by $\tilde{C}_T$ then gives
\begin{equation}
  C_T = \frac{\int\! \tilde{f}_T(x)\, dx} 
    {\int_{\cal O}\! \tilde{f}_T(x)\, dx}.
\label{eq:cut-2}
\end{equation}
In this case there are a factor $C_T$ more universes described by $T$ 
in the multiverse than in ${\cal O}$.  Environmental selection solves 
observer naturalness problems at a cost of removing a factor $C_T$ 
of these universes --- we call $C_T$ the cut factor for theory $T$.

Consider, for example, the following two theories --- the Standard 
Model where only the Higgs mass-squared parameter scans with 
a distribution function constant in $|m_h^2|$ and a theory 
beyond the Standard Model where only the weak scale scans with 
a distribution function constant in $1/\ln v$.  Considering selection 
by Eq.~(\ref{eq:cond-v}), the cut factors in these theories are 
$O(10^{32})$ and $O(100)$, respectively.  If we take this naively, 
it seems to suggest that the latter theory is preferred over the 
former, and we might say that the Standard Model is less likely 
since it involves the cost of a much larger cut factor.%
\footnote{Here we do not consider the cut factor arising from selection 
 of the cosmological constant, $\Lambda \simlt Q^3 T_{\rm eq}^4$. 
 In fact, the size of the cut factor associated with the cosmological 
 constant does not depend, e.g., on the scale of supersymmetry breaking 
 if the constant term in the superpotential, $W_0$, scans up to the 
 fundamental scale, $M_*$, with the distribution flat in $|W_0|^2$, 
 as suggest by string theory~\cite{Denef:2004ze}.  This part of the 
 cut factor is then of order $M_*^4/\Lambda \approx 10^{120}$ regardless 
 of the theory.  (An exception for this is given by supersymmetric 
 theories in which both supersymmetry and $R$ symmetry are broken 
 at low energies.)}
(This result is similar to that of the conventional naturalness argument, 
although here observable universes always have $v \ll M_{\rm Pl}$ due 
to selection effects.)  Of course this is not rigorous --- our ignorance 
of ${\cal N}_T$, as well as other possible selection effects, makes 
the argument based only on the cut factor unreliable; without knowledge 
of ${\cal N}_T$ there is no real reason to prefer a theory beyond the 
Standard Model over the Standard Model (other than considerations based 
on other physics such as gauge coupling unification and dark matter). 
Nevertheless, the relative cut factors for two theories do contribute 
to the relative number of observers described by each theory, as in 
Eq.~(\ref{eq:n_TT'}), hence we may describe a large cut factor for some 
theory as a cost in its solution to an observer naturalness problem.

\subsection{Evidence for environmental selection}
\label{subsec:evid}

We have seen that observer naturalness problems, which appear in the 
form of (a), (b) or (c) of Fig.~\ref{fig:obs_prob}, can be solved in 
general by environmental selection.  Turning this argument around, 
we find that effects of environmental selection can show up in one 
or both of the following forms:
\begin{enumerate}
\item[(i)]
{\it Amazing coincidences:} we are living in a region of parameter space 
 that admits complex observers, which, however, is only a (very) small 
 portion of the entire parameter space.
\item[(ii)]
{\it Living on the edge:} we are living (very) near an ``edge'' of the 
 parameter region ${\cal O}$ that admits complex observers, i.e. physical 
 parameters take values that correspond to a point very close to the 
 observer boundary, compared with the extent of the region ${\cal O}$.
\end{enumerate}
In fact, how environmental selection manifests itself depends on the 
basis --- we have seen that a phenomenon that appears as (ii) in one 
basis can appear as (i) in another basis, or vice versa.  However, 
since we are often presented with a natural basis in which the physical 
meaning of parameters is most direct and/or intuitive, it is meaningful 
to consider the manifestation of environmental selection in that particular 
basis (physical basis).  As we have seen, the manifestation then takes 
the form of (i) if the effective distribution function $f_{\rm eff}(x)$ 
is (nearly) constant over the observer region ${\cal O}$, while it takes 
the form of (ii) or a combination of (i) and (ii) if $f_{\rm eff}(x)$ 
is peaked towards a boundary of ${\cal O}$.  There are many possible 
origins for a nontrivial form of $f_{\rm eff}(x)$: an $x$ dependence 
of the landscape vacua, the dynamics of the population mechanism, 
the observer factor $n(x)$, and environmental selection acting 
on variables other than $x$.

Can the observation of one or both of the above two phenomena, (i) and 
(ii), be viewed as evidence for environmental selection?  In a given 
theory, $T$, the answer is yes, since otherwise it is very difficult 
to explain these features.  Even though there is an observer naturalness 
problem associated with a very small value for $P_T$, after environmental 
selection a typical point within ${\cal O}$ results, i.e.
\begin{equation}
  P_T \rightarrow P_{{\cal O},T} \approx 1.
\label{eq:ES}
\end{equation}
The smaller the original $P_T$, the more amazing the coincidence, 
or the closer to the edge we are living.  The significance for the 
evidence for environmental selection is then quantified by the size 
of the naturalness probability $P_T$, with a smaller value of $P_T$ 
corresponding to stronger evidence.

In practice, however, we do not know beforehand the correct theory 
describing our universe.  Suppose we encounter a situation described 
by (i) or (ii) above.  Do we conclude that we have found evidence for 
environmental selection, or that the theory we are considering is simply 
wrong?  Whether an observation of the phenomenon described in (i) or 
(ii) -- an observer naturalness problem -- can be viewed as evidence 
for environmental selection depends on whether one can find an alternative 
theory in which the problem does not arise.  A simple alternative theory 
without the naturalness problem may provide a better description of our 
universe.  On the other hand, it is possible that we cannot find such 
a theory, or can find only theories that are significantly more complicated. 
Then we may conclude that environmental selection provides the best 
explanation of the phenomenon, so that we provisionally accept the 
multiverse theory.

In principle, with competing theories to describe nature, the evidence 
for environmental selection at a particular observer boundary is given 
by the largest value of $P_T$ associated with that boundary.  However, 
it may be that a theory with large $P_T$ is very complicated or ad hoc, 
so that it does not appear to be an adequate solution to the observer 
naturalness problem.  In this case it may be judged that the evidence 
for environmental selection is better represented by a smaller value 
of $P_T$ associated with some simpler theory.

The discussion given above illustrates why the cosmological constant is 
a powerful argument for environmental selection: in all known quantum 
field theories the naturalness probability is extremely small ($P_T 
\approx 10^{-120}$~--~$10^{-60}$ depending on the presence of weak 
scale supersymmetry and the nature of its breaking), implying that 
the largest value of $P_T$ is of order $10^{-60}$ or smaller.  The 
absence of a simple theory with $P_T = O(1)$ is crucial in this argument. 
A similar argument may also be made for the quark and lepton masses 
discussed in sections~\ref{sec:evid-prob} and \ref{sec:flavor}.  We 
have argued that all known theories of flavor are quite inadequate 
to explain the relevant observer naturalness problems, leading to 
$P_T \simlt (10^{-3}$~--~$10^{-2})$.  Although this is numerically less 
impressive than the case of the cosmological constant, it is nevertheless 
very important.  A single piece of evidence for environmental selection, 
no matter how significant, could be completely erased by the discovery 
of a single new theory.  The more independent pieces of evidence for 
environmental selection, the more convincing the overall picture becomes.

One might think it always difficult to ``confirm'' the absence of 
alternative theories which do not have the problem.  Indeed, the absence 
of such theories can in general be inferred only from negative results 
of theoretical search.  However, in the case that the observer naturalness 
problem is related to a fine-tuning naturalness problem, it is possible 
that we can be convinced rather firmly that the problem does in fact 
exist.  This will be the case, for example, if we do not see any 
deviation of gravity from Newton's law down to a scale (much) smaller 
than $O(100~{\rm \mu m})$, since it will tell us the absence of a physical 
threshold that can control the observed value of the cosmological constant. 
In sections~\ref{sec:EWSB} and \ref{sec:EWSB-alt}, we will also argue 
that the observation (or non-observation) of new physics at the TeV 
scale may also be viewed, depending on what we will see, as evidence 
for the existence of an observer naturalness problem, and hence 
environmental selection.

We stress that environmental selection can provide numerical predictions 
that are difficult to obtain in other ways.  Specifically, this occurs 
if the effective distribution function $f_{\rm eff}$ has a nontrivial 
form in the physical basis.  In this case the physical parameters take 
values corresponding to a point close to the observer boundary, giving 
nontrivial predictions.

An example of predictions made possible by environmental selection is 
given by the stability boundary of the desired electroweak phase of the 
Standard Model~\cite{Feldstein:2006ce}.  Suppose that the Standard Model 
is valid up to some high scale $M_*$ near the Planck scale and that the 
weak scale $v$ results from environmental selection.  Suppose further 
that the Higgs quartic coupling at $M_*$, $\lambda_{h,*}$, and the top 
quark Yukawa coupling at $M_*$, $y_{t,*}$, vary from one universe to 
another.  There is then an observer boundary $O(\lambda_{h,*}, y_{t,*})=1$ 
corresponding to sufficient stability of the desired electroweak phase. 
Now, if the distribution function $f_{\rm eff}(\lambda_{h,*}, y_{t,*})$ 
is strongly peaked towards the phase boundary then our universe is expected 
to be close to this edge.  In particular, if the peaking is stronger in 
$\lambda_{h,*}$ than in $y_{t,*}$ then the most probable point on the 
phase boundary has $M_{\rm Higgs} \simeq 107~{\rm GeV}$ and $m_t \simeq 
175~{\rm GeV}$.  Discovery of the Higgs boson near this mass, together 
with the absence of any new electroweak physics, would then provide 
evidence that our universe is near the edge of this observer phase 
boundary, and hence of environmental selection.  In general, if we 
find ourselves living close to an observer boundary and if we do not 
have a simple (alternative) theory explaining that fact, then we may 
regard it as evidence for environmental selection.

\section{Predictions for {\boldmath $m_u$}, {\boldmath $m_d$} and 
 {\boldmath $m_e$} from Environmental Selection}
\label{sec:u-d-e}

The observer naturalness problem associated with the stability 
of neutrons, deuterons and complex nuclei was introduced in 
section~\ref{sec:evid-prob}, and is illustrated in Fig.~\ref{fig:region}. 
We showed that, no matter what the theory of flavor, there is always 
a factor in the naturalness probability, $\tilde{P}$, from the 
dependence on $v/\Lambda_{\rm QCD}$.  In section~\ref{sec:flavor} 
we argued that there is a factor, $P_F$, in the naturalness probability 
that is highly dependent on the theory of flavor.  In all known theories 
$P= P_F \tilde{P} \simlt (10^{-3}$~--~$10^{-2}$).  In this section 
we argue that this naturalness problem can be solved by environmental 
selection; furthermore there are several possible solutions with 
different consequences.

We begin by restating the naturalness problem in terms of Standard Model 
parameters.  For the neutron and complex nuclei stability boundaries 
a crucial quantity is
\begin{equation}
  m_n - m_p - m_e = C_I (y_d - y_u) v - y_e v 
    - C_\alpha \alpha \Lambda_{\rm QCD},
\label{eq:n-p-e}
\end{equation}
where $C_I \approx (0.5$~--~$2)$ and $C_\alpha \approx (0.5$~--~$2)$ 
are strong interaction coefficients, and we choose a definition 
of the QCD scale such that $\Lambda_{\rm QCD} = 100~{\rm MeV}$. 
Throughout we use approximate ranges for $m_{u,o}$ and $m_{d,o}$ 
from Ref.~\cite{Yao:2006px}.  For complex nuclei the binding 
energy per nucleon is $E_{\rm bin} = C_B \Lambda_{\rm QCD}$, with 
$C_B \approx 0.08$ another strong interaction coefficient.  The 
neutron and complex nuclei boundaries are then described by the 
inequalities
\begin{equation}
  0 < C_I (y_d - y_u) \frac{v}{\Lambda_{\rm QCD}} 
    - y_e \frac{v}{\Lambda_{\rm QCD}} - C_\alpha \alpha < C_B.
\label{eq:H-N}
\end{equation}
In the approximation that the deuteron binding energy, $B_D$, is 
linear in $m_u + m_d$ in the region of interest, the stability 
boundary for the deuteron can be written as
\begin{equation}
  \frac{B_D}{\Lambda_{\rm QCD}} 
    = C_1 - C_2 (y_d + y_u) \frac{v}{\Lambda_{\rm QCD}} > 0,
\label{eq:D}
\end{equation}
where $C_{1,2}$ are two further strong interaction coefficients. 
These three boundaries involve just four independent combinations 
of Standard Model parameters, $y_{u,d,e} v/\Lambda_{\rm QCD}$ 
and $\alpha$.

\begin{figure}
\begin{center}
\begin{picture}(280,190)(-20,0)
  \Line(0,0)(0,190) \Line(0,190)(-4,183) \Line(0,190)(4,183)
  \Text(-8,190)[r]{$Q$}
  \DashLine(0,130)(210,130){3} \Text(-6,130)[r]{$0$}
  \DashLine(30,100)(90,100){3} \LongArrow(60,130)(60,100)
  \Text(63,115)[l]{\footnotesize $m_e$}
  \Text(57,115)[r]{\tiny $0.5~{\rm MeV}$}
  \DashLine(30,40)(150,40){3} \LongArrow(60,100)(60,40)
  \DashLine(83,70)(97,70){1} \DashLine(83,10)(97,10){1}
  \DashLine(90,10)(90,70){1} \Vertex(90,40){3}
  \Text(63,70)[l]{\footnotesize $\delta_{\rm EM}$}
  \Text(57,70)[r]{\tiny $1.0 \mp 0.5~{\rm MeV}$}
  \DashLine(90,178)(210,178){3} \LongArrow(120,40)(120,178)
  \Text(118,109)[r]{\footnotesize $\delta_{d-u}$}
  \Text(124,109)[l]{\tiny $2.3 \pm 0.5~{\rm MeV}$}
  \Line(177,130)(177,173) \Line(183,130)(183,173)
  \Line(180,178)(174,168) \Line(180,178)(186,168)
  \Line(178.6,175.6)(180.4,177.4) \Line(177,172)(181.2,176.2)
  \Line(177,170)(181.9,174.9) \Line(177,168)(182.7,173.7)
  \Line(177,166)(183,172) \Line(177,164)(183,170) \Line(177,162)(183,168)
  \Line(177,160)(183,166) \Line(177,158)(183,164) \Line(177,156)(183,162)
  \Line(177,154)(183,160) \Line(177,152)(183,158) \Line(177,150)(183,156)
  \Line(177,148)(183,154) \Line(177,146)(183,152) \Line(177,144)(183,150)
  \Line(177,142)(183,148) \Line(177,140)(183,146) \Line(177,138)(183,144)
  \Line(177,136)(183,142) \Line(177,134)(183,140) \Line(177,132)(183,138)
  \Line(177,130)(183,136) \Line(179,130)(183,134) \Line(181,130)(183,132)
  \Text(190,154)[l]{$Q \simeq 0.8~{\rm MeV}$}
\end{picture}
\caption{$Q$ value for the reaction $n \rightarrow p e^- \bar{\nu}$.}
\label{fig:Q}
\end{center}
\end{figure}
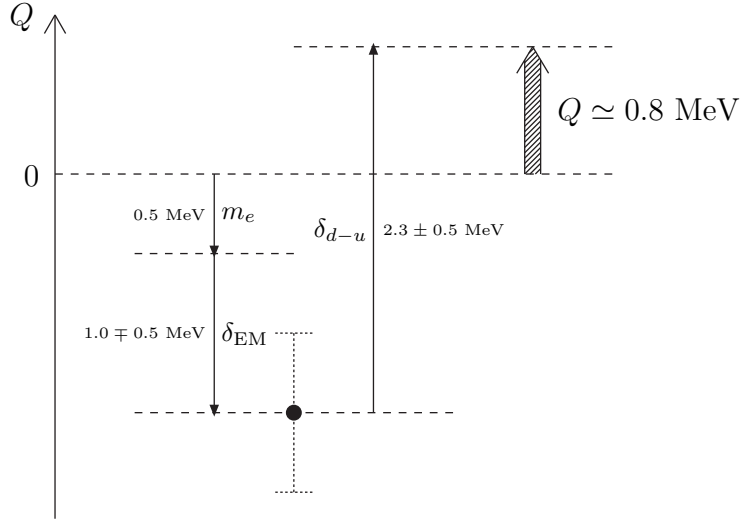

Much of the observer naturalness problem arises because the three 
terms that contribute to $m_n - m_p - m_e$ in Eq.~(\ref{eq:n-p-e}) 
are comparable, as shown in Fig.~\ref{fig:Q}, even though Yukawa 
couplings and the ratio of mass scales $v/\Lambda_{\rm QCD}$ are 
expected to range over many orders of magnitude.
The electron mass, $y_e v \simeq 0.5~{\rm MeV}$ and the electromagnetic 
mass splitting $\delta_{\rm EM} \equiv C_\alpha \alpha \Lambda_{\rm QCD} 
\simeq 1.0 \mp 0.5~{\rm MeV}$ differ by only about a factor of $2$. 
They both stabilize the neutron, and hence an unstable neutron requires 
$y_d > y_u$.  Not only is this the case, but the isospin breaking term 
$\delta_{d-u} \equiv C_I (m_d - m_u) = 2.3 \pm 0.5~{\rm MeV}$ only 
just over-compensates the negative terms to give a net available energy 
for neutron beta decay of just $0.8~{\rm MeV}$, an amount that is also 
very close to the electron mass and electromagnetic terms.  In any theory 
of flavor where $y_{u,d,e}$ are determined by symmetries, these numerical 
coincidences, allowing neutron instability and the existence of complex 
nuclei, must simply be accidents.  Furthermore, these accidents also 
involves $v/\Lambda_{\rm QCD}$ and $\alpha$.  The other Standard Model 
combination, $(y_d + y_u) v/\Lambda_{\rm QCD}$, which could also range 
over many orders of magnitude, is numerically within about a factor of 
$2$ of the deuteron stability boundary.  Finally, as is apparent from 
Fig.~\ref{fig:region}, the observed parameters are actually much closer 
to the neutron stability boundary than a factor of $2$.  The perpendicular 
distance in coupling parameter space is $(10$~--~$30)\%$, depending on 
the values of $C_I$ and $y_{u,o}/y_{d,o}$.  As discussed in previous 
sections, some aspects of these accidents could arise from a theory 
of flavor that determines $y_{u,d,e}$ to be small and in the approximate 
ratio $1:2:1$ at a unified scale; but the accidents associated with 
the value of $v/\Lambda_{\rm QCD}$ and the closeness to the neutron 
stability boundary remain.

Environmental selection cuts out large region of parameter space where 
neutrons are stable, or deuterons or complex nuclei are unstable, greatly 
reducing the observer naturalness problem.  The only remaining question 
is whether our observed universe is a typical universe within the observer 
region ${\cal O}$ shown in Fig.~\ref{fig:region}.  This depends on the 
probability distribution for $y_{u,d,e}v/\Lambda_{\rm QCD}$ and $\alpha$, 
and on whether ${\cal O}$ has been correctly identified, allowing several 
solutions of the problem with different implications.  We begin by taking 
$\alpha$ fixed to its observed value while allowing $y_{u,d,e}$, $v$ and 
$\Lambda_{\rm QCD}$ to scan.  Since there are only two scales $v$ and 
$\Lambda_{\rm QCD}$ appearing in the nuclear stability observer boundaries, 
we can take $\Lambda_{\rm QCD} = 100~{\rm MeV}$ without loss of generality, 
setting the unit of mass.  Furthermore, since $v$ always multiplies 
a Yukawa coupling, environmental selection for the light fermion masses 
can be discussed in terms of a distribution function $f(m_{u,d,e})$ 
without loss of generality.%
\footnote{The distribution function $f(m_{u,d,e})$ here and 
 below really means $f(m_{u,d,e}/\Lambda_{\rm QCD})$, but we omit 
 $\Lambda_{\rm QCD}$ for notational simplicity.  In the language 
 of section~\ref{subsec:force}, this is the effective distribution 
 for $m_{u,d,e}/\Lambda_{\rm QCD}$ after integrating out the other 
 parameters.  Specifically, $f(m_{u,d,e}/\Lambda_{\rm QCD}) = 
 \int_{\cal O} \delta(m_u/\Lambda_{\rm QCD}-y_u v/\Lambda_{\rm QCD})\, 
 \delta(m_d/\Lambda_{\rm QCD}-y_d v/\Lambda_{\rm QCD})\, 
 \delta(m_e/\Lambda_{\rm QCD}-y_e v/\Lambda_{\rm QCD})\, 
 f(y_u,y_d,y_e,v,\Lambda_{\rm QCD})\, dy_u\, dy_d\, dy_e\, dv\, 
 d\Lambda_{\rm QCD}$.  Using $f(m_{u,d,e})$, we are able to discuss 
 environmental selection at nuclear boundaries without answering 
 how the degeneracy inside $m_{u,d,e}$ (scaling $v/\Lambda_{\rm QCD}$ 
 and $y_{u,d,e}$ oppositely keeping $m_{u,d,e}$ fixed) is determined.}
For a given $f$ we can compute $\langle m_u \rangle$, $\langle m_d 
\rangle$ and $\langle m_e \rangle$ and compare them with the values 
observed in our universe, $m_{u,o}$, $m_{d,o}$ and $m_{e,o}$.  We 
can also compute how close a typical universe in ${\cal O}$ is to 
the neutron stability boundary.

Predictions for $\langle m_{u,d,e} \rangle$ involve the mass scales 
that arise in the observer boundaries, namely the maximum value of 
$m_+ = m_d + m_u$ allowed by deuteron stability
\begin{equation}
  m_{+{\rm max}} = \frac{C_1}{C_2} \Lambda_{\rm QCD} 
    \simeq (1.4~\mbox{--}~2.7)\, m_{+,o} 
    \simeq (1.4~\mbox{--}~2.7) \times (5~\mbox{--}~12)~{\rm MeV},
\label{eq:m_+max}
\end{equation}
for $a = (1.3$~--~$5.5)~{\rm MeV}$, the binding energy per nucleon 
in stable complex nuclei
\begin{equation}
  E_{\rm bin} = C_B \Lambda_{\rm QCD} 
    \approx 8~{\rm MeV},
\label{eq:E_bin}
\end{equation}
and the electromagnetic contribution to the proton mass
\begin{equation}
  \delta_{\rm EM} = C_\alpha \alpha \Lambda_{\rm QCD} 
    \simeq (0.5~\mbox{--}~1.5)~{\rm MeV}.
\label{eq:delta_EM}
\end{equation}
While all three mass scales are proportional to $\Lambda_{\rm QCD}$, 
$\delta_{\rm EM}$ is significantly smaller than $m_{+{\rm max}}$ and 
$E_{\rm bin}$.  Hence it is important to see which of these mass scales 
enter the predictions for $\langle m_{u,d,e} \rangle$.

We consider three situations that solve the observer naturalness problem:
\begin{description}
\item[I.] An important part of the observer boundary is missing.
\item[II.] The probability distribution is flat in mass space, 
 $d{\cal N} = A\, dm_u\, dm_d\, dm_e$.  In this case the closeness 
 to the neutron stability boundary discussed above is accidental.
\item[III.] The probability distribution $d{\cal N} = f(m_u,m_d,m_e)\, 
 dm_u\, dm_d\, dm_e$ yields a nontrivial probability force towards the 
 neutron stability boundary.
\end{description}

For case~I, consider a multiverse with $m_u$, $m_d$ and $m_e$ uniformly 
distributed on logarithmic scales so that the relevant diagrams are 
shown in the left panels of Fig.~\ref{fig:region}.  Although the original 
naturalness problem $P \ll 1$ has been ameliorated by a large cut factor, 
the naturalness problem is not entirely removed by environmental selection 
for neutron instability and deuteron and complex nuclei stability, since 
$P_{\cal O} \ll 1$.  For example, there are large regions of ${\cal O}$ 
at small $m_{u,e}$ that are distant from the observer boundary.  A complete 
solution follows if there are additional relevant boundaries which we 
have failed to identify, that reduce ${\cal O}$ to the point where our 
universe becomes typical within ${\cal O}$.  This would certainly require 
new physical constraints to remove the large regions with low values of 
$m_u/m_{u,o}$ and $m_e/m_{e,o}$.  A complete solution may need further 
cuts to remove large values of $m_e/m_{e,o}$ not already excluded, for 
example using the threshold for the $pp$ reaction.  Furthermore, to 
understand our closeness to the neutron stability boundary, it would 
be necessary for other cuts to approach our universe with a similar 
closeness on a logarithmic scale.  While we can certainly identify 
physical processes that would introduce extra boundaries, we are unable 
to argue that they induce catastrophic changes rather than just the 
substantial changes discussed in section~\ref{sec:evid-prob}.

For the second case, II, suppose that the distributions of $m_u$, $m_d$ 
and $m_e$ are flat on linear scales, so that the effects of environmental 
selection can be understood from the cuts of the observer boundaries 
drawn in the right panels of Fig.~\ref{fig:region}.  The observed masses 
are relatively typical within ${\cal O}$, so that $P_{\cal O}$ is not 
much smaller than unity, and the naturalness problem is largely solved. 
In fact, the observed masses are still close to the neutron stability 
boundary, even in the right panels of Fig.~\ref{fig:region}, which 
in this example is accidental.  Having assumed a simple form for the 
multiverse distribution, i.e. that $f(m_u,m_d,m_e)$ is constant, we 
are able to use the precise form of the observer boundary to compute 
the average observed values of the electron, up quark and down quark 
masses by integrating over ${\cal O}$
\begin{equation}
  \langle m_e \rangle 
  = \frac{1}{4} C_I\, m_{+{\rm max}} 
  \approx C_I \left(\frac{m_{+,o}}{5~{\rm MeV}}\right) 
    (2~\mbox{--}~3)~{\rm MeV},
\label{eq:me_flat}
\end{equation}
\begin{equation}
  \langle m_+ \rangle 
  = \frac{3}{4} m_{+{\rm max}} 
  \approx (1~\mbox{--}~2)\, m_{+,o},
\label{eq:mu_flat}
\end{equation}
\begin{equation}
  \langle m_- \rangle 
  = \frac{1}{2} m_{+{\rm max}} 
  \approx (0.7~\mbox{--}~1.4)\, m_{+,o},
\label{eq:md_flat}
\end{equation}
where $m_\pm = m_d \pm m_u$.  In these equations the analytic expressions 
are obtained without including the effect of the complex nuclei boundary, 
and the numerical range corresponds to $a = (1.3$~--~$5.5)~{\rm MeV}$. 
(Including the effect of the complex nuclei boundary changes the 
numerical values only up to about $30\%$.)  These results demonstrate 
that environmental selection yields predictions for $m_{u,d,e}$ once 
a simple form for the probability distribution has been assumed.  The 
predictions for $m_u$ and $m_d$ are good.  The prediction for $m_e$ 
is quite uncertain.  For many values of the strong interaction parameters 
it is somewhat large; for example, for central values of $C_I$, $a$ 
and $m_{+,o}$, $\langle m_e \rangle \approx 6 m_{e,o}$, so that low 
values of $C_I$ and $m_{+,o}$ as well as a high value of $a$ are 
preferred.  Nevertheless, we stress that {\it a major part of the 
observer naturalness problem is solved if $f(m_u,m_d,m_e)$ is relatively 
flat within ${\cal O}$.}  On linear scales for $m_{u,d,e}$, our universe 
is quite typical of ${\cal O}$.  Of course, case~II does imply that 
the closeness to the neutron boundary is accidental, and the rest of 
this section is devoted to understanding this closeness.  A peaked 
distribution function can also give $\langle m_e \rangle \propto 
\delta_{\rm EM}$, rather than $\propto m_{+{\rm max}}$, leading 
immediately to an understanding of the lightness of the electron.

For case~III, inside ${\cal O}$ the distribution $f$ is peaked towards 
the neutron stability boundary, allowing us to explore the consequences 
of an environmental explanation for why $m_{u,d,e}$ are so close to this 
boundary.  We call this stability boundary the $n$ surface --- it is a 
2-dimensional plane in the 3-dimensional space of masses, $m_{u,d,e}$. 
Within ${\cal O}$, the probability force ${\bf F} = \nabla f$ can be 
resolved into components parallel, $F_\parallel$, and perpendicular, 
$F_\perp$, to the $n$ surface.  We assume that $F_\perp$ points towards 
the $n$ surface.  The $F_\parallel$ field will determine the most 
probable location on the $n$ surface.

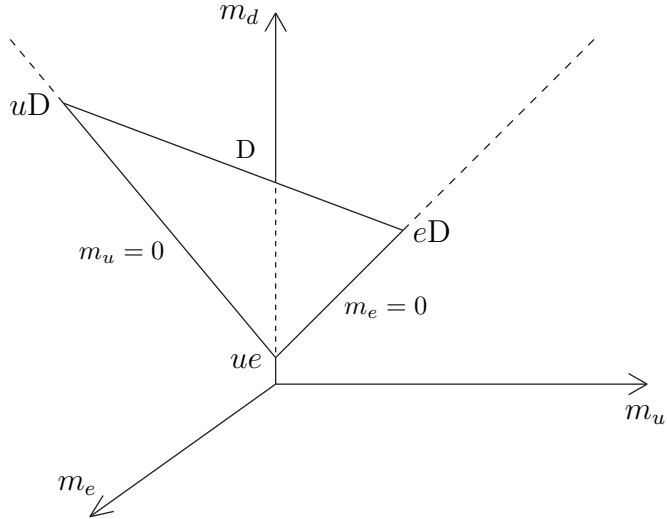
\begin{figure}
\begin{center}
\begin{picture}(240,190)(-100,0)
  \Line(0,50)(140,50) \Line(140,50)(133,54) \Line(140,50)(133,46)
  \Text(140,42)[t]{$m_u$}
  \Line(0,50)(0,58) \DashLine(0,58)(0,128){2} \Line(0,128)(0,190)
  \Line(0,190)(-4,183) \Line(0,190)(4,183)
  \Text(-6,190)[r]{$m_d$}
  \Line(0,50)(-70,0) \Line(-70,0)(-65,8) \Line(-70,0)(-61,2)
  \Text(-75,10)[b]{$m_e$}
  \Line(0,60)(48,108)  \DashLine(48,108)(120,180){3}
  \Text(26,83)[tl]{\footnotesize $m_e = 0$}
  \Line(0,60)(-80,156) \DashLine(-80,156)(-100,180){3}
  \Text(-42,104)[tr]{\footnotesize $m_u = 0$}
  \Line(-80,156)(48,108)
  \Text(-15,136)[bl]{\footnotesize D}
  \Text(-5,59)[r]{$ue$}
  \Text(52,108)[l]{$e$D}
  \Text(-85,156)[r]{$u$D}
\end{picture}
\caption{2-dimensional $n$ surface in the 3-dimensional space of 
 masses, $m_{u,d,e}$.}
\label{fig:n-surface}
\end{center}
\end{figure}

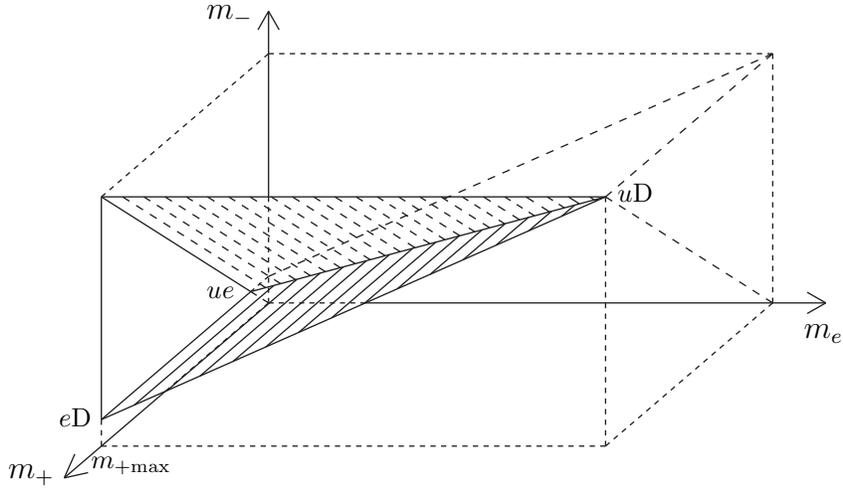
\begin{figure}
\begin{center}
\begin{picture}(290,180)(-80,-20)
  \DashLine(0,50)(36.5238,50){2} \Line(36.5238,50)(210,50)
  \Line(210,50)(203,54) \Line(210,50)(203,46)
  \Text(210,42)[t]{$m_e$}
  \DashLine(0,50)(0,90){2} \Line(0,90)(0,160)
  \Line(0,160)(-4,153) \Line(0,160)(4,153)
  \Text(-6,159)[r]{$m_-$}
  \DashLine(0,50)(-38.9058,16.6522){2} \Line(-38.9058,16.6522)(-77,-16)
  \Line(-77,-16)(-73,-7) \Line(-77,-16)(-68,-14)
  \Text(-81,-15)[r]{$m_+$}
  \Line(-63,6)(-6.70213,54.2553) \DashLine(-6.70213,54.2553)(0,60){2}
  \Line(-63,6)(127,90)
  \DashLine(0,60)(190,144){3} \DashLine(127,90)(190,144){3}
  \DashLine(0,50)(-6.70213,54.2553){3} \Line(-6.70213,54.2553)(-63,90)
  \Line(-63,90)(127,90) \DashLine(127,90)(190,50){3}
  \Line(-6.70213,54.2553)(127,90)
  \Line(-63,6)(-63,90) \Text(-52,-8)[t]{\footnotesize $m_{+{\rm max}}$}
  \DashLine(-63,90)(0,144){2} \DashLine(0,144)(190,144){2}
  \DashLine(-63,-4)(-63,6){2} \DashLine(127,90)(127,-4){2}
  \DashLine(127,-4)(190,50){2} \DashLine(190,50)(190,144){2}
  \DashLine(-63,-4)(127,-4){2}
  \Line(-50.6087,11.4783)(2.01755,56.5865)
  \Line(-38.2174,16.9565)(10.7373,58.9176)
  \Line(-25.8261,22.4348)(19.4570,61.2488)
  \Line(-13.4348,27.9130)(28.1767,63.5800)
  \Line(-1.04348,33.3913)(36.8964,65.9112)
  \Line(11.3478,38.8696)(45.6161,68.2424)
  \Line(23.7391,44.3478)(54.3358,70.5735)
  \Line(36.1304,49.8261)(63.0555,72.9047)
  \Line(48.5217,55.3043)(71.7752,75.2359)
  \Line(60.9130,60.7826)(80.4949,77.5671)
  \Line(73.3043,66.2609)(89.2146,79.8982)
  \Line(85.6957,71.7391)(97.9343,82.2294)
  \Line(98.0870,77.2174)(106.654,84.5606)
  \Line(110.478,82.6957)(115.374,86.8918)
  \Line(122.870,88.1739)(124.093,89.2229)
  \DashLine(-55,90)(-1.07254,55.7603){3}
  \DashLine(-47,90)(4.55702,57.2654){3}
  \DashLine(-39,90)(10.1866,58.7704){3}
  \DashLine(-31,90)(15.8161,60.2755){3}
  \DashLine(-23,90)(21.4457,61.7805){3}
  \DashLine(-15,90)(27.0753,63.2855){3}
  \DashLine(-7,90)(32.7048,64.7906){3}
  \DashLine(1,90)(38.3344,66.2956){3}
  \DashLine(9,90)(43.9640,67.8007){3}
  \DashLine(17,90)(49.5935,69.3057){3}
  \DashLine(25,90)(55.2231,70.8107){3}
  \DashLine(33,90)(60.8526,72.3158){3}
  \DashLine(41,90)(66.4822,73.8208){3}
  \DashLine(49,90)(72.1118,75.3259){3}
  \DashLine(57,90)(77.7413,76.8309){3}
  \DashLine(65,90)(83.3709,78.3359){3}
  \DashLine(73,90)(89.0005,79.8410){3}
  \DashLine(81,90)(94.6300,81.3460){3}
  \DashLine(89,90)(100.260,82.8511){3}
  \DashLine(97,90)(105.889,84.3561){3}
  \DashLine(105,90)(111.519,85.8611){3}
  \DashLine(113,90)(117.148,87.3662){3}
  \DashLine(121,90)(122.778,88.8712){3}
  \Text(-13,54)[r]{\footnotesize $ue$}
  \Text(-67,7)[r]{\footnotesize $e$D}
  \Text(132,92)[l]{\footnotesize $u$D}
\end{picture}
\caption{The $n$ surface in the 3-dimensional space of $m_+$, $m_-$ and 
 $m_e$ (shaded by solid lines).  The $m_e$ axis is stretched relative to 
 the $m_\pm$ axes to make the figure more visible.  The observer region 
 ${\cal O}$ is the region surrounded by the $n$ surface, the $m_u = 0$ 
 surface (shaded by dashed lines), the $m_e = 0$ surface, and the 
 deuteron stability surface at $m_+ = m_{+{\rm max}}$.  The complex 
 nuclear stability boundary is not shown.}
\label{fig:n-surface_2}
\end{center}
\end{figure}

The $n$ surface has a triangular shape, as shown in Fig.~\ref{fig:n-surface}. 
Two of the edges of the triangle correspond to edges of physical space, 
$m_{u,e} = 0$, while the other edge corresponds to the intersection 
of the $n$ surface with the D surface, the boundary for deuteron 
stability.  We call these three edges the $u,e$ and D edges, and 
label the three vertices of the triangle as $ue$, $u$D and $e$D, 
as shown in Fig.~\ref{fig:n-surface}.  The $F_\parallel$ field will 
determine where on the $n$ surface triangle the distribution $f$ 
is maximized; there are three possibilities:
\begin{description}
\item[III-1.] In the interior, not close to an edge.
\item[III-2.] On an edge, not close to a vertex.
\item[III-3.] On a vertex.
\end{description}
On a linear scale, the $n$ surface is small compared to the 
expected available parameter space, so that it is unlikely that 
$\tilde{f}$ would have a sharp peak in the interior of the triangle. 
However, the observer factor, $n$, may vary significantly across the 
triangle inducing a maximum in the interior.  For case~III-2, as we 
move along an edge, $f$ rises to reach a maximum and then falls. 
As an example consider the distribution in ${\cal O}$ to be
\begin{equation}
  f(m_u,m_d,m_e) = A\, (m_d - m_u)^{-p},
\label{eq:f_m-}
\end{equation}
where $A$ is a normalization constant and $p$ is positive.  This 
situation would arise if the populated landscape has a low probability 
to yield a universe with large breaking of isospin, or if there are 
more observers in a universe as isospin is restored.  In a slice 
through parameter space at constant $m_e$, the resulting probability 
force, ${\bf F}_- \equiv \nabla f$, is towards and perpendicular to 
the $n$ surface.  However, in the full 3-dimensional space ${\bf F}_-$ 
is not perpendicular to the $n$ surface.  This is apparent in 
Fig.~\ref{fig:n-surface_2} which shows the $n$ surface in the 
3-dimensional space of $m_+$, $m_-$ and $m_e$. 
Clearly the force will lead to a preference for low values of $m_e$, 
so that if $p$ is large enough the most probable universes will have 
$m_e$ close to zero, i.e. close to the $e$ edge of the $n$ surface 
triangle of Fig.~\ref{fig:n-surface}.  From Fig.~\ref{fig:n-surface_2}, 
all points on the $e$ edge are equally probable, so that there is no 
expectation of being close to either the $ue$ or $e$D vertices.  We 
can compute the expectation values of $m_{u,d,e}$ by integrating over 
the 3-dimensional region ${\cal O}$.  For $p > 3$ the regions in 
Fig.~\ref{fig:n-surface_2} at low $m_-$ and low $m_e$ dominate the 
integrals, and one sees that $\delta_{\rm EM}$ sets the scale for 
both $m_e$ and $m_-$, giving
\begin{equation}
  \langle m_e \rangle = \frac{1}{p-3}\, \delta_{\rm EM},
\label{eq:me_-}
\end{equation}
and
\begin{equation}
  C_I\, \langle m_- \rangle = \frac{p-1}{p-3}\, \delta_{\rm EM}.
\label{eq:m-_-}
\end{equation}
This explains why the three contributions to $m_n - m_p - m_e$ 
shown in Eq.~(\ref{eq:n-p-e}) are comparable.  On the other hand, 
Fig.~\ref{fig:n-surface_2} shows that $m_+$ is uniformly distributed 
along the $n$ surface from small values to $m_{+{\rm max}}$, and 
hence
\begin{equation}
  \langle m_+ \rangle = \frac{1}{2}\, m_{+{\rm max}}.
\label{eq:m+_-}
\end{equation}
Notice that, of the three mass scales that enter the observer boundaries, 
Eqs.~(\ref{eq:m_+max},~\ref{eq:E_bin},~\ref{eq:delta_EM}), $E_{\rm bin}$ 
does not appear.  This is because the probability distribution is peaked 
away from the complex nuclei boundary, which therefore becomes irrelevant 
in determining the averages.%
\footnote{The appearance of the mass scales in Eqs.~(\ref{eq:me_-},%
 ~\ref{eq:m-_-},~\ref{eq:m+_-}) depends only on a probability force 
 towards low $m_-$, and not on the particular choice of the power law 
 $f$ in Eq.~(\ref{eq:f_m-}).}

For case~III-3 there are three possible vertices of the $n$ surface 
triangle to consider.  Our universe is certainly not close to the $u$D 
vertex, since this gives $m_e$ too large, hence we study the remaining 
two vertices.  There are many $f$ that lead to these vertices; we begin 
by identifying and studying the simplest cases.  The $ue$ vertex can 
be reached by a preference for low values of $m_{u,d,e}$.  However, 
low values of $m_u$ or $m_e$ are not sufficient as they lead to the 
$u$ or $e$ edges, hence as a simple example we study the distribution
\begin{equation}
  f(m_u,m_d,m_e) = A\, m_d^{-q},
\label{eq:f_md}
\end{equation}
with $q$ positive.  At the $ue$ vertex the value of $m_d$ is determined 
by $\delta_{\rm EM}$.  For sufficiently large $q$ this is the only 
relevant scale so that the average values of $m_{u,d,e}$ over ${\cal O}$ 
will all be determined by $\delta_{\rm EM}$:
\begin{equation}
  \langle m_e \rangle = \frac{1}{q-4}\, \delta_{\rm EM},
\label{eq:me_md-}
\end{equation}
\begin{equation}
  C_I\, \langle m_+ \rangle = \frac{q}{q-4}\, \delta_{\rm EM},
\label{eq:m+_md-}
\end{equation}
\begin{equation}
  C_I\, \langle m_- \rangle = \frac{q-2}{q-4}\, \delta_{\rm EM}.
\label{eq:m-_md-}
\end{equation}
While choices of $q$ and $C_I$ can lead to a hierarchy between these 
averages (for example $\langle m_e\rangle \simeq \delta_{\rm EM}$, 
$\langle m_+ \rangle \simeq 10\,\delta_{\rm EM}$ and $\langle m_- 
\rangle \simeq 6\,\delta_{\rm EM}$ for $q = 5$ and $C_I = 0.5$), 
the hierarchy is more readily generated when $\langle m_e \rangle 
\propto \delta_{\rm EM}$ and $\langle m_+ \rangle \propto m_{+{\rm 
max}}$, as in Eqs.~(\ref{eq:me_-},~\ref{eq:m+_-}).

The $e$D vertex is favored by large $m_u$ and small $m_e$; however 
large or small $m_d$ both lead away from the $e$D vertex.  A simple 
distribution peaking towards the $e$D vertex is
\begin{equation}
  f(m_u,m_d,m_e) = A\, m_u^r,
\label{eq:f_mu}
\end{equation}
with $r \simgt m_{+{\rm max}}/\delta_{\rm EM}$.  Any distribution 
peaking near this vertex will lead to $\langle m_- \rangle \propto 
\delta_{\rm EM}$ and $\langle m_+ \rangle \simeq m_{+{\rm max}}$, 
which agree with observations.  The prediction for the electron 
mass depends on the strength and direction of the force.  In the 
example of Eq.~(\ref{eq:f_mu}), $\langle m_e \rangle \approx 
m_{+{\rm max}}/r$.

In two Higgs doublet theories a probability distribution for the 
ratio of vacuum expectation values, $\tan\beta$, contributes to the 
distributions for $m_{u,d,e}$.  Suppose that this distribution favors 
large $\tan\beta$ and that $\tan\beta_o \gg 1$.  In this case low 
$m_d$ and $m_e$ are preferred, so that within the observer region 
the probability distribution is peaked towards the $ue$ vertex of 
the $n$ surface, leading to $\langle m_e \rangle, \langle m_+ \rangle, 
\langle m_- \rangle \propto \delta_{\rm EM}$.  The probability 
distribution in the electroweak symmetry breaking sector may be 
the origin of the force determining $\langle m_{u,d,e} \rangle$.

Finally we consider the possibility that $\alpha$ also scans.  In this 
case the size of the electromagnetic mass difference, $\delta_{\rm EM}$, 
scans relative to the purely QCD scales of $E_{\rm bin}$ and $m_{+{\rm 
max}}$.  This means that there is a shift in the position of the allowed 
window for $(C_I m_- - m_e)$ from neutron and complex nuclei stability. 
If $\alpha$ increases too much this window shifts to a non-physical 
region where $m_- > m_+$, leading to an upper limit on $\alpha$
\begin{equation}
  \alpha < \frac{C_I C_1}{C_\alpha C_2},
\label{eq:alpha_max}
\end{equation}
which is numerically about $0.2$, with large uncertainties.  Thus 
$\alpha$ is about an order of magnitude away from the maximum value 
that it may take anywhere in ${\cal O}$.

If $\alpha$ is very small then the range for $(C_I m_- - m_e)$ 
increases.  The amount of increase is negligible on a linear scale, 
but is sizable on a logarithmic scale.  For example, with $\alpha 
= 10^{-4}$, $(C_I m_- - m_e)$ can range from $10^{-2}~{\rm MeV}$ to 
$8~{\rm MeV}$.  However, if gauge couplings unify then $\alpha$ and 
$\Lambda_{\rm QCD}$ become related.  As $\alpha$ is decreased, so 
$\Lambda_{\rm QCD}$ becomes exponentially smaller.  It could be that 
selection effects on the size of $\Lambda_{\rm QCD}$ compared with 
the unified scale and/or the electroweak scale dominate over selection 
effects of $\alpha$ in nuclear physics.

In many circumstances, precise predictions for environmental selection 
follow from assuming sharply varying distribution functions $f$. 
However, in the present example of nuclear physics in the parameter 
space $m_{u,d,e}$ and $\alpha$, it is possible that $\tilde{f}$ is 
sufficiently slowly varying over ${\cal O}$ that physical arguments 
based on the observer factor, $n$, could lead to predictions without 
any assumptions of sharply varying $f$.  For example, consider 
variations in the parameters within ${\cal O}$ that lead from the 
neutron surface to the observer boundary with no complex stable nuclei. 
Moving towards the boundary of no complex stable nuclei, the observer 
factor may be reduced by successive nuclei becoming unstable.  On the 
other hand, moving closer to the neutron surface than our universe 
will lead to a longer neutron lifetime and therefore to more primordial 
helium production; the reduction in primordial hydrogen will result 
in fewer hydrogen burning stars.  The competition of these (and other) 
effects may determine the location of our universe within ${\cal O}$. 
Of course, even in this case, some assumption about $\tilde{f}$ is 
still needed.

\section{Electroweak Symmetry Breaking selected by Nuclear Stability}
\label{sec:EWSB}

The origin of electroweak symmetry breaking is one of the largest 
mysteries remaining in the Standard Model.  The quadratic divergence 
of the Higgs mass-squared parameter in the Standard Model implies 
that if the scale of new physics $M$ is (much) larger than the weak 
scale $v$, the theory requires fine-tuning.  This hierarchy problem was 
a key motivation for much of the model building in the last 30~years. 
What do we know about the scale $M$?  In many (non-supersymmetric) 
theories beyond the Standard Model, precision electroweak data 
indicates a ``little hierarchy problem'': $v/M$ is uncomfortably 
small, typically of order $(10^{-2}$~--~$10^{-1})$ or smaller (see 
e.g.~\cite{Barbieri:2000gf}).  There is also a similar fine-tuning 
problem in supersymmetric theories, although its origin is different. 
A sufficiently heavy Higgs boson typically requires superparticles 
to be somewhat heavier than the weak scale, leading to some amount 
of fine-tuning (see e.g.~\cite{Kitano:2006gv}).  In either case, we 
find that some amount of unnaturalness is present, at least for the 
simplest theories, suggesting that environmental selection may be 
playing a role.  In this section we investigate whether a hierarchy 
between $v$ and $M$ is to be expected from environmental selection, 
the size of any such hierarchy, and how the hierarchy depends on 
which parameters are assumed to scan.

To address the question of the environmental selection of $v/M$, several 
issues must be addressed: what is the theory under discussion, which 
parameters of that theory scan, and what observer boundaries implement 
the selection?  Below we formulate a fairly general class of theories 
that describes electroweak symmetry breaking, and the nuclear stability 
boundaries of section~\ref{sec:evid-prob} are used to implement 
selection.  In section~\ref{sec:EWSB-alt} we consider an alternative 
possibility that selection occurs at the electroweak phase boundary.

The mass scale of the new physics that generates electroweak symmetry 
breaking is defined to be $M$, and we assume that the effective theory 
below $M$ is the Standard Model with the Higgs potential
\begin{equation}
  V = m_h^2\, h^\dagger h + \frac{\lambda_h}{2} (h^\dagger h)^2.
\label{eq:SM-pot}
\end{equation}
Integrating out the physics of the electroweak symmetry breaking 
sector at scale $M$ in general leads to several contributions 
to $m_h^2$, some positive and some negative, which we write as
\begin{equation}
  m_h^2 = \left( g_1(x_i) - g_2(x_i) \right) M^2,
\label{eq:g_i}
\end{equation}
where the functions $g_{1,2}$ are both positive.  The dimensionless 
parameters $x_i$ are the set of parameters of the theory above $M$ 
that substantially affect electroweak symmetry breaking.  These 
parameters are evaluated at the scale $M$, so that $g_{1,2}$ are 
also functions of $M$ through the logarithmic sensitivity of $x_i$ 
on $M$: $g_{1,2}(x_i(M))$.  Note that $m_h^2$ here includes the 
quadratically divergent radiative corrections in the Standard Model 
that are regulated by the theory above $M$.  The parameters $x_i$ 
thus include the Standard Model gauge and Yukawa couplings.

The precise nature of the couplings $x_i$ and the functional form 
of $g_{1,2}$ are unimportant for our discussion, so that we write 
$g_1(x_i) = Ax$ and $g_2(x_i) = Ay$, giving
\begin{equation}
  m_h^2 = (x-y) A M^2,
\label{eq:x-y}
\end{equation}
where $x,y,A >0$.  The numerical constant $A$ is chosen such that 
typical values for $y$ in the observer region are of order unity. 
The parameters $x$ and $y$ depend logarithmically on $M$ through 
renormalization group evolution.  ($A$ is a one-loop factor in 
many theories beyond the Standard Model.  Since $m_h^2$ contains 
quadratically divergent contributions in the Standard Model, and 
thus $x$ and $y$ contain pieces proportional to the $SU(2)$ gauge 
and top Yukawa couplings squared, respectively, the value of $A$ 
should not be much smaller than the one-loop factor.)  In the context 
of conventional naturalness criteria, electroweak symmetry breaking 
is unnatural if $x$ is typically larger than $y$ in an ensemble (only 
a small fraction of members in the ensemble leads to electroweak 
symmetry breaking), whereas if $x$ is typically of order $y$ or less 
then the natural value of the weak scale is $\sqrt{A}\, M$ for a 
quartic coupling of order unity.

\begin{figure}[t]
\begin{center}
\begin{picture}(180,150)(0,-13)
  \Line(-10,0)(180,0) \Line(180,0)(173,4) \Line(180,0)(173,-4)
  \Line(0,-10)(0,140) \Line(0,140)(-4,133) \Line(0,140)(4,133)
  \Text(180,-8)[t]{$x_1$} \Text(-8,138)[r]{$x_2$}
  \CArc(260,-165)(300,106.0,144.1)
  \DashCArc(260,-165)(296,106.2,143.5){3}
  \DashCArc(260,-165)(289,106.6,142.5){3}
  \DashCArc(267,-174)(289,108.0,140.2){3}
  \DashCArc(281,-192)(289,110.8,135.4){3}
  \DashCArc(310,-220)(289,117.1,127.0){3}
  \Line(176,79)(164,114) \Text(177,93)[tl]{$M/v$}
  \Line(164,114)(162.1,104.2) \Line(164,114)(171.5,107.4)
  \Text(54,103)[]{\large $\langle h \rangle = 0$}
\end{picture}
\caption{The (supersymmetric) little hierarchy problem as an observer 
 naturalness problem.  Contours of $M/v$ are drawn in a 2-dimensional 
 slice of parameter space of a generic electroweak symmetry breaking 
 sector.  As the experimental limit on $M$ increases, so the allowed 
 region of parameter space shrinks to values of larger $M/v$ close 
 to the phase boundary.}
\label{fig:LHP}
\end{center}
\end{figure}
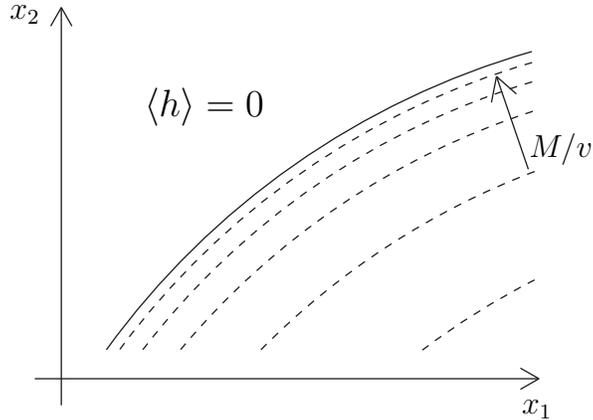

A key observation is that as experimental limits on physics beyond 
the Standard Model get stronger so the mass scale $M$ is constrained 
to be larger.  In some (non-supersymmetric) theories of electroweak 
symmetry breaking, this typically arises from contributions of 
particles of mass $M$ to the precision electroweak observables. 
In many supersymmetric models, the increased lower limits on the 
Higgs boson and superparticle masses have pushed up the mass scale 
of some superparticles to about a TeV or larger.  These little 
hierarchies between $v$ and $M$ imply that the parameters $x_i$ 
are constrained towards the phase boundary of electroweak symmetry 
breaking.  As experiments push up $M$, so $g_1$ and $g_2$ of 
Eq.~(\ref{eq:g_i}) cancel to give $v/\sqrt{A}M$ (much) smaller 
than unity.  This is illustrated in Fig.~\ref{fig:LHP} for 
a 2-dimensional slice through the parameter space.
The unusual closeness of the parameters to this boundary can be 
viewed as an observer naturalness problem.

\begin{figure}
  \center{\includegraphics[width=.5\textwidth]{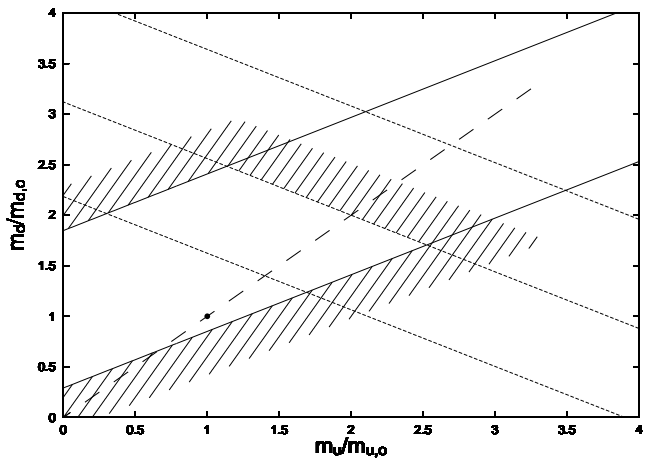}}
\caption{The observer region in $m_u$-$m_d$ space (inside the shading). 
 The dashed line drawn from the lower left to the upper right represents 
 the trajectory followed when only the electroweak vacuum expectation 
 value $v$ is varied (neglecting the effects of a variation of $m_e$).}
\label{fig:obs-mu-md}
\end{figure}

In this section, we investigate the various possible ways in which 
environmental selection on a multiverse can play a role in electroweak 
symmetry breaking.  We use the class of theories described above, 
which contains most non-supersymmetric and supersymmetric theories 
beyond the Standard Model.  The relevant observer boundaries are those 
of nuclear stability, shown in Fig.~\ref{fig:obs-mu-md}, so that in 
general one must consider scanning parameters that affect nuclear physics 
as well as the parameters of the electroweak symmetry breaking sector. 
We start by considering only a few parameters scanning, and then 
progress to more general situations.

First of all, it is natural to expect that some parameters of the 
theory at $M$ scan in the multiverse, so that there are universes 
with different values of the weak scale $v$.  Universes with $x > y$ 
have $v = 0$, while those with $y > x$ have
\begin{equation}
  v^2 = (y-x) \frac{A M^2}{\lambda_h(v)}.
\label{eq:v2}
\end{equation}
Scanning of parameters in the electroweak symmetry breaking sector 
at $M$ may also lead to a nontrivial effective distribution for 
$\lambda_h(v)$.  Since a light Higgs boson has not been discovered, 
however, this distribution should lead to $\lambda_h(v)$ typically 
having an order unity value in the observer region.  This implies
\begin{equation}
  \frac{v^2}{A M^2} \approx y - x,
\label{eq:v2-AM2}
\end{equation}
in our universe, where $y$ is of order unity due to definition of $A$. 
The crucial question is if some cancellation between $y$ and $x$ is 
expected, which cannot be understood in the conventional symmetry 
approach.  If not, then $v \approx \sqrt{A} M$ as in the conventional 
case.  If so, however, we obtain an extra hierarchy between $v$ 
and $M$ that cannot be explained by a symmetry, and the question 
becomes: is $y-x$ typically of order $10^{-2}$--$10^{-1}$, with the 
multiverse generating a little hierarchy, or is it extremely small, 
for example $10^{-30}$, giving a large hierarchy?

In addition to $v$, the nuclear stability observer boundaries depend 
on parameters $y_{u,d,e}$, $\alpha$ and $\Lambda_{\rm QCD}$.  In 
sections~\ref{subsec:M-scan} and \ref{subsec:EWS-scan} we assume 
that these additional parameters do not scan, so that the only scanning 
in the Standard Model at low energies relevant for electroweak symmetry 
breaking is that of $v$.  In this case, environmental selection for 
neutron instability and deuteron stability defines a fixed observer 
window for $v$
\begin{equation}
  v_- < v < v_+,
\label{eq:v-range}
\end{equation}
where, from  Eq.~(\ref{eq:cond-v}), $v_- \simeq 0.5\, v_o$ and 
$v_+ \simeq 2 v_o$.  (The value of $v_+$ depends on the parameter 
$a$ describing the strength of the deuteron binding; here we take 
$a = 2.2~{\rm MeV}$.)  This is illustrated in the $m_u$-$m_d$ plane 
in Fig~\ref{fig:obs-mu-md}.  As $v$ is varied about $v_o$ so $m_u$ 
and $m_d$ vary, but with a fixed ratio, as shown by the dashed line. 
(The corresponding variation of $m_e$ gives only small effects.) 
Since $v_o$ lies centrally in the observer window, it is consistent 
with a distribution $f_v(v)$ that is slowly varying.  Can environmental 
selection for $v$ generate a little or large hierarchy, and if so is 
the effective distribution for $v$ mildly varying?

Another possibility, studied in section~\ref{subsec:nucl-scan}, is 
that the entire set $y_{u,d,e}$, $\alpha$, $\Lambda_{\rm QCD}$ and $v$ 
scans, so that there is a non-zero probability distribution throughout 
the nuclear observer region.  A 2-dimensional slice through the observer 
region is shown in Fig~\ref{fig:obs-mu-md}; the scanning is no longer 
restricted to the dashed line.  An interesting question then is whether 
the closeness to the neutron stability boundary is a statistical accident, 
or whether it results with high probability due to a strongly varying 
distribution function.  As stressed in section~\ref{sec:evid-prob}, 
the nuclear observer region depends on only four combinations of these 
quantities, $m_{u,d,e}/\Lambda_{\rm QCD}$ and $\alpha$.  For example, 
a common scanning of $v$ and $\Lambda_{\rm QCD}$ does not affect 
nuclear physics.  This implies that a numerical value for $v$ is 
no longer determined by the nuclear observer region alone.  Can 
environmental selection from the nuclear observer boundaries, shown 
in Fig~\ref{fig:obs-mu-md}, lead to a little or large hierarchy?

\subsection{Scanning the mass scale {\boldmath $M$}}
\label{subsec:M-scan}

The fundamental field theory at the cutoff scale $M_*$ ($\gg M$) will 
have a certain set of parameters.  We assume that the scanning at 
$M_*$ is limited to those parameters that affect the scale $M$ (which 
may be only $M$ itself), leading to a distribution function $f(M)$. 
In particular, the dimensionless parameters $x_i(M_*)$ do not scan. 
Assuming that effects on $x_i$ from the parameters controlling $M$ 
are small, this implies that the scanning of the parameters $x$ and 
$y$ in Eq.~(\ref{eq:x-y}) comes only through a calculable logarithmic 
dependence on $M$, which arises from nontrivial scaling of these 
parameters under renormalization group evolution.

Suppose now that $y-x$ is positive and of order unity throughout the 
multiverse (which will be the case if $y-x$ is positive at $M_*$ and 
becomes larger as the renormalization scale is lowered).  Environmental 
selection requires that $v$ lies in the fixed range $v_- < v < v_+$. 
This selects $M$ to be in the range
\begin{equation}
  v_- \sqrt{\frac{\lambda_h}{(y-x)A}} < M 
    < v_+ \sqrt{\frac{\lambda_h}{(y-x)A}}.
\label{eq:Mrange}
\end{equation}
The ratio between $v$ and $M$ is then the same as if there were no 
selection, $v/M \approx \sqrt{A}$.

There is, however, another possibility.  It could be that as $M$ varies 
so $y-x$ passes through zero.  In this case, there is a critical value 
$M = M_c$ corresponding to the electroweak phase boundary, $y-x = 0$. 
By assumption, neither the critical value $M_c$, nor the nuclear window 
parameters, $v_\pm$, are scanning.  It would be an accident for $M_c$ 
to be close to $v_\pm$, so we assume that they are distant.  A physically 
interesting new possibility arises if $y-x$ is small in our universe, 
which is possible if $M_c \gg v_\pm$.  This new possibility corresponds 
to the environmental selection of a large hierarchy.

\begin{figure}
\begin{center}
  \center{\includegraphics[width=.41\textwidth]{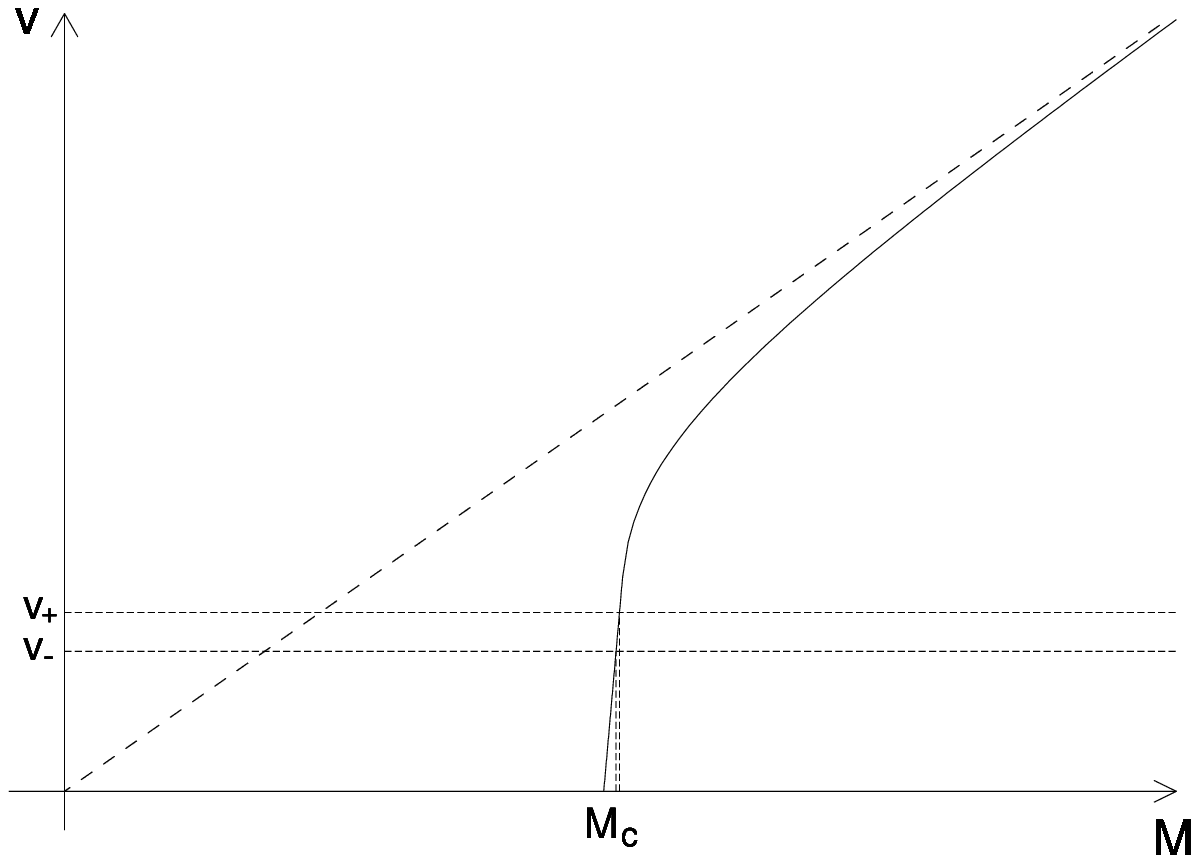}
  \hspace{1.8cm} \includegraphics[width=.41\textwidth]{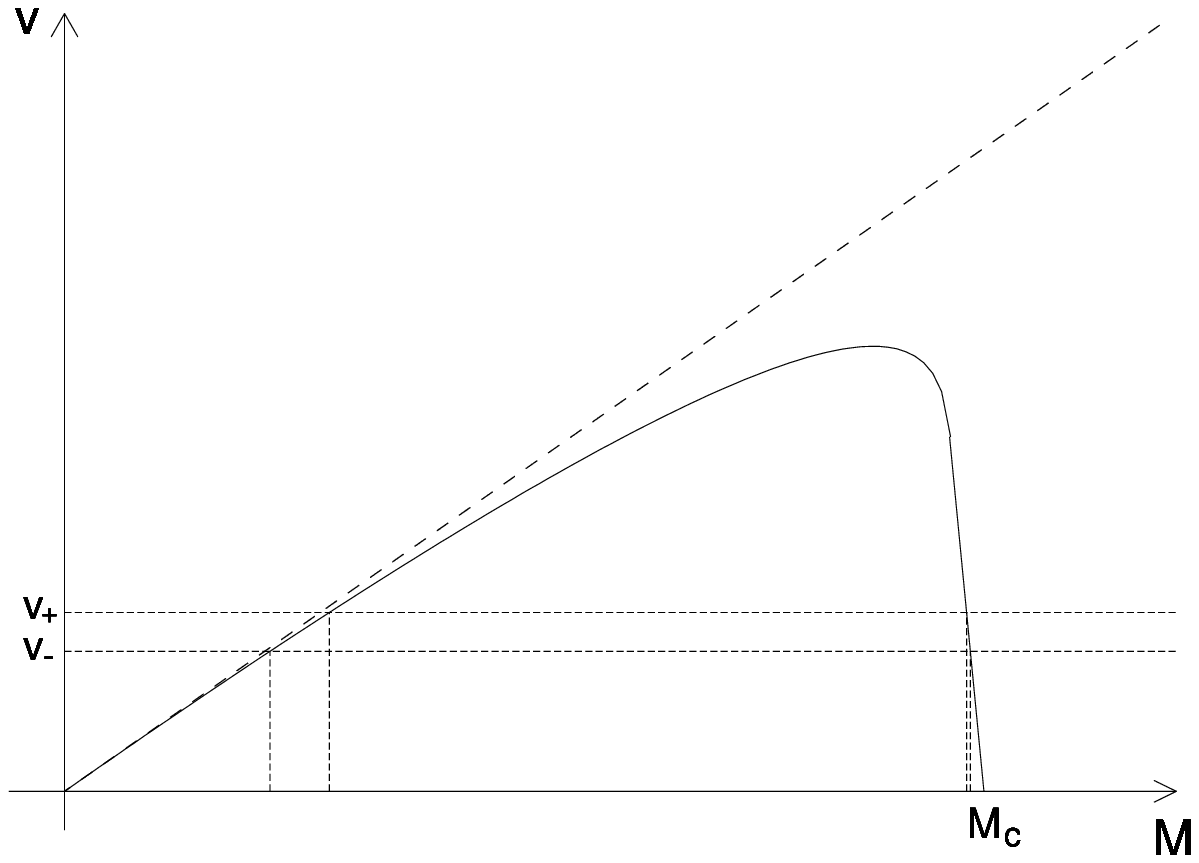}}
\caption{Illustration of the ranges for $M$ selected by the observer 
 condition $v_- < v < v_+$.  The solid curve gives $v(M)$ in regions 
 of $M$ where electroweak symmetry is broken.}
\label{fig:v-M}
\end{center}
\end{figure}

Such a large hierarchy can occur in two ways, depending on the sign 
of the beta function, $\beta$, for $y-x$, as illustrated in the two 
panels in Fig.~\ref{fig:v-M}.  (The definition of $\beta$ here is 
given by $d(y-x)/d\ln\mu = \beta$.) 
For $\beta > 0$ (the left panel), electroweak symmetry breaking is 
only possible for universes with $M > M_c$ and, since $M_c \gg v_\pm$, 
the observer condition Eq.~(\ref{eq:v-range}) selects $M$ to be just 
above $M_c$ in the range $M_c (1+\delta_-) < M < M_c (1+\delta_+)$, 
where $\delta_\pm = \lambda_h v_\pm^2 / |\beta| A M_c^2$.  The value 
of $y-x$ is of order $v^2/A M_c^2 \ll 1$, i.e. the cancellation of 
order $v^2/A M_c^2$ is forced by environmental selection and the 
hierarchy between the weak scale and the scale of new physics is very 
large, $M_c/v \gg 1$.  (Note that $x$ and $y$ themselves are of order 
unity.)  On the other hand, in the case that $\beta < 0$ (the right 
panel) the broken phase has $M < M_c$, and environmental selection gives 
$M_c (1-\delta_+) < M < M_c (1-\delta_-)$; the value of $y-x$ in this 
case is, again, of order $v^2/A M_c^2 \ll 1$.  For $\beta > 0$ only 
a large hierarchy is possible.  For $\beta < 0$, however, universes 
with $v_- < v < v_+$ are also possible with $y-x \approx 1$ and 
$M \approx v/\sqrt{A}$, the case corresponding to conventional natural 
theories.  The case of a large hierarchy is more probable if the 
distribution function for $M$, $f(M)$, is sufficiently weighted 
towards large $M$.

One possibility is that $M$ is the overall scale of superparticle 
masses, so that the theory above $M$ is supersymmetric.  If the hierarchy 
is large, the situation described above then corresponds to the split 
supersymmetry scenario discussed in Refs.~\cite{ArkaniHamed:2004fb,%
Giudice:2004tc}.  However, the theory above $M$ does not have to be 
supersymmetric.  It may, for example, be a strongly interacting theory 
leading to a composite Higgs boson below $M$, possibly as a pseudo 
Nambu-Goldstone boson~\cite{Kaplan:1983fs,ArkaniHamed:2001nc,Contino:2003ve}. 
In this case we obtain a prediction on the Higgs boson mass as a function 
of $M$, by setting the Higgs quartic coupling to be either very large 
or small at the scale $M$ (compared with the logarithmically enhanced 
contribution between $M$ and $v$).  This could be interesting because 
for $M \ll M_{\rm Pl}$ predicted values of the Higgs boson mass are 
outside the range of $\approx (130$~--~$180)~{\rm GeV}$, expected 
if the theory above $v$ is the Standard Model up to a high scale of 
order the Planck scale with a (absolutely) stable electroweak symmetry 
breaking vacuum~\cite{Cabibbo:1979ay,Altarelli:1994rb}.

We stress that the situation described here is very special: for 
example, if $\Lambda_{\rm QCD}$ scans in the multiverse, it is 
possible that the resulting hierarchy is ``little,'' e.g. $v/M 
\approx (10^{-2}$~--~$10^{-1})$, since the value of $M$ may be 
set close to the observed weak scale by environmental selection, 
as we show below.

\subsection{Full scan of the electroweak symmetry breaking sector}
\label{subsec:EWS-scan}

\begin{figure}[t]
  \center{\includegraphics[width=.6\textwidth]{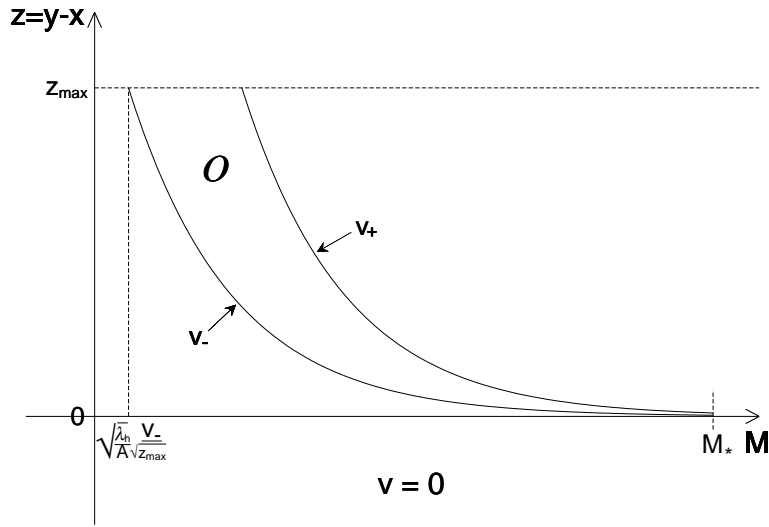}}
\caption{Sketch of the 2-dimensional observer region in the $M$-$z$ 
 plane.}
\label{fig:M-z_obs_1}
\end{figure}

Next we consider a general scanning of the electroweak symmetry breaking 
sector, so that all of $x,y,\lambda_h$ and $M$ vary independently, but 
we keep $y_{u,d,e}$, $\alpha$ and $\Lambda_{\rm QCD}$ fixed.  In this 
case it is useful to consider a 2-dimensional observer region in the 
$M$-$z$ plane, as shown in Fig.~\ref{fig:M-z_obs_1}, where $z \equiv y-x$. 
As before, we assume that $\lambda_h(v)$ has a typical value of order 
unity, e.g. with its distribution function being strongly peaked at 
$\sim 1$, and hereafter we neglect the effect of its scanning.%
\footnote{Our conclusions are not affected by the scanning of 
 $\lambda_h(v)$.  (The distribution of $\lambda_h(v)$ should, of 
 course, be consistent with the bound on the Higgs boson mass.) 
 In some cases, for example in the minimal supersymmetric standard 
 model, $\lambda_h(M)$ is determined by the theory at $M$ and has 
 very little ability to scan.}
Electroweak symmetry breaking occurs in universes with $z > 0$, and the 
observer boundaries $v = v_\pm$ are shown in Fig.~\ref{fig:M-z_obs_1}. 
Other parts of the observer boundary correspond to $M = M_*$, the maximum 
value of $M$ (the cutoff scale), and $z = z_{\rm max}$, the maximum 
value of $z$ determined by the ranges for $x$ and $y$ in the landscape, 
which we take to be of order unity.

The effective probability distribution in this 2-dimensional observer 
region ${\cal O}$ is obtained from the distribution function for 
$x,y,M$ as
\begin{equation}
  f_{\rm eff}(z,M)
    = \int_{\cal O} dx\, dy\, f(x,y,M)\, \delta\left(z-(y-x)\right).
\label{eq:f_eff-z-M}
\end{equation}
With $x$ and $y$ scanning over the multiverse, the logarithmic 
dependence of these parameters on $M$ frequently does not give 
a major effect, and hence we neglect it here.  When these effects 
are important they lead to very interesting results, and we 
defer a discussion of this until section~\ref{subsec:M_c}.  If 
$f_{\rm eff}(z,M)$ were constant over ${\cal O}$, it is clear 
that our universe would be expected to have $z \approx 1$ and 
$v \approx \sqrt{A} M$, since this corresponds to most of the area 
of ${\cal O}$.  However, it is also clear that, as $f_{\rm eff}(z,M)$ 
becomes progressively more peaked towards low $z$ and/or high $M$, 
so the typical hierarchy will first grow to a little hierarchy, 
$z \approx O(10^{-2}$~--~$10^{-1})$, and finally to a large hierarchy, 
$z \ll O(10^{-2})$.  A power distribution is easily able to overcome 
the narrowing of the observer region at large $M$.  The probability 
distribution for the hierarchy, $z$, is given by
\begin{equation}
  f_z(z) = \int_{\cal O} dM\, f_{\rm eff}(z,M),
\label{eq:f_z}
\end{equation}
and a general scanning of the electroweak symmetry breaking sector 
of the theory could lead to a large probability for either a little 
hierarchy or a large hierarchy.  Indeed, since a strong dependence of 
$f_{\rm eff}(z,M)$ on $z$ and $M$ could result from many sources --- 
the landscape of vacua, the population mechanism, integrating out other 
parameters and the observer distribution $n$ --- it would certainly 
not be surprising if either a little or large hierarchy resulted. 
{\it Environmental selection of the weak scale can provide a simple 
generic explanation for our present difficulties in constructing 
natural theories of electroweak symmetry breaking.}

Note that this mechanism works both in the context of 
non-supersymmetric and supersymmetric theories,%
\footnote{In supersymmetric theories there are at least two Higgs 
 doublets, but our analysis can be applied to these cases by 
 identifying $h$ as the linear combination causing electroweak 
 symmetry breaking.}
and that the small value of $z$ implies a cancellation between the 
positive and negative contributions in Eq.~(\ref{eq:g_i}) ($x$ and 
$y$) that cannot be explained in the conventional symmetry viewpoint. 
In section~\ref{subsec:y=1-v}, we study in detail the simplified case 
that $y$ does not scan, and $x$ and $M$ have power-law or logarithmic 
distributions.  We compute $f_z(z)$, find conditions for large and 
little hierarchies, and obtain an analytic result for the size of 
the hierarchy when it is little.  We also compute the distribution 
for $v^2$, $f_v(v^2)$, in the observer window.

Before closing this subsection, let us consider the case that only 
the dimensionless variables in the electroweak symmetry breaking sector, 
$x_i$, scan, with the value of $M$ fixed.  In this case experiments 
constrain the value of $M$ to be larger than about a TeV.  What value 
should we expect for $M$?  Without any special reason, we expect $M$ 
to take some ``random'' value between TeV and fundamental scales; it 
is unlikely that $M$ is close to the observed weak scale, since it 
requires an accident of order $O(0.01$~--~$0.1)$, as discussed in 
section~\ref{subsec:obs_nat_prob}.  Hence we typically expect a large 
hierarchy with $M \gg v$.  The values of $x_i$ are environmentally 
selected by Eq.~(\ref{eq:v-range}) to a very small range:
\begin{equation}
  \frac{\lambda_h v_-^2}{A M^2} < y-x < \frac{\lambda_h v_+^2}{A M^2},
\label{eq:range}
\end{equation}
in which there is a large cancellation between $x$ and $y$.

\subsection{Scanning over the entire nuclear stability observer region}
\label{subsec:nucl-scan}

We continue to study theories of electroweak symmetry breaking that 
lead to the Standard Model as an effective theory below $M$, with 
$m_h^2 = (x-y) A M^2$.  However, as well as having $v$ scan, via the 
scanning of $x$, $y$ and $M$, we now allow the other parameters of 
the nuclear stability observer boundaries to also scan.  This means 
that instead of exploring a distribution through the observer region 
corresponding to only varying $v$, as shown by the dashed line of 
Fig.~\ref{fig:obs-mu-md}, we are now exploring a distribution over 
the entire observer region discussed in section~\ref{sec:evid-prob}. 
With $y_{u,d,e}$, $\alpha$ and $\Lambda_{\rm QCD}$ all scanning, 
the problem is apparently very complex.  However, here we are only 
interested in two aspects of the problem: the probability distribution 
for $z = y-x$ that determines the size of the hierarchy between $v$ 
and $M$, and the probability force perpendicular to the neutron surface 
that determines how close typical observers are to the neutron stability 
boundary.  These questions can be addressed by studying an effective 
distribution over a reduced 2-dimensional projection of the 
parameter space.

\begin{figure}[t]
  \center{\includegraphics[width=.6\textwidth]{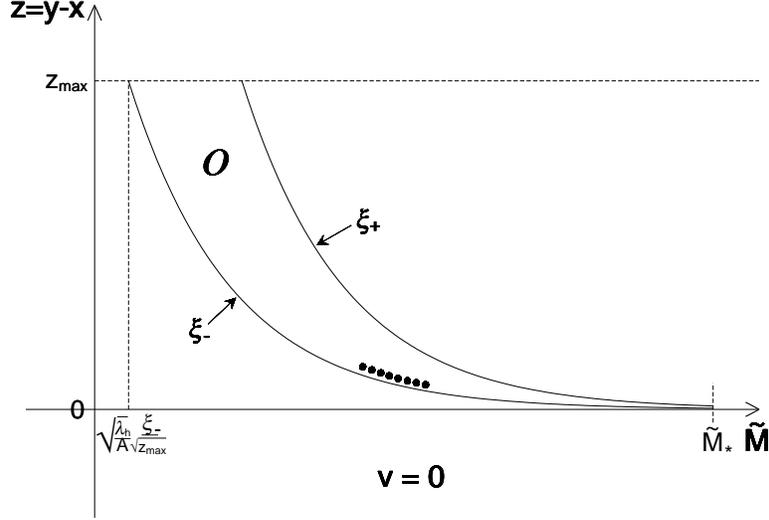}}
\caption{Sketch of the 2-dimensional observer region in the 
 $\tilde{M}$-$z$ plane.}
\label{fig:tilM-z_obs_1}
\end{figure}

The neutron and complex nuclei stability boundaries of Eq.~(\ref{eq:H-N}) 
can be written in the form
\begin{equation}
  \xi_- < \xi < \xi_+,
\label{eq:xi-range}
\end{equation}
where
\begin{equation}
  \xi^2 = z \,  \frac{A \tilde{M}^2}{\lambda_h},
\label{eq:xi}
\end{equation}
with
\begin{equation}
  \tilde{M} = \frac{C_I(y_d - y_u) - y_e}{\alpha \Lambda_{\rm QCD}} M,
\label{eq:coeff}
\end{equation}
and $\xi_- = C_\alpha$ and $\xi_+ = C_\alpha + C_B/\alpha$.  Note the 
similarity in the form of the equations for $\xi$, Eqs.~(\ref{eq:xi}) 
and (\ref{eq:xi-range}), to those for $v$, Eqs.~(\ref{eq:v2}) and 
(\ref{eq:v-range}).  This means that the shape of the observer 
region in the 2-dimensional projection of the parameter space on 
the $\tilde{M}$-$z$ plane, shown in Fig.~\ref{fig:tilM-z_obs_1}, 
is the same as in the last subsection on the $M$-$z$ plane. 
The curved observer boundaries at $\xi = \xi_-$ and $\xi = \xi_+$, 
however, now correspond to neutron and complex nuclei stability, 
respectively.  Our universe thus lies very close to $\xi = \xi_-$. 
How probable this is will be determined by the effective distribution 
for $\xi^2$, $f_\xi(\xi^2)$.  Similarly the expected size for the 
hierarchy will be determined by the distribution for $z$, $f_z(z)$.

In general, the probability distribution in our whole parameter space 
is given by the multi-dimensional distribution function $f(x, y, M, 
\lambda_h, y_u, y_d, y_e, \alpha, \Lambda_{\rm QCD})$, which is not 
easy to deal with.  A crucial point, however, is that {\it as far 
as questions of the size of the hierarchy, $z$, and the closeness 
to the neutron stability boundary are concerned, we only need to 
study the effective distribution function $f_{\rm eff}(z,\tilde{M})$ 
obtained after integrating out all the other variables}.  In particular, 
we can parameterize our ignorance of the (potentially) complicated 
distribution function $f$ and the shape of the observer region in 
this multi-dimensional space by the effective distribution function 
$f_{\rm eff}$ in the 2-dimensional $\tilde{M}$-$z$ space.  Note 
that as long as the form of $f_{\rm eff}$ is kept arbitrary, there 
is no loss of generality.

The effective distribution function $f_{\rm eff}(z,\tilde{M})$ is 
formally given by
\begin{eqnarray}
  f_{\rm eff}(z,\tilde{M}) 
  &=& \int_{\cal O} dx\, dy\, dM\, d\lambda_h\, 
    dy_u\, dy_d\, dy_e\, d\alpha\, d\Lambda_{\rm QCD}\, 
    f(x,y,M,\lambda_h,y_u,y_d,y_e,\alpha,\Lambda_{\rm QCD}) 
\nonumber\\
  && \qquad \times\, 
    \delta \left( z-(y-x) \right)\, \delta \biggl( \tilde{M} 
      - \frac{C_I(y_d-y_u)-y_e}{\alpha\Lambda_{\rm QCD}}M \biggr).
\label{eq:f_eff-z-Mtilde}
\end{eqnarray}
The problem of studying selection in the $\tilde{M}$-$z$ plane then 
becomes identical to that in the $M$-$z$ plane discussed in the 
previous subsection with the replacement $M \rightarrow \tilde{M}$ 
and $v \rightarrow \xi$, except that the upper boundary $\xi_+$ 
now depends on a scanning parameter $\alpha$.  The effect of the 
$\alpha$ scanning on the analysis, however, is small as long as 
the scanning is mild, e.g. the range of the $\alpha$ scanning 
does not span many orders of magnitudes.  We assume this to be 
the case, and hereafter we neglect the effect from this scanning.
The effective distributions for $z$ and $\xi^2$ are given in terms 
of $f_{\rm eff}(z,\tilde{M})$ by
\begin{equation}
  f_z(z) = \int_{\cal O} d\tilde{M}\, f_{\rm eff}(z,\tilde{M}),
\label{eq:f_z_til}
\end{equation}
and
\begin{equation}
  f_\xi(\xi^2) 
  = \int_{\cal O} dz\, d\tilde{M}\, f_{\rm eff}(z,\tilde{M})\, 
    \delta\biggl( \xi^2 - \frac{z A \tilde{M}^2}{\lambda_h} \biggr),
\label{eq:f_xi}
\end{equation}
respectively.

We can identify three very different situations.  The first is that 
$f_{\rm eff}(z,\tilde{M})$ is very mildly varying.  In this case, 
from Fig.~\ref{fig:tilM-z_obs_1} we see that a typical universe will 
lie in the middle of the observer region and hence environmental 
selection will lead to neither a hierarchy nor a closeness to 
the neutron boundary.  A second situation has a strongly varying 
$f_{\rm eff}(z,\tilde{M})$, but with the $\tilde{M}$ component of 
the probability force field unable to overcome the narrowing of the 
observer region at large $\tilde{M}$, shown in Fig.~\ref{fig:tilM-z_obs_1}. 
In this case, if there is a strong force to low values of $z$, the 
multiverse yields a little hierarchy by making low values of $z \approx 
O(10^{-2}$~--~$10^{-1})$ typical.  If in addition there is a significant 
probability force to low $\tilde{M}$, {\it the combined effects of the 
$z$ and $\tilde{M}$ distributions lead to a closeness to the neutron 
stability boundary as well as to a little hierarchy}, as shown by the 
dots in Fig.~\ref{fig:tilM-z_obs_1}.  Finally, the probability force 
towards large $\tilde{M}$ may be strong enough to give a large hierarchy 
with extremely small $z$.  In this case the observer boundaries at 
$\xi_-$ and $\xi_+$ are very close to each other.  At such low values 
of $z$, is it reasonable to have $f_{\rm eff}$ sufficiently different 
on these boundaries to favor the boundary at $\xi_-$?  Here we must 
recall that $f$ is a product of a populated landscape distribution, 
$\tilde{f}$, and an observer distribution, $n$.  It is certainly 
unreasonable for $\tilde{f}$ to have such a large variation over 
such a small region of parameter space.  However, as we move from 
the $\xi_-$ boundary to the $\xi_+$ boundary, nuclear physics changes 
very significantly, which is in a way independent of how small $z$ is. 
Hence a closeness to the neutron boundary may be typical with a large 
hierarchy, {\it but only if it is induced by $n$}, through such 
arguments as appeared in the last paragraph of section~\ref{sec:u-d-e}.

The closeness to the neutron boundary may be a hint that 
$f_{\rm eff}(z,\tilde{M})$ is not flat.  As argued previously, 
there are many origins for strong probability forces, so that 
a little or large hierarchy is quite natural in the multiverse. 
In the next subsection we explore in some detail a subclass of the 
theories specified by Eq.~(\ref{eq:g_i}), allowing explicit formulae 
for the effective distributions and the size of the hierarchy.

\subsection{An explicit example with power law distributions}
\label{subsec:y=1-v}

In the previous two subsections we assumed that the positive and 
negative terms in $m_h^2/M^2$ scan independently.  To provide a simple 
explicit illustration of our results, in this subsection we choose 
to scan only the positive term, i.e. we fix $y=1$ so that
\begin{equation}
  m_h^2 = (x - 1) A M^2,
\label{eq:y=1}
\end{equation}
and $z=1-x$.  The analysis with $x$ fixed and $y$ scanning is very 
similar.  For ease of calculation we assume a polynomial distribution 
function
\begin{equation}
  f(x,M)\, dx\, dM 
    \propto x^n\, dx\; M^q\, d\ln M,
\label{eq:f_x-M}
\end{equation}
and the range of $x$ is given by $0 \leq x \leq x_{\rm max}$, with 
$x_{\rm max}$ some number larger than 1, while $M$ has some very 
large maximum value $M_*$.

\begin{figure}[t]
  \center{\includegraphics[width=.6\textwidth]{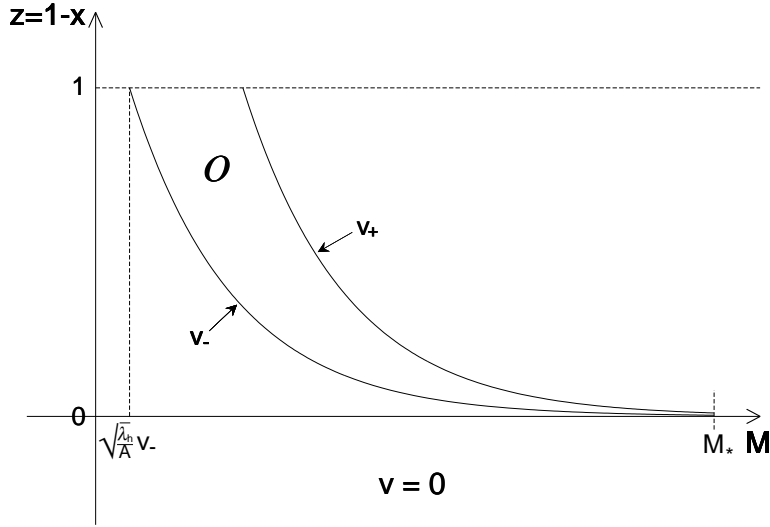}}
\caption{Sketch of the 2-dimensional observer region in the $M$-$z$ 
 plane, where $z = 1-x$.}
\label{fig:M-z_obs_2}
\end{figure}

We begin by assuming that the only parameter appearing in the nuclear 
stability boundary that scans is $v$, so that the condition of 
Eq.~(\ref{eq:v-range}) gives the observer region
\begin{equation}
 v_-^2 < v^2 = (1 -x) \frac{A M^2}{\lambda_h} < v_+^2.
\label{eq:v-range-2}
\end{equation}
The observer region ${\cal O}$ is sketched in the $M$-$z$ plane in 
Fig.~\ref{fig:M-z_obs_2}, which is obtained simply by setting $y=1$ 
in Fig.~\ref{fig:M-z_obs_1}. 
The region is separated by the nuclear physics observer boundaries 
from regions with $v < v_-$ and $v > v_+$, and approaches very close 
to $x = 1$, i.e. $z = 0$, as $M$ approaches $M_* \gg v_o$.  The 
narrowing of the observer region at large $M$ is as we have seen 
in Fig.~\ref{fig:M-z_obs_1}.

What is the typical value of the scale $M$?  For a logarithmic 
distribution, $q=0$, one might guess that all decades of $M$ are 
equally probable; but this is not the case, because at large $M$ 
the observer region narrows.  It is useful to compute an effective 
probability distribution for the hierarchy $z = 1-x = |m_h^2|/A M^2$ 
over the observer region
\begin{equation}
  f_z(z) = \int_{\cal O} dx\, dM\, f(x,M)\, \delta\left(z-(1-x)\right) 
    \sim (1-z)^n\, \frac{1}{z^{q/2}},
\label{eq:f_z-2}
\end{equation}
for sufficiently large $M_*$.  From the viewpoint of the hierarchy, 
this shows that the critical value of $q$ is $2$ not zero: when 
$q=2$ the distribution for the hierarchy $z$ is flat on logarithmic 
scales, at least for large $M$ where $z$ is small.  Thus $q \geq 
2$ gives a large hierarchy, shown by the narrow wedge at large 
$M$ in Fig.~\ref{fig:M-z_obs_2}, and $q<2$ gives a little (or no) 
hierarchy, corresponding to the bulk of the observer region in 
Fig.~\ref{fig:M-z_obs_2}.  For example, if the a priori distribution 
for $M$ is logarithmic, as might happen if the scale is triggered by 
dimensional transmutation associated with some gauge dynamics, the 
most probable observed value of $M$ is small because of the weighting 
from the narrowing of the observer region ${\cal O}$.%
\footnote{In the case that $M$ arises as a dimensional transmutation, 
 the expected distribution of $M$ below $M_*$ is not exactly $\propto 
 d\ln M$, since it would lead to an unphysical conclusion that the 
 probabilities of having $M$ between $10^{n}M_*$ and $10^{n+1}M_*$ 
 ($n \leq -1$) are equal for all $n$ down to $n \rightarrow -\infty$. 
 (The distribution should, for example, be cut off at some 
 value $M_{\rm min}$ or in fact be $\propto d(1/\ln M)$; see 
 discussions at the end of section~\ref{subsec:illust} and 
 in section~\ref{subsec:obs_nat_prob}.)  The assumption/approximation 
 of the exact logarithmic distribution, however, is sufficient for 
 our purposes here.}
On the other hand, if $d{\cal N} \sim dM^2$, as expected in a typical 
non-supersymmetric perturbative theory with mass scale $M$, we find 
$q=2$, so that all scales are equally probable, implying a large 
hierarchy, but not one with $M$ near $M_*$.

\begin{figure}
  \center{\includegraphics[width=.5\textwidth]{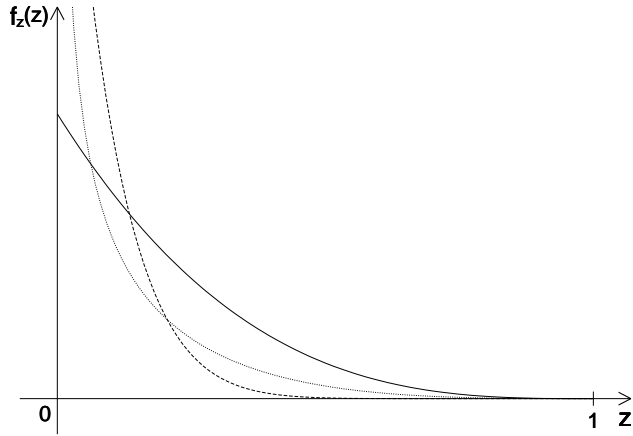}}
\caption{The distribution function $f_z(z)$ for $(n,q) =(3,0)$ (solid), 
 $(10,0)$ (dashed), and $(3,1)$ (dotted).  Each function is normalized 
 such that $\int_0^1\! f_z(z)\, dz = 1$.}
\label{fig:fz}
\end{figure}

If $q<2$, how large is the little hierarchy?  For $n>0$, the first 
factor in Eq.~(\ref{eq:f_z-2}) suppresses the probability of having 
$z \sim 1$, i.e. the probability of having the lowest values of $M$ 
gets suppressed, so that some amount of hierarchy between $v$ and $M$ 
arises from the scanning in the multiverse.  For $q<2$ the distribution 
of Eq.~(\ref{eq:f_z-2}) leads to the average value of $z$:
\begin{equation}
  \langle z \rangle 
    = \biggl\langle \biggl| \frac{m_h^2}{A M^2} \biggr| \biggr\rangle 
    = \frac{2-q}{2n+4-q},
\label{eq:z-av}
\end{equation}
so that for $2n \gg 2-q$ we obtain an extra hierarchy of a factor of 
$\simeq (2-q)/2n$ from the multiverse, which cannot be explained in 
the conventional symmetry approach.  Since $|q|$ is expected to be 
small, this is an appreciable effect for large $n$.  In Fig.~\ref{fig:fz} 
we plot $f_z(z)$ of Eq.~(\ref{eq:f_z-2}) for $(n,q)=(3,0),(10,0),(3,1)$. 
We clearly see that the distribution is peaked towards $z \ll 1$. 
The probability of us observing fine-tuning of $\Delta^{-1}$ or more 
that cannot be understood by the symmetry approach is given by
\begin{equation}
  {\cal P}_\Delta 
  = \frac{\int_0^{\Delta^{-1}}\! f_z(z)\, dz}{\int_0^1 f_z(z)\, dz},
\label{eq:P_Delta}
\end{equation}
which for $\Delta \gg n \gg 1$ reduces to
\begin{equation}
  {\cal P}_\Delta \approx \frac{n}{\Delta},
\label{eq:P_Delta_q=0}
\end{equation}
for $q=0$ (the case with dimensional transmutation) and
\begin{equation}
  {\cal P}_\Delta \approx \sqrt{\frac{4n}{\pi\Delta}},
\label{eq:P_Delta_q=1}
\end{equation}
for $q=1$.  For example, for $n=10$ and $q=0$ ($n=10$ and $q=1$) we have 
a $\simeq 20\%$ ($\simeq 50\%$) probability of observing fine-tuning 
$\Delta^{-1} < 2\%$ which cannot be explained in the conventional 
symmetry picture.

The observer boundaries of Fig.~\ref{fig:M-z_obs_2} are the contours 
for $v=v_\pm$ so that, by construction, $v$ can only be observed between 
$v_-$ and $v_+$. What are the most probable values of $v$ to be observed? 
The effective distribution for $v^2$ is defined by
\begin{equation}
  f_v(v^2) = \int_{\cal O} dx\, dM\, f(x,M)\, 
    \delta\left(v^2 -(1-x)\frac{A M^2}{\lambda_h}\right).
\label{eq:f_v}
\end{equation}
For a large hierarchy with $q \geq 2$, we find a flat distribution in $v^2$
\begin{equation}
  f_v(v^2) \sim {\rm const.}
\label{eq:f_v-large}
\end{equation}
When $y_{u,d,e}$, $\alpha$ and $\Lambda_{\rm QCD}$ do not scan, the 
observed value of the weak scale $v_o$ is roughly midway between 
$v_-$ and $v_+$, as shown in Eq.~(\ref{eq:cond-v}), which is consistent 
with Eq.~(\ref{eq:f_v-large}).%
\footnote{Caution, however, is needed in interpreting 
 Eq.~(\ref{eq:f_v-large}).  This result arises because at large 
 $M$ the contours of $v_-$ and $v_+$ are extremely close in $M$-$x$ 
 space, so that the assumed form for $f$, Eq.~(\ref{eq:f_x-M}), 
 implies little variation between the contours.  While this is 
 expected for the multiverse distribution $\tilde{f}$, it may 
 not be true for the observer distribution $n$.  In this case 
 Eq.~(\ref{eq:f_x-M}) needs to be modified to incorporate the 
 effect represented by $n$.}
For a little (or no) hierarchy, with $q<2$, we find a distribution
\begin{equation}
  f_v(v^2) \sim v^{q-2}.
\label{eq:f_v-little}
\end{equation}
This is again consistent with $v_o$ centrally located in its observer 
window, provided that $q$ is not too negative.

Next, we allow the other parameters that appear in the nuclear 
stability observer boundaries to also scan.  As described in 
section~\ref{subsec:nucl-scan}, the issues of the size of the 
hierarchy and the closeness to the neutron surface can be addressed 
by studying a 2-dimensional projection of parameter space.  Since 
$y$ does not scan, $z=1-x$ and so here we discuss the $\tilde{M}$-$x$ 
plane.  We assume the effective probability distribution in this 
plane has a power law behavior
\begin{equation}
  f_{\rm eff}(x,\tilde{M})\, dx\, d\tilde{M} 
    \propto x^n\, dx\; \tilde{M}^{\tilde{q}}\, d\ln \tilde{M}.
\label{eq:f_eff-x-Mtilde}
\end{equation}
The neutron and complex nuclei boundaries are again given by 
Eqs.~(\ref{eq:xi-range},~\ref{eq:xi},~\ref{eq:coeff}), but with 
$z=1-x$, and are represented in Fig. \ref{fig:tilM-z_obs_1} 
in the $\tilde{M}$-$x$ plane (with $y$ now set to unity).  The 
observer region as $x \rightarrow 1$ again becomes a narrow 
wedge where the hierarchy is large: $M \gg v$.

The equations describing this setup are very similar to those in the 
case with only $v$ scanning, except that the observer boundaries are 
now at $\xi_\pm$ rather than $v_\pm$, and $\tilde{M}$ is not the scale 
of the new physics, see Eqs.~(\ref{eq:xi-range}~--~\ref{eq:coeff}). 
The size of the hierarchy is governed by the effective distribution 
for $z$ and is given by the analogue of Eq.~(\ref{eq:f_z-2}):
\begin{equation}
  f_z(z) \sim (1-z)^n\, \frac{1}{z^{\tilde{q}/2}}.
\label{eq:f_z-3}
\end{equation}
Thus $\tilde{q} \geq 2$ gives a large hierarchy and $\tilde{q} < 2$ 
gives a little (or no) hierarchy.  The size of the little hierarchy 
depends on $n$, and using the distribution Eq.~(\ref{eq:f_eff-x-Mtilde}) 
we find
\begin{equation}
  \biggl\langle \biggl| \frac{m_h^2}{A M^2} \biggr| \biggr\rangle 
    = \frac{2-\tilde{q}}{2n+4-\tilde{q}}.
\label{eq:z-av-2}
\end{equation}
The little hierarchy increases with $n$, as shown by the sequence of 
dots in Fig.~\ref{fig:tilM-z_obs_1} near the neutron stability boundary.

What makes the proximity to the neutron boundary typical? 
The effective distribution for $\xi$, which determines the 
probability force perpendicular to the neutron surface, follows 
immediately from a calculation analogous to that which led to 
Eqs.~(\ref{eq:f_v-large},~\ref{eq:f_v-little}) giving
\begin{equation}
  f_\xi(\xi^2) \sim {\rm const.} 
    \qquad {\rm for}\;\; \tilde{q} \geq 2,
\label{eq:f_xi-large}
\end{equation}
and
\begin{equation}
  f_\xi(\xi^2) \sim \xi^{\tilde{q}-2} 
    \qquad {\rm for}\;\; \tilde{q} < 2.
\label{eq:f_xi-little}
\end{equation}
As with Eq.~(\ref{eq:f_v-large}), caution is necessary in interpreting 
Eq.~(\ref{eq:f_xi-large}).  At large $\tilde{M}$ our assumed distribution, 
Eq.~(\ref{eq:f_eff-x-Mtilde}), may not adequately account for variations 
in the observer factor $n(\xi)$ between $\xi_-$ and $\xi_+$.  In 
particular, we cannot conclude from Eq.~(\ref{eq:f_xi-large}) that a 
large hierarchy is incompatible with a multiverse explanation for the 
closeness to the neutron stability boundary.  With a little hierarchy, 
sufficient closeness to the neutron stability boundary results with 
the distribution of Eq.~(\ref{eq:f_eff-x-Mtilde}) for $-10 \simlt 
\tilde{q} \simlt -2$.

\subsection{Landscapes with a critical value for {\boldmath $M$}}
\label{subsec:M_c}

\begin{figure}[t]
  \center{\includegraphics[width=.6\textwidth]{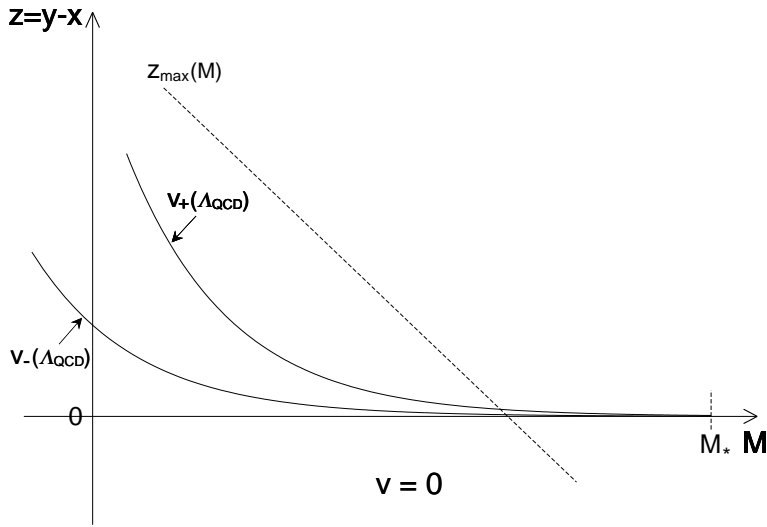}}
\caption{Sketch of the 2-dimensional observer region in the $M$-$z$ 
 plane.  The maximum value of $z$, $z_{\rm max}(M)$, is depicted 
 by the dashed line.}
\label{fig:M_c}
\end{figure}

In the previous subsections we have ignored the logarithmic evolution 
of $x$ and $y$.  For many landscapes this is permissible, but for some 
this evolution plays a critical role.  Consider a landscape such that 
$x_* - y_*$ scans only in a restricted region with $x_* - y_* > 0$, 
where $x_* \equiv x(M_*)$ and $y_* \equiv y(M_*)$, which are the 
fundamental scanning parameters.  In this case, electroweak symmetry 
breaking is only possible because of the evolution of $x-y$ to lower 
energies according to the beta function $\beta = d(x-y)/d\ln\mu > 0$. 
For each value of $(x_*,y_*)$, the quantity $x(M) - y(M)$ passes 
through zero at some $M_c(x_*,y_*)$, so that in these universes 
electroweak symmetry breaking is only possible if $M < M_c(x_*,y_*)$. 
As $(x_*,y_*)$ vary over the entire landscape, suppose that the largest 
value of $M_c(x_*,y_*)$ is $M_{c,{\rm max}}$, which is much less than 
$M_*$.  This maximum critical mass is clearly a property of the particular 
landscape under consideration; it defines a maximum possible value for 
the electroweak scale anywhere in the multiverse, and has an important 
effect on the observer region of the electroweak symmetry breaking 
sector.  For the critical universes with $M_c(x_*,y_*) = M_{c,{\rm max}}$ 
the evolution equation for $z = y-x$ can be solved as
\begin{equation}
  z_{\rm max}(M) = \beta\, \ln\frac{M_{c,{\rm max}}}{M},
\label{eq:z_max}
\end{equation}
where we have approximated that $\beta$ is constant.  The trajectory 
of Eq.~(\ref{eq:z_max}) is sketched in Fig.~\ref{fig:M_c} for the case 
that $M_{c,{\rm max}} \gg v_\pm$, which provides a new boundary to the 
observer region, since $z_{\rm max}$ represents the maximal value of 
$z$ in the multiverse for a given $M$.
In this subsection, we assume that the distribution for $M$ and $z$ 
leads to most universes in the observer region having large $M$, close 
to $M_{c,{\rm max}}$, so that this new part of the observer boundary 
plays an important role.%
\footnote{For many landscapes the evolution of $x-y$ from $M_*$ to 
 $M$ can be ignored.  It induces a shift in the origin of $z=y-x$ 
 by an amount $\beta \ln(M_*/M)$, which is less than unity since 
 $\beta \ln(M_*/M)$ is expected to be of $O(1)$ or smaller.  In many 
 cases the local property of $f_{\rm eff}(z,M)$ is not significantly 
 changed by this shift.}

If $y_{u,d,e}$, $\alpha$ and $\Lambda_{\rm QCD}$ do not scan then 
$v_\pm$ are fixed numbers that arise from nuclear selection.  Since 
there is no reason for the critical mass $M_{c,{\rm max}}$ of the 
landscape to be close to $v_\pm$, a large hierarchy is expected. 
Here we are more interested in the case that $v_\pm$ varies in the 
landscape, and for simplicity we accomplish this via a scanning of 
$\Lambda_{\rm QCD}$, which we take to have a mild distribution, for 
example flat on a logarithmic scale.  In this case $v_\pm \propto 
\Lambda_{\rm QCD}$, so that the observer boundaries in Fig.~\ref{fig:M_c} 
labeled by $v_\pm$ have $z_\pm \propto \Lambda_{\rm QCD}^2$, and hence 
move up and down as $\Lambda_{\rm QCD}$ scans.  For a large hierarchy, 
$M$ is so much larger than $v$ that the interval of $z$ that can 
contribute to universes in the observer region is very small $\Delta z 
\approx v^2/M^2$.  Since the distribution for $\Lambda_{\rm QCD}$ 
is mild, a gain in probability is accomplished by having the $v_\pm$ 
boundary curves of Fig.~\ref{fig:M_c} move upwards, allowing a larger 
interval of $z$ to contribute.  As the $v_\pm$ curves move up so the 
values of $M$ that are on the $z_{\rm max}$ observer boundary decrease. 
At some point this will imply a significant loss of probability via 
the distribution for $M$.  For example, for $f(M)dM \propto M^q d\ln M$ 
with large $q$, this loss of probability begins to set in once $M \sim 
(1-1/q) M_{c,{\rm max}}$ at which point $z_{\rm max} \sim \beta/q$. 
The mild distribution for $\Lambda_{\rm QCD}$ allows the $v_\pm$ 
curves of Fig.~\ref{fig:M_c} to rise so that, at this value of $M$, 
$z_\pm$ will be close to $z_{\rm max}$.  A further drop in $M$ causes 
a significant drop in probability from $f(M)$, which is not offset by 
a sufficient growth in $\Delta z$, since $z_c(M)$ grows logarithmically. 
Hence we predict that a little hierarchy develops.  Specifically, 
for a distribution function
\begin{equation}
  f(z,M,\Lambda_{\rm QCD}) \, dz\, dM\, d\Lambda_{\rm QCD}
    \propto f(z)\, M^q\, dz\, d\ln M\, d\ln \Lambda_{\rm QCD},
\label{eq:f_z-M-Lambda}
\end{equation}
integrating over the observer region we obtain a typical size of the 
hierarchy
\begin{equation}
  \langle z \rangle 
    = \frac{\int z\, f(z)\, e^{-\frac{qz}{\beta}}dz}
        {\int f(z)\, e^{-\frac{qz}{\beta}}dz}
    \approx \frac{\beta}{q},
\label{eq:z-ave}
\end{equation}
where the range of the $z$ integration is from $0$ to the maximum value 
that $z$ takes in the energy interval between $\approx v$ and $M_*$, 
and we have taken $f(z)$ to be constant in the last expression for 
illustration, although the result is not very sensitive to the form 
of $f(z)$.  Note that this result applies to an arbitrary positive 
value of $q \simgt O(\beta)$.

Integrating over $\Lambda_{\rm QCD}$ effectively ``removes'' the $v_\pm$ 
boundaries in the 2-dimensional parameter space of $\{ M,z \}$, and 
the resulting effective distribution in the $M$-$z$ plane is unaltered 
for a prior distribution flat in $\ln \Lambda_{\rm QCD}$.  Allowing 
any mild prior distribution for $\Lambda_{\rm QCD}$, or allowing other 
Standard Model parameters such as the Yukawa couplings $y_{u,d,e}$ to 
scan with mild distributions, would lead to an effective distribution 
$f_{\rm eff}(z,M)$ that would lead to the basic 
result of Eq.~(\ref{eq:z-ave}) being unaffected.  The physics behind 
this result is both simple and general: the $1/q$ factor results from 
a probability force in the $M$ direction making the most probable 
universes in the observer region those close to $M_{c,{\rm max}}$, 
while the factor of $\beta$ arises from the shape of the observer 
boundary near $M_{c,{\rm max}}$, as given in Eq.~(\ref{eq:z_max}).

We conclude that landscapes having a critical value of $M$, above 
which electroweak symmetry breaking is impossible, very easily lead 
to a little hierarchy with $z$ suppressed by both a loop factor, $\beta$, 
and a probability force factor, for example $1/q$.  Unlike the case of 
sections~\ref{subsec:EWS-scan} and \ref{subsec:nucl-scan}, large $q$ 
does not lead to a large hierarchy.  Instead it acts to make $M$ close 
to $M_{c,{\rm max}}$ and, since $z_{\rm max}(M_{c,{\rm max}}) = 0$, 
this leads to a little hierarchy.  Note that the loop factor can provide 
an extra hierarchy corresponding to $1$~--~$2$ orders of magnitude of 
fine-tuning, on top of the $1/q$ factor.  Hence, this mechanism typically 
yields $M$ in the TeV region or larger.

If the physics at $M$ is supersymmetric, this mechanism is the one 
considered in Ref.~\cite{Giudice:2006sn}.   If the only scanning 
parameters are a uniform scaling of the overall scale of supersymmetry 
breaking and of the supersymmetric Higgs mass parameter $\mu$, the 
scale of $M_c$ corresponds to the scale where the determinant of the 
Higgs boson mass matrix, ${\rm det}({\cal M}_H^2)$, passes through 
zero.  To avoid a large hierarchy, some other parameter, such as 
$\Lambda_{\rm QCD}$ should scan.  One again needs to assume that, 
for the landscape as a whole, there is some maximum value of the scale 
where ${\rm det}({\cal M}_H^2)$ passes through zero, and this is taken 
to be $M_{c,{\rm max}}$.  In this subsection, we have stressed the 
generality of the idea, and that the loop suppression factor appearing 
in Eq.~(\ref{eq:z-ave}) does not depend on whether the physics at 
$M$ is supersymmetric or not.  In the supersymmetric case, $\beta$ 
is determined by the specific form of the renormalization group 
equations for the Higgs boson mass parameters, and one finds that 
$\beta \approx O(0.1)$.  Nevertheless this is an important accomplishment, 
since without the landscape the ratio between $v$ and $M$ involves 
a large logarithm $\ln(M_*/v)$ (in the case of high scale supersymmetry 
breaking) so that $A \approx \beta \ln(M_*/v) \approx O(1)$, giving 
$M \approx \sqrt{\lambda_h} v \approx M_Z/\sqrt{2}$, where $M_Z$ is 
the $Z$ boson mass.  The factor of $\beta$, together with $1/q$, can 
easily make $M$ at the TeV scale or larger.

\subsection{Summary and discussion}
\label{subsec:sum_dis}

The conventional naturalness argument implies that the mass scale $M$ 
of new physics beyond the Standard Model that generates electroweak 
symmetry breaking should not be much larger than the Higgs vacuum 
expectation value $v$.  In contrast, we have shown that environmental 
selection for nuclear stability can lead to $M$ substantially larger 
than $v$.  We have considered a very general framework that is 
independent of the model of the new physics, assuming only that 
it generates both positive and negative contributions to the Higgs 
mass-squared parameter, $m_h^2 = (x-y) A M^2$.  If the landscape 
does not possess a maximum critical mass $M_{c,{\rm max}}$ 
significantly less than $M_*$, the relevant observer boundaries 
are shown in Fig.~\ref{fig:M-z_obs_1} when only $x,y$ and $M$ 
scan, and in Fig.~\ref{fig:tilM-z_obs_1} when $y_{u,d,e}$, $\alpha$ 
and $\Lambda_{\rm QCD}$ also scan.  In both cases we find
\begin{itemize}
\item
A large hierarchy is generated by a distribution for $M$ ($\tilde{M}$) 
that grows sufficient with $M$ ($\tilde{M}$) to overcome the narrowing 
observer region of Fig.~\ref{fig:M-z_obs_1} (Fig.~\ref{fig:tilM-z_obs_1}). 
While this does not require a very strong peaking, the distribution 
must continue growing over many orders of magnitude in $M$ ($\tilde{M}$).
\item
A little hierarchy is generated when the distribution for $M$ 
($\tilde{M}$) does not grow sufficiently to generate a large hierarchy, 
and the distribution for $z$ is strongly peaked to low values of $z$. 
This strong peaking need only persist for one or two orders of magnitude 
in $z$.
\end{itemize}
In both cases there is necessarily a large cut factor, either from 
a mild growth in the distribution for $M$ ($\tilde{M}$) over many 
decades for a large hierarchy, or from a strong distribution for $z$ 
over a much more limited range for a little hierarchy.  In general, 
a significant variation of a distribution function can arise from many 
sources --- the distribution of vacua in the landscape, the population 
of these vacua, integrating out parameters, and the number density 
of observers --- so that in the multiverse the existence of a hierarchy, 
either little or large, is not surprising.

For landscapes with a maximum critical mass $M_{c,{\rm max}}$
\begin{itemize}
\item
A little hierarchy is generated by a distribution that favors large 
$M$ near $M_{c,{\rm max}}$.  The size of the little hierarchy depends 
on both the strength of this distribution, and also on a loop factor 
that arises from the beta function for $x-y$.  The combination of 
these factors makes it probable that $M$ is as large as several TeV.
\end{itemize}
In section~\ref{sec:LH-cc} we argue that, in the presence of WIMP dark 
matter, a strong distribution for $M$ can arise from the probability 
distribution for the cosmological constant.
 
Until now we have not specified the physics behind $x$ and $y$; here 
we consider a few simple schemes, stressing how a cancellation between 
$x$ and $y$ could lead to a strong distribution for $z$ and hence 
a little hierarchy.  The quadratic divergences of the Standard Model 
contribute to both $x$ (e.g. the $SU(2)$ gauge contribution) and $y$ 
(e.g. the top quark contribution).  At mass scale $M$ suppose that these 
quadratic divergences are cut off by particles of mass $M_{W'}$ and 
$M_{t'}$, respectively.  If the theory at $M$ is supersymmetric, these 
are the wino and top squark masses, while if the theory at $M$ is 
non-supersymmetric (for example, with composite Higgs dynamics) they 
are some states of the model.  In all of these schemes, integrating 
out the physics at $M$ gives a low energy theory with
\begin{equation}
  m_h^2 \sim 
    \frac{g^2}{16\pi^2}M_{W'}^2 - \frac{3y_t^2}{16\pi^2}M_{t'}^2,
\label{eq:typ-model}
\end{equation}
where $g$ is the Standard Model $SU(2)$ gauge coupling.  The numerical 
coefficients are model dependent and extra contributions to $m_h^2$ 
are expected, for example those that regulate the hypercharge and 
Higgs-quartic divergences; but neither of these affects the arguments 
below.

Whether the theory at $M$ is supersymmetric or not, it may well be 
that $M$ arises as a dimensional transmutation, in which case it 
is reasonable that the distribution for $\ln M$ is sufficiently 
flat that a large hierarchy does not develop.  In this case, what 
distribution for the parameters in Eq.~(\ref{eq:typ-model}) would 
lead to a little hierarchy?

We need a distribution $f_z(z)$ peaked at low values, i.e. in 
Fig.~\ref{fig:M-z_obs_1} or Fig.~\ref{fig:tilM-z_obs_1} the probability 
force must have a large component downwards.  Since $z=0$ is not 
a special point from the fundamental theory point of view (see the 
discussion in section~\ref{subsec:illust}), this implies that most 
universes will be in the phase with $z < 0$, i.e. $m_h^2 > 0$.  In 
the multiverse, the positive term in Eq.~(\ref{eq:typ-model}) typically 
dominates over the negative term.  Such a distribution could arise 
in several ways.  For example, suppose that $M_{W'}/M_{t'}$ does 
not scan, but $g$ and $y_t$ do.  If $g$ has a distribution peaked 
at a (much) larger value than $y_t$, then most universes will have 
negative $z$, and near the observer boundary the probability force 
will be towards smaller values of $z$.   The shapes of the distributions 
for $g$ and $y_t$ need not be power law, as assumed for simplicity 
in section~\ref{subsec:y=1-v}.  For example, they could be Gaussians 
with peaks at $\bar{g}$ and $\bar{y}_t$, with $\bar{g}$ sufficiently 
larger than $\bar{y}_t$, so that most universes have $m_h^2 > 0$. 
The few universes that have $m_h^2$ negative will typically have 
low $z$ and therefore a little hierarchy.  For narrow Gaussians 
for $g^2$ and $y_t^2$ with the standard deviations $\delta_g$ 
and $\delta_y$, respectively, we find
\begin{equation}
  \langle z \rangle 
  = \biggl\langle \biggl| \frac{m_h^2}{A_y M^2} \biggr| \biggr\rangle 
  \approx \frac{\int_0^\infty z\, e^{-\frac{(z+z_c)^2}{2 \delta^2}} dz} 
    {\int_0^\infty e^{-\frac{(z+z_c)^2}{2 \delta^2}} dz} 
  \sim \frac{\delta^2}{z_c},
\label{eq:typ-model-2}
\end{equation}
where $z_c \equiv (A_g/A_y) \bar{g}^2 - \bar{y}_t^2 > 0$ and $\delta^2 
\equiv (A_g/A_y)^2 \delta_g^2 + \delta_y^2$, with $A_{g,y}$ positive 
coefficients defined by $m_h^2 = (A_g g^2 - A_y y_t^2)M^2$.  Alternatively, 
it could be that the scanning of the masses $M_{W'}$ and $M_{t'}$ is 
more important than that of the couplings, and that the little hierarchy 
results because the distributions typically give $M_{W'}$ (much) larger 
than $M_{t'}$.

In particular models, it is possible to see other situations that 
lead to a little hierarchy.  For example, in the minimal supersymmetric 
standard model it could be that the distributions for the supersymmetric 
Higgs mass parameter $\mu$ and the scale of the soft supersymmetry breaking 
mass parameters $\tilde{m}$ differ.  If $\mu$ is typically (much) larger 
than $\tilde{m}$, then most universes do not have electroweak symmetry 
broken by $\langle h \rangle \neq 0$.  This generically leads to a 
strong distribution preferring low $z$, and universes in the observer 
region will have a little hierarchy.  More generally, in the minimal 
supersymmetric standard model the electroweak phase boundary takes 
the form
\begin{equation}
  (|\mu|^2 + m_{H_1}^2)(|\mu|^2 + m_{H_2}^2) = |\mu B|^2,
\label{eq:mssm}
\end{equation}
with the soft Higgs mass-squared parameters $m_{H_1}^2$ and $m_{H_2}^2$ 
depending on other parameters of the theory via renormalization group 
scaling.  Any multiverse distribution for the parameters of the model 
that typically makes the left-hand-side of Eq.~(\ref{eq:mssm}) larger 
than the right-hand-side will generically lead to a little hierarchy. 
The scale of the superparticle masses are then raised significantly 
above $v$ and the measured values of the parameters should be close 
to satisfying the critical condition, Eq.~(\ref{eq:mssm}).

\section{Electroweak Symmetry Breaking as Observer Boundary}
\label{sec:EWSB-alt}

In the last section we assumed that the relevant observer 
boundaries for selecting the electroweak vacuum expectation value, 
$\langle h \rangle = v$, were those of neutron, deuteron and complex 
nuclei stability.  Of these three boundaries, the requirement that 
some complex nuclei are stable seems clearly to be the most robust 
requirement for observers.  For example, if the neutron is stable, 
nuclear energy is produced in diffuse protogalaxies rather than in 
stars.  While this is a drastic change from our universe, some form 
of observers might still be possible.  On the other hand, it is 
harder to imagine that some complex observers develop in the world 
in which the only stable nucleus is $p$ or $\Delta^{++}$.  In this 
section, we retain only the complex stable nuclei boundary, dropping 
the neutron and deuteron (in)stability requirements from the observer 
boundary.  This allows much smaller values for $v$.  How small can 
$v$ become while remaining in the observer region?  If electroweak 
symmetry is broken dominantly by the QCD condensate, there is a strong 
washout of the baryon asymmetry of the universe due to sphaleron effects. 
This implies that complex structures involving baryons do not arise 
if $v \simlt \Lambda_{\rm QCD}$, so that there is a complexity boundary 
for $v$ near $\Lambda_{\rm QCD}$ which is therefore a candidate for 
being part of the observer boundary.  Since $\Lambda_{\rm QCD}$ is much 
smaller than the scale of the electroweak symmetry breaking sector, 
$M$, we can speak of this boundary as the phase boundary between 
$\langle h \rangle = 0$ and $\langle h \rangle \neq 0$ phases.  In 
this section, we use only two boundaries: complex nuclear stability 
and the Higgs breaking of electroweak symmetry, which were both 
classified as catastrophic boundaries in section~\ref{sec:evid-prob}.

\begin{figure}[t]
  \center{\includegraphics[width=.6\textwidth]{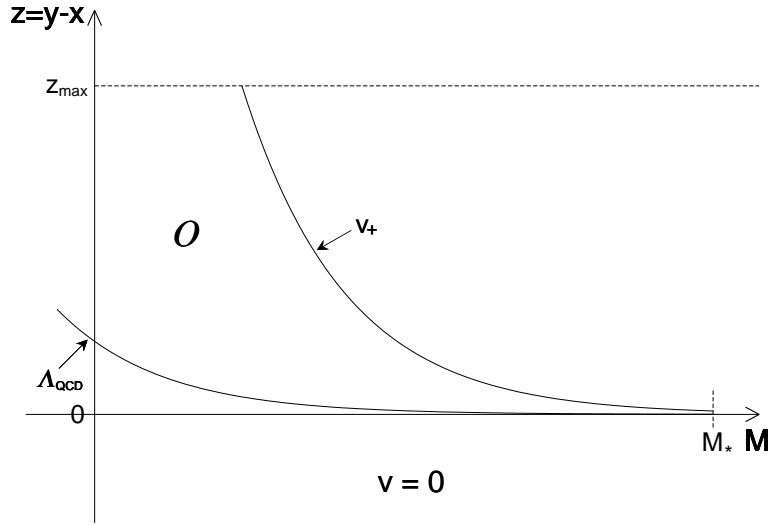}}
\caption{Sketch of the 2-dimensional observer region in the $M$-$z$ 
 plane.  The two boundaries of the observer region depicted are 
 $v = v_+$ and $v \approx \Lambda_{\rm QCD}$.}
\label{fig:M-z_phase}
\end{figure}

We consider the generic electroweak symmetry breaking sector discussed 
in the previous section, with relevant scanning parameters $x$, $y$ 
and $M$ and a hierarchy
\begin{equation}
  z \equiv y-x = \frac{|m_h^2|}{A M^2}.
\label{eq:z-phase}
\end{equation}
With $y_{u,d,e}$, $\alpha$ and $\Lambda_{\rm QCD}$ fixed, the relevant 
observer region is now as shown in Fig.~\ref{fig:M-z_phase}. 
Compared to Fig.~\ref{fig:M-z_obs_1}, the observer boundary at $v_+$ 
remains while that at $v_-$ has been replace by one at $v \approx 
\Lambda_{\rm QCD}$.  This greatly enlarges the observer region, with 
$v$ varying over about $4$~orders of magnitude compared to a factor~$4$ 
in section~\ref{sec:EWSB}.  Despite this, we find that there is very 
little change in the physical picture of environmental selection.

For example, in the case that $y$ is fixed and $x$ and $M$ have 
polynomial distributions as in Eq.~(\ref{eq:f_x-M}), the effective 
distributions for $z$ and $v$ are not changed and given as before (see 
Eqs.~(\ref{eq:f_z-2},~\ref{eq:f_v-large},~\ref{eq:f_v-little})) by
\begin{equation}
  d{\cal N} = f_z(z)\, dz 
    \sim (1-z)^n\, \frac{1}{z^{q/2}}\, dz,
\label{eq:f_z-phase}
\end{equation}
and
\begin{equation}
  d{\cal N} = f_v(v^2)\, dv^2 
    \sim \left\{
    \begin{array}{ll}
      v\, dv     & {\rm for}\;\; q \geq 2, \\
      v^{q-1} dv & {\rm for}\;\; q < 2.
    \end{array} \right.
\label{eq:f_v-phase}
\end{equation}
Once again, $q \geq 2$ gives a large hierarchy and $q < 2$ a little 
(or no) hierarchy.  Furthermore the size of the little hierarchy is 
largely governed by $n$, as discussed in section~\ref{subsec:y=1-v}. 
One difference is that the range of $v$ in the observer region is now 
much larger; how reasonable is it that our universe is about a factor 
of $5$ from $v_+$ and $3$~orders of magnitude from $\Lambda_{\rm QCD}$? 
For $0 \simlt q \simlt 1$ the observed value of $v$ is quite typical, 
but for larger $q$, $v$ becomes progressively more peaked near $v_+$. 
However, this cannot be viewed as evidence against a large hierarchy. 
For large $q$ it is unlikely that Eq.~(\ref{eq:f_v-phase}) is the 
correct distribution near $v_+$: as $v$ approaches this observer 
boundary so successively more nuclei become unstable, so that the 
observer factor, $n$, is likely to become smaller.

Given the observer region of Fig.~\ref{fig:M-z_phase} we can 
understand the origin of large or little hierarchies from the 
qualitative features of the distribution $f(z,M)$, without recourse 
to any particular functional form.  The observer boundary at $v_+$ 
causes a narrowing of the observer region at large $M$, so that 
a large hierarchy results only if the distribution gives a sufficient 
preference to large $M$ to overcome this narrowing.  If the probability 
force in the $M$ direction is insufficient, then the size of the 
little hierarchy depends on the strength of the probability force 
in the negative $z$ direction.  At first sight a force in the negative 
$z$ direction apparently pushes the electroweak vacuum expectation 
value close to the value $\Lambda_{\rm QCD}$ at the baryon washout 
boundary.  This is incorrect: such a force makes the most probable 
region that with low $z$, but the value of $v$ that this corresponds 
to changes with $M$.  This is why Eq.~(\ref{eq:f_v-phase}) involves 
$q$ and not $n$.

We again conclude that the little hierarchy problem (supersymmetric 
or not) is very easily solved by environmental selection.  Strongly 
varying distribution functions of the electroweak symmetry breaking 
sector are able to generate either a large or little hierarchy; 
furthermore this is not sensitive to which observer boundary is 
used to limit the lower value of $v$.

\section{Connections to the Cosmological Constant}
\label{sec:LH-cc}

In this paper we argue that evidence for the multiverse can be found 
in unnaturalness in the cosmological constant, nuclear physics, 
and electroweak symmetry breaking.  This implies that scanning of 
parameters is occurring in all three different arenas.  Is it correct 
to consider each scanning problem separately, or should there be 
a combined treatment?  In section~\ref{subsec:nucl-scan} we provided 
an analysis of the combined scanning in the Standard Model and 
electroweak symmetry breaking sectors.  In this section we study 
connections with the scanning of the cosmological constant.

In section~\ref{subsec:force}, we argued that ``integrating out'' a set 
of parameters $x_b$ modifies the distribution function for a set $x_a$
\begin{equation}
  f_{\rm eff}(x_a) = \int_{{\cal O}(x_a)} f_{\rm prior}(x_a, x_b)\, dx_b,
\label{eq:eff_xa}
\end{equation}
if the observer region ${\cal O}$ for $x_b$ depends on $x_a$.  In 
the present case we consider $x_b = \Lambda$ and $x_a$ to be the 
set of scanning parameters of the Standard Model and the electroweak 
symmetry breaking sector.  Assuming the multiverse probability 
distribution for $\Lambda$ to be flat on a linear scale
\begin{equation}
  f_{\rm eff}(x_a) 
  \propto f_{\rm prior}(x_a) \int^{\rho_{\rm NL}(x_a)} d\Lambda 
  = f_{\rm prior}(x_a) \, \rho_{\rm NL}(x_a),
\label{eq:eff_xa-2}
\end{equation}
where $\rho_{\rm NL}$ is the energy density of the universe when it 
goes non-linear.  Hence the distribution function for the parameters 
of the Standard Model and the electroweak symmetry breaking sector is 
modified by the scanning of the cosmological constant if $\rho_{\rm NL}$ 
depends on the parameters $x_a$.  In our previous analyses, the assumed 
form of the distribution should apply to $f_{\rm eff}$, rather than 
to $f_{\rm prior}$.  This implies that the scanning of the cosmological 
constant can affect the form of the distribution functions that 
appeared in the analyses in previous sections.

In this section, we explore some consequences of a nontrivial dependence 
of $\rho_{\rm NL}$ on $x_a$.  Perhaps the simplest origin for such 
a dependence is the case of WIMP dark matter.  The temperature of 
matter-radiation equality, $T_{\rm eq}$, depends on the WIMP mass 
via its annihilation cross section, $T_{\rm eq} \propto 1/\sigma_A 
\sim m_{\rm WIMP}^2$.  Hence $\rho_{\rm NL} \sim Q^3 T_{\rm eq}^4 \sim 
Q^3 m_{\rm WIMP}^8$, where $Q$ is the primordial density perturbation. 
Working with the generic electroweak symmetry breaking sector introduced 
in section~\ref{sec:EWSB}, with mass scale $M$ and dimensionless 
parameters $x_i$, the WIMP mass is proportional to $M$ so that
\begin{equation}
  T_{\rm eq} = g(x_i)\, M^2,
\label{eq:T_eq}
\end{equation}
where $g(x_i) > 0$ is a model dependent function, depending on the 
mass and interactions of the WIMP.  The effective distribution function 
for the electroweak symmetry breaking sector now becomes
\begin{equation}
  f_{\rm eff}(x_i,M) 
  \propto f_{\rm prior}(x_i,M)\, \int^{\rho_{\rm NL}(x_i,M)} d\Lambda 
  \sim f_{\rm prior}(x_i,M)\, g(x_i)^4 M^8 Q^3.
\label{eq:eff_xi-M}
\end{equation}
Note that here we have assumed that the parameter $Q$ does not scan. 
If $Q$ also scans, the result of integrating out cosmological parameters 
is altered from Eq.~(\ref{eq:eff_xi-M}), as we will see below.

If $Q$ does not scan, Eq.~(\ref{eq:eff_xi-M}) shows 
that the effective distribution for the parameters of the electroweak 
symmetry breaking sector, $\{ x_i, M \}$, can acquire an important 
component from integrating out the cosmological constant.  In particular, 
the cosmological constant may provide a strong probability force to 
larger values of $M$ through the $M^8$ factor.  At first sight it 
appears that this factor drives a large hierarchy, $M \gg v$, but 
this is not the case.  A crucial issue is: what stops the runaway 
to large $M$?  It is important to separate two cases.

In the first case, the factor $T_{\rm eq}^4 \sim \{ g(x_i) M^2 \}^4$ 
in $f_{\rm eff}$ pushes the amount of dark matter up to the maximum 
allowed by some astrophysical observer boundary, for example that 
of stellar collisions in galaxies~\cite{Tegmark:2005dy}, so that 
$T_{\rm eq}$ is essentially fixed to the boundary value $T_{\rm eq} 
= T_{{\rm eq},*}$.  The distribution function for the parameters 
relevant for electroweak symmetry breaking, $\{ x, y, \lambda_h, 
M \}$ where $x,y,\lambda_h \subset x_i$, is obtained by integrating 
out parameters that appear in $g(x_i)$ but not in $m_h^2$ or 
$\lambda_h$, within the observer region $T_{\rm eq} < T_{{\rm eq},*}$. 
This in general leads to a complicated dependence of $f_{\rm eff}$ 
on $\{ x, y, \lambda_h, M \}$, not in the simple form of 
Eq.~(\ref{eq:eff_xi-M}).  Note that as $M$ grows beyond the 
TeV scale, since $T_{\rm eq}$ stays to be $T_{{\rm eq},*}$, 
cancellations must occur in $g(x_i)$ so that $m_{\rm WIMP} \ll M$. 
This implies that a large hierarchy can develop only if $f_{\rm prior}$ 
has a strong preference towards larger values of $M$.

In the second case, the most probable region of scanning parameter 
space does not lead to the dark matter density being on the edge 
of its maximal value determined by astrophysics.  The values of the 
parameters $\{ x_i, M \}$ are determined by other physics, including 
electroweak symmetry breaking.  The distribution of the parameters 
relevant for electroweak symmetry breaking may take a form $f_{\rm eff} 
\sim f\, T_{\rm eq}^4$, so that the factor of $g(x_i)^4 M^8$ can play 
an important role in electroweak symmetry breaking.  Some examples 
of this are given in the following subsection.

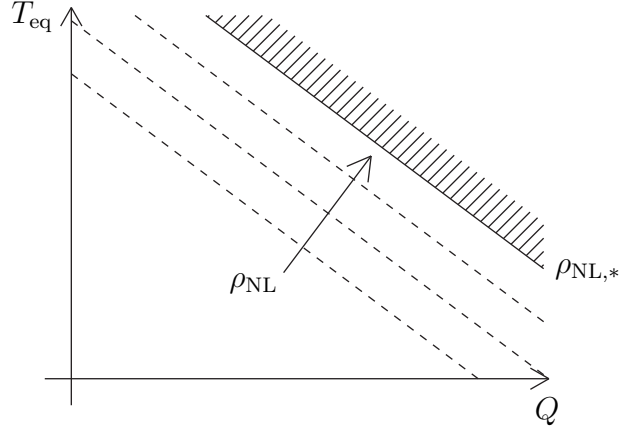
\begin{figure}[t]
\begin{center}
\begin{picture}(180,151)(0,-14)
  \Line(-10,0)(180,0) \Line(180,0)(173,4) \Line(180,0)(173,-4)
  \Line(0,-10)(0,140) \Line(0,140)(-4,133) \Line(0,140)(4,133)
  \Text(180,-7)[t]{$Q$} \Text(-7,137)[r]{$T_{\rm eq}$}
  \Line(50.6667,137)(178,41.5)
  \DashLine(24,137)(178,21.5){3}
  \DashLine(0,135)(178,1.5){3}
  \DashLine(0,115)(153.333,0){3}
  \Line(51.4286,136.429)(51.9996,137)
  \Line(54.2857,134.286)(56.9997,137)
  \Line(57.1429,132.143)(61.9999,137)
  \Line(60.0000,130.000)(67.0000,137)
  \Line(62.8571,127.857)(72.0001,137)
  \Line(65.7143,125.714)(75.7143,135.714)
  \Line(68.5714,123.571)(78.5714,133.571)
  \Line(71.4286,121.429)(81.4286,131.429)
  \Line(74.2857,119.286)(84.2857,129.286)
  \Line(77.1429,117.143)(87.1429,127.143)
  \Line(80.0000,115.000)(90.0000,125.000)
  \Line(82.8571,112.857)(92.8571,122.857)
  \Line(85.7143,110.714)(95.7143,120.714)
  \Line(88.5714,108.571)(98.5714,118.571)
  \Line(91.4286,106.429)(101.429,116.429)
  \Line(94.2857,104.286)(104.286,114.286)
  \Line(97.1429,102.143)(107.143,112.143)
  \Line(100.000,100.000)(110.000,110.000)
  \Line(102.857,97.8571)(112.857,107.857)
  \Line(105.714,95.7143)(115.714,105.714)
  \Line(108.571,93.5714)(118.571,103.571)
  \Line(111.429,91.4286)(121.429,101.429)
  \Line(114.286,89.2857)(124.286,99.2857)
  \Line(117.143,87.1429)(127.143,97.1429)
  \Line(120.000,85.0000)(130.000,95.0000)
  \Line(122.857,82.8571)(132.857,92.8571)
  \Line(125.714,80.7143)(135.714,90.7143)
  \Line(128.571,78.5714)(138.571,88.5714)
  \Line(131.429,76.4286)(141.429,86.4286)
  \Line(134.286,74.2857)(144.286,84.2857)
  \Line(137.143,72.1429)(147.143,82.1429)
  \Line(140.000,70.0000)(150.000,80.0000)
  \Line(142.857,67.8571)(152.857,77.8571)
  \Line(145.714,65.7143)(155.714,75.7143)
  \Line(148.571,63.5714)(158.571,73.5714)
  \Line(151.429,61.4286)(161.429,71.4286)
  \Line(154.286,59.2857)(164.286,69.2857)
  \Line(157.143,57.1429)(167.143,67.1429)
  \Line(160.000,55.0000)(170.000,65.0000)
  \Line(162.857,52.8571)(172.857,62.8571)
  \Line(165.714,50.7143)(175.714,60.7143)
  \Line(168.571,48.5714)(178,58.0004)
  \Line(171.429,46.4286)(178,52.9996)
  \Line(174.286,44.2857)(178,47.9997)
  \Line(177.143,42.1429)(178,42.9999)
  \Text(183,41)[l]{$\rho_{{\rm NL},*}$}
  \Line(80,40)(113,84) \Text(78,40)[tr]{$\rho_{\rm NL}$}
  \Line(113,84)(103.804,80.0718) \Line(113,84)(111.804,74.0718)
\end{picture}
\caption{The probability force from the cosmological constant in the 
 $Q$-$T_{\rm eq}$ plane.  The contours of $\rho_{\rm NL}$ are drawn 
 by dashed lines, while the observer boundary from astrophysics, 
 $\rho_{\rm NL} = \rho_{{\rm NL},*}$ is depicted in the solid line.}
\label{fig:Q-Teq}
\end{center}
\end{figure}

A completely different situation arises if the space of scanning 
parameters is increased to include $Q$.  As illustrated in 
Fig.~\ref{fig:Q-Teq}, the probability force from the cosmological 
constant no longer acts in the direction of increasing $m_{\rm WIMP}$, 
but in the direction of increasing $\rho_{\rm NL}$, so that now the 
issue becomes: what stops the runaway in the $\rho_{\rm NL}$ direction? 
It is again an astrophysical limit on the amount of dark matter, but 
this is now a limit on $\rho_{\rm NL}$ rather than on $T_{\rm eq}$: 
$\rho_{\rm NL} < \rho_{{\rm NL},*}$.  The probability distribution 
along this astrophysical boundary now depends on the distribution 
$f_Q(Q)$.  Integrating first $\Lambda$ and then $Q$ over the observer 
region, we obtain
\begin{equation}
  f_{\rm eff}(x_i,M) 
  \propto f_{\rm prior}(x_i,M)\, T_{\rm eq}^4 
    \int^{(\rho_{{\rm NL},*}/T_{\rm eq}^4)^{1/3}}\! Q^3 f_Q(Q)\, dQ,
\label{eq:eff_xi-M_Q-scan}
\end{equation}
where $T_{\rm eq}$ is given by Eq.~(\ref{eq:T_eq}).  The induced 
contribution to the effective distribution now depends on the form of 
the original distribution for the density perturbation $Q$, since the 
observer boundary for $Q$ depends on $T_{\rm eq} = T_{\rm eq}(x_i,M)$. 
However, in the simple case that $f_Q \propto 1/Q$, so that the $Q$ 
distribution is flat on a logarithmic scale, $f_{\rm eff}(x_i,M) 
\sim f_{\rm prior}(x_i,M)$, and the $T_{\rm eq}^4$ contribution is 
removed.  Hence, a nontrivial contribution to $f_{\rm eff}$ is generated 
by a nontrivial distribution for $\ln Q$, not by the probability force 
from the cosmological constant.

\subsection{Examples of the cosmological constant affecting electroweak 
 symmetry breaking}
\label{subsec:LH-cc}

We have seen that integrating out cosmological parameters with WIMP 
dark matter can generate a nontrivial component in the distribution 
of parameters relevant for electroweak symmetry breaking.  In general 
the induced distribution takes a complicated form depending on what 
parameters scan and how they are determined.  In the case that $Q$ 
does not scan and the dark matter density is not determined by an 
astrophysical bound, the effective distribution function receives 
a factor of $g(x_i)^4 M^8$, as shown in Eq.~(\ref{eq:eff_xi-M}). 
Assuming that parameters appearing in $g(x_i)$ but not in $m_h^2$ 
or $\lambda_h$ are determined independently of $M$, the distribution 
for the parameters $x$, $y$ and $M$, appearing in $m_h^2 = (x-y) A M$, 
takes the form
\begin{equation}
  f_{\rm eff}(x,y,M) 
  \sim f_{\rm prior}(x,y,M)\, \tilde{g}(x,y)^4 M^8,
\label{eq:eff_xyM}
\end{equation}
where $\tilde{g}(x,y)$ is related to $g(x_i)$ in Eq.~(\ref{eq:T_eq}). 
In this subsection we illustrate how the scanning of cosmological 
parameters can affect electroweak symmetry breaking, using the example 
of Eq.~(\ref{eq:eff_xyM}).

Suppose that $f_{\rm prior}$ is a rather mild function of $x$, $y$ 
and $M$.  The effective distribution $f_{\rm eff}$ is then determined 
essentially by the factor $\tilde{g}(x,y)^4 M^8$.  The fact that the 
dark matter density does not saturate the astrophysical bound implies 
that the runaway to larger $M$ due to the $M^8$ factor should be 
stopped not by the condition $T_{\rm eq} < T_{{\rm eq},*}$ but by 
some other physics.  The simplest possibility is that $M$ does not 
scan in the multiverse.  At the end of section~\ref{subsec:EWS-scan} 
we argued that if $M$ is fixed then a large hierarchy is to be expected, 
as in Eq.~(\ref{eq:range}).  However, this conclusion is reversed 
if $\Lambda_{\rm QCD}$ scans.  Since the nuclear physics boundaries 
depend on $M$ only through the ratio $M/\Lambda_{\rm QCD}$, it 
makes no difference to the argument on the size of the hierarchy 
whether $M$ scans or $\Lambda_{\rm QCD}$ scans.  Allowing $x$, $y$ 
and $\Lambda_{\rm QCD}$ to scan, the observer region is that of 
Fig.~\ref{fig:M-z_obs_1}, with $M$ replaced by $1/\Lambda_{\rm QCD}$. 
Taking a mild distribution for $\Lambda_{\rm QCD}$ avoids a large 
hierarchy.%
\footnote{In fact, consistency of the setup requires that the hierarchy 
 is not very large, since a value of $M$ much larger than the weak 
 scale makes the natural size of $T_{\rm eq} \sim 8\pi M^2/M_{\rm Pl}$ 
 much larger than $T_{{\rm eq},*}$, implying that the astrophysical 
 bound is saturated.}
The amount of the little hierarchy is then determined by the factor 
$\tilde{g}(x,y)^4$.  For example, for $\tilde{g} \sim x^m$ and 
a logarithmic distribution for $\Lambda_{\rm QCD}$, $\langle 
z \rangle \approx 1/(4m+2)$, so that a little hierarchy of order 
$10^{-2}$~--~$10^{-1}$ can be easily obtained for $m$ a factor of 
a few.  In general, with $\tilde{g}^4$ strongly preferring $x > y$, 
we obtain a little hierarchy.  The crucial physics here is that 
the large scale structure observer boundary, $\Lambda \approx 
\rho_{\rm NL}$, depends sensitively on the electroweak symmetry 
breaking parameters $x$ and $y$ through WIMP dark matter.  The 
probability distribution for the cosmological constant then generates 
an effective distribution for $x$ and $y$, which can be sharply 
varying.

Another possibility of preventing the runaway to large $M$ arises 
if the range for the scanning of $x$ and $y$ are such that there is 
a maximum energy, $M_{c,{\rm max}}$, above which the $m_h^2$ parameter 
is always positive.  As shown in detail in section~\ref{subsec:M_c}, 
a strong probability force to large $M$ leads to a little hierarchy, 
providing the nuclear observer boundaries of $v_\pm$ are able to scan, 
for example via a scanning of $\Lambda_{\rm QCD}$.  Here we simply 
point out that the strong force in the $M$ direction can arise from 
integrating out the cosmological constant, i.e. from the $M^8$ factor 
in Eq.~(\ref{eq:eff_xyM}).  This leads to a little hierarchy without 
the need for any cut factor beyond that for the cosmological constant. 
In the case of supersymmetry, with $d{\cal N} \sim M^8 dM$ ($q \approx 
O(10)$) the multiverse allows an improvement in the naturalness by 
a factor of $\approx q/\beta \approx O(100)$.

\section{Conclusions}
\label{sec:concl}

Environmental selection on a multiverse is a radical departure from 
conventional methods of fundamental physics for explaining physical 
phenomena.  Nevertheless, in light of the cosmological constant 
problem and the discovery of dark energy, it warrants further 
exploration.  Will sufficient evidence emerge to convince us 
that the multiverse exists?

Since we cannot directly explore other universes, it may be questioned 
whether evidence for the multiverse can be found at all.  However, 
theories in physics and cosmology that cannot be directly tested in 
the laboratory are far from new --- one only has to think of unified 
theories and inflation.  The inability to make direct laboratory 
tests of the new particles and interactions does not put these theories 
beyond the realm of science; rather, it leads to careful investigations 
of whether they make successful indirect numerical predictions for 
data that cannot be satisfactorily explained by other means.  Thus 
for any new theory or framework that cannot be directly probed, 
two questions are important
\begin{enumerate}
\item[(a)]
In the absence of the new theory, is there numerical data that cannot 
be adequately explained using other known theories?  Indeed, does the 
data represent a problem for existing theories?
\item[(b)]
Does the new theory provide a numerical understanding of the data, 
thus solving the previous problem?
\end{enumerate}

For inflation, the data of (a) includes the flatness and isotropy 
of the universe, and the spectrum of density perturbations.  So far, 
these highly significant cosmological problems have only been solved 
by inflation.  Similarly, unified theories provide an understanding 
of gauge coupling constant unification.  If the numerical understanding 
of the data is sufficiently precise, and thought to be better than 
competing theories, then the new theory may become the provisional 
standard view.  As long as the numerical significance of its predictions 
is not overwhelming, one must stress the provisional nature of the 
understanding and the importance of seeking both further arenas in 
which it can be tested and new competing theories.  For example, 
proton decay would provide further evidence for unified theories, 
but despite intensive searches, such evidence is still lacking.

Seeking evidence for the multiverse is no different in principle than 
seeking evidence for other theories that cannot be directly probed in 
the laboratory.  In this paper we have argued that evidence for the 
multiverse can be found in three different arenas: the cosmological 
constant, nuclear physics and electroweak symmetry breaking.  In all 
three cases the conventional approach based on symmetries has not 
provided a numerical understanding of the data, rather in each case 
it leads to naturalness problems.  The observed values of parameters 
are very close to special values that are critical for the formation 
of some complex structure, yet this closeness is not adequately 
explained by any symmetry.  An observer region in the parameter 
space of some theory is defined by requiring the existence of certain 
complex structures necessary for observers.  Unnaturalness results 
if the observer region is very small compared to the entire parameter 
space, or if we observe values of the parameters very close to the 
boundary of the observer region.  We have introduced a naturalness 
probability, $P$, that allows a numerical evaluation of these problems. 
The value of $P$ can be highly dependent on the theory $T$ under 
consideration.  The evidence for unnaturalness, (a), is governed 
by the maximal value of $P_T$ that can be obtained in {\it simple} 
theories: the lower the maximal $P_T$, the more severe the naturalness 
problem.

In each arena the multiverse easily and generically solves the 
naturalness problem.  Each arena is somewhat different, and we 
summarize our results for each below; but there are also some common 
features.  In all three cases, environmental selection elegantly 
explains why our universe is not to be found in the largest region 
of parameter space.  Furthermore, the multiverse distribution is 
likely to have a strong dependence on the parameters of the low energy 
effective theory, through the landscape of vacua of the fundamental 
theory, the population mechanism, integrating out parameters, and 
from the physics that determines the density of observers.  With 
a strongly varying distribution, it is most probable to observe 
a universe close to the observer boundary, solving naturalness 
problems and leading to predictions.  Given the current theoretical 
status, the multiverse appears to us to provide the most elegant 
and plausible prediction for several parameters, including 
$\Lambda$ and $m_{u,d,e}$.

The cosmological constant problem is the most severe naturalness 
problem, with a naturalness probability in the range
\begin{equation}
  P_\Lambda \approx (10^{-120}~\mbox{--}~10^{-60}).
\label{eq:P_Lambda}
\end{equation}
The robustness of the observer boundary is particularly convincing 
--- how could observers form in a dilute gas of inflating elementary 
particles?  The interesting open questions are whether our universe is 
sufficiently close to the observer boundary, and how runaway behavior 
can be prevented if $T_{\rm eq}$ and $Q$ scan.  These questions, however, 
are secondary: a notoriously intractable problem has an elegant solution 
that predicts dark energy.

Astrophysicists have often remarked on the special values of parameters 
required for a variety of phenomena in nuclear physics.  We have 
obtained the observer region in the 4-dimensional parameter space 
$m_{u,d,e}/\Lambda_{\rm QCD}$, $\alpha$ resulting from the stability 
boundaries for neutrons, deuterons and complex nuclei, as shown in 
Fig.~\ref{fig:region}.  We find this observer region to be small, 
with our universe within $(10$~--~$30)\%$ of the neutron stability 
boundary, so that in the Standard Model the naturalness probability 
is $P_{{\rm nuc},{\rm SM}} \approx (10^{-16}$~--~$10^{-4})$.  Even 
if we could construct a theory of flavor with successful, precise 
predictions for the Yukawa couplings $y_{u,d,e}$ and at the same time 
find a theory that correctly predicted the weak scale, $v$, there 
would still be a naturalness probability associated with the value of 
$\Lambda_{\rm QCD}$ that we estimate to be $\approx (0.01$~--~$0.04)$. 
Despite decades of experiments on flavor physics, with recent increasing 
levels of accuracy in $B$ meson and neutrino physics, progress on 
a theory of flavor has been limited.  The most promising theories 
appear to be based on flavor symmetries, with a sequential pattern 
of transferring symmetry breaking to successive generations of quarks 
and charged leptons.  There are a great number of candidate theories, 
but none is sufficiently promising to be widely recognized as the 
standard.  We have estimated that such a lack of progress in flavor 
physics, especially in the first generation masses, decreases the 
naturalness probability of the nuclear observer region by about 
an order of magnitude, leading to
\begin{equation}
  P_{\rm nuc} \simlt (10^{-3}~\mbox{--}~10^{-2}).
\label{eq:P_nuc}
\end{equation}
We think that this estimate is conservative.  For example, in theories 
with Abelian flavor symmetries this estimate does not take into account 
that there are many simple ways of assigning charges to the three 
generations.  Given the state of theories of flavor, we suspect that 
the naturalness probability of the nuclear observer boundary is less 
than this conservative estimate.  Suppose that $m_{u,d,e}$ had each 
been significantly different, for example by an order of magnitude. 
It would still be possible to accommodate this in theories with Abelian 
flavor symmetries, with about the same level of success as for the 
actual observed values, by changing the charges of the first generation.

The nuclear naturalness problem certainly requires more than $1\%$ 
fine-tuning, and has not received sufficient recognition.  If symmetries 
rule flavor physics, it must be viewed as purely accidental; while 
in the multiverse it can be viewed as a prediction.  After decades of 
studying theories of flavor, the current status is that we are far from 
a convincing explanation for the masses of the electron, the up quark 
and the down quark.  The multiverse allows simple arguments that 
relate these masses to the QCD scale.  For example, if the multiverse 
distribution strongly favors isospin restoration, then $m_e$ and 
$m_d - m_u$ are expected to be close to $\delta_{\rm EM} \simeq 
1.0 \pm 0.5~{\rm MeV}$, the electromagnetic mass difference of the 
proton and neutron.  It is true that such multiverse predictions 
require an assumption on the form of the distribution function; 
but such assumptions may be much simpler than the choice of flavor 
group, representations and sequential symmetry breaking of the 
standard approach.

The multiverse predictions for $m_{u,d,e}/\Lambda_{\rm QCD}$ can be 
retained even if unified and/or flavor symmetries describe the overall 
pattern of quark and lepton masses and mixings.  This allows us to 
preserve particular successful relations, such as $\theta_C \sim 
\sqrt{m_d/m_s}$ or $m_b = m_\tau$ at the unified scale.  The only 
requirement is that the theory contain three independent scanning 
parameters that allow $m_{u,d,e}/\Lambda_{\rm QCD}$ to be 
environmentally selected.

A simple, natural theory of electroweak symmetry breaking is lacking. 
The natural regions of simple technicolor, supersymmetric and composite 
Higgs models have been excluded.  Precision measurements of electroweak 
observables, together with direct searches for the Higgs boson, have 
led to successive increases in unnaturalness, leading to
\begin{equation}
  P_{\rm EWSB} \simlt (10^{-2}~\mbox{--}~10^{-1}),
\label{eq:P_EWSB}
\end{equation}
in simple models.  This problem in the data is of type (a) and is widely 
appreciated, so that much recent research has focused on building models 
that alleviate the problem.  We have shown that, no matter what the 
ultimate physics at mass scale $M$ behind electroweak symmetry breaking, 
environmental selection on a multiverse leads very easily to $M \gg v$, 
with either a little or large hierarchy.  This is a very robust result 
requiring only that some parameters of electroweak symmetry breaking 
scan, and that the multiverse distribution is strongly varying.  A 
distribution $f(M) \sim M^q$ gives a large hierarchy if $q \geq 2$ 
and a little hierarchy (or no hierarchy) for $q < 2$, which includes 
the important case of $M$ being induced by a dimensional transmutation 
$(q=0)$.  For $q<2$ the little hierarchy, $v^2/M^2$, gains a factor of 
$\approx 1/n$ from the multiverse, where $n$ describes the peaking of 
a distribution in some other parameter of the electroweak symmetry 
breaking sector.  In multiverses where electroweak symmetry breaking 
is only possible if $M$ is below some critical value, $M_{c,{\rm max}}$, 
a little hierarchy develops from a distribution favoring large $M$. 
The size of the hierarchy, $v^2/M^2$, is enhanced by a loop factor 
$\beta$, gaining a factor of $\approx \beta/q$, so that $M$ is 
typically in the TeV region or larger.  These results are independent 
of what other Standard Model parameters are scanning, including Yukawa 
couplings and $\Lambda_{\rm QCD}$, and do not even depend on whether 
selection is happening at the nuclear stability boundaries, or at the 
phase boundary for electroweak symmetry breaking itself.  Hence we 
stress that the multiverse provides a very general solution to the 
hierarchy problem, whether little or large.

All three naturalness problems have the common feature of being solved 
by a multiverse distribution that makes observers typically close to an 
observer boundary.  There may be connections between the three problems, 
arising from integrating out certain parameters in a more fundamental 
theory.  For example, the probability force driving a little hierarchy 
may originates from the probability distribution for the cosmological 
constant, with WIMP dark matter acting as a mediator.

The current status of naturalness in electroweak symmetry breaking, 
Eq.~(\ref{eq:P_EWSB}), indicates that either we have not yet arrived 
at the right theory or that environmental selection is playing an 
important role.  The LHC will determine the correct interpretation 
of Eq.~(\ref{eq:P_EWSB}), leading us either to a new natural theory, 
or to a third arena for multiverse evidence.  Even if the naturalness 
probability is much larger than for the cosmological constant, the 
pervasive pattern of a finely tuned universe will make the multiverse 
much harder to dismiss.  The importance of the LHC in this regard 
cannot be overemphasized: for the cosmological constant and nuclear 
naturalness problems we are stuck --- we may not be able to experimentally 
determine the relevant theory beyond what we already know, in which 
case increasing the naturalness probability is a theoretical enterprise. 
However, the LHC will teach us a great deal about the theory of 
electroweak symmetry breaking and hence will lead to a better 
determination of $P_{\rm EWSB}$.  One possibility is that the LHC 
will reveal a completely natural theory that we have not been able 
to invent.  Below we mention a few examples where LHC data could 
determine a small value for $P_{\rm EWSB}$.

If the LHC discovers a light Higgs boson of mass $m_{\rm Higgs}$ and 
sets a limit of $M_{\rm col}$ on new colored particles, then
\begin{equation}
  P_{\rm EWSB} 
    \simlt 0.08 \left(\frac{m_{\rm Higgs}}{150~{\rm GeV}}\right)^2 
      \left(\frac{2~{\rm TeV}}{M_{\rm col}}\right)^2.
\label{eq:P_light-Higgs}
\end{equation}
This result is true in the vast majority of theories, although some 
counterexamples are known.  Another possibility is that only a light 
Higgs boson is discovered, with a mass very close to the vacuum instability 
limit of the Standard Model.  While Eq.~(\ref{eq:P_light-Higgs}) still 
applies as a direct consequence, there is the additional implication 
that the hierarchy is large with a very much smaller $P_{\rm EWSB}$. 
Evidence for a large hierarchy could also emerge from the discovery 
of split supersymmetry.   Alternatively, the discovery of weak scale 
supersymmetry, with a light Higgs boson and a top squark heavier 
than $\approx 1~{\rm TeV}$ would indicate a little hierarchy with 
a naturalness probability
\begin{equation}
  P_{\rm EWSB} 
    \simlt 0.05 \left(\frac{m_{\rm Higgs}}{130~{\rm GeV}}\right)^2 
      \left(\frac{1~{\rm TeV}}{m_{\tilde{t}}} \right)^2 
      \left(\ln\frac{M_{\rm mess}/m_{\tilde{t}}}{10}\right)^{-1},
\label{eq:P_TeV-stop}
\end{equation}
where $M_{\rm mess}$ is the messenger scale of supersymmetry breaking.

The discovery of dark energy has verified a remarkable prediction of 
the multiverse; but this could be undermined by the discovery of an 
alternative solution to the cosmological constant problem.  The nuclear 
stability boundaries imply at least $1\%$ fine-tuning in any known 
theory, and the multiverse allows a striking understanding of $m_{u,d,e}$. 
Our current theories of electroweak symmetry breaking are unnatural; 
a confirmation by the LHC would solidify evidence for the multiverse 
in a third arena.  Even with $P_{\rm nuc}$ and $P_{\rm EWSB}$ much 
larger than $P_\Lambda$, the three arenas together would provide 
significant, robust evidence for a multiverse.

\section*{Acknowledgments}

We thank Jesse Thaler for useful conversations.  This work was supported 
in part by the Director, Office of Science, Office of High Energy 
and Nuclear Physics, of the US Department of Energy under Contract 
DE-AC02-05CH11231, and in part by the National Science Foundation 
under grant PHY-0457315.  The work of Y.N. was also supported by 
the National Science Foundation under grant PHY-0555661, by a DOE 
OJI, and by an Alfred P. Sloan Foundation.

\end{document}